\documentclass[11pt,a4paper]{article}
\usepackage{jheppub}
\usepackage{amsmath,amssymb,amsfonts,bm,amscd}
\usepackage{mathtools}
\usepackage{graphicx}
\usepackage{multirow}
\usepackage{verbatim}
\usepackage{appendix}
\usepackage{fancybox}
\usepackage{array}

\newcommand{\bea}{\begin{eqnarray}}
\newcommand{\eea}{\end{eqnarray}}
\newcommand{\be}{\begin{equation}}
\newcommand{\ee}{\end{equation}}

\newcommand{\Z}{{\mathbb Z}}
\newcommand{\R}{{\mathbb R}}
\newcommand{\C}{{\mathbb C}}

\def\Tr{{\rm Tr \,}}

\def\frak{\mathfrak}

\def\tilde{\widetilde}
\def\hat{\widehat}
\def\bar{\overline}

\def\CA{{\mathcal A}}
\def\CB{{\mathcal B}}

\def\CE{{\mathcal E}}

\def\CH{{\mathcal H}}
\def\CI{{\mathcal I}}

\def\CL{{\mathcal L}}
\def\CM{{\mathcal M}}
\def\CN{{\mathcal N}}
\def\CO{{\mathcal O}}

\def\CR{{\mathcal R}}

\def\CW{{\mathcal W}}

\newcommand{\cp}{{\mathbb{C}}{\mathbf{P}}}

\renewcommand{\bar}{\overline}
\renewcommand{\hat}{\widehat}

\def\Tr{{\mathrm{Tr}}}

\title{Trisecting non-Lagrangian theories}

\author{Sergei Gukov}

\affiliation{Walter Burke Institute for Theoretical Physics, California Institute of Technology, \\ Pasadena, CA 91125, USA}

\abstract{We propose a way to define and compute invariants of general smooth 4-manifolds based on topological twists of
non-Lagrangian 4d $\CN=2$ and $\CN=3$ theories in which the problem is reduced to a fairly standard computation in topological A-model,
albeit with rather unusual targets, such as compact and non-compact Gepner models, asymmetric orbifolds, $\CN=(2,2)$ linear dilaton theories,
``self-mirror'' geometries, varieties with complex multiplication, {\it etc.}

\vspace{2cm}

\vspace{2cm}

\texttt{CALT-TH-2017-034}
}

\begin{document}
\maketitle



\section{Introduction}

The topological twist of four-dimensional super-Yang-Mills theory, proposed nearly 30 years ago \cite{Witten:1988ze},
motivated many important developments in physics and in pure mathematics, as well as connections between the two subjects.
If $SO(4) \cong SU(2)_1 \times SU(2)_2$ denotes the holonomy group of the four-dimensional Euclidean ``space-time''
(which must be a smooth 4-manifold) and $SU(2)_R \times U(1)_r$ is the R-symmetry of the 4d $\CN=2$ theory,
then the topological twist replaces $SU(2)_1$ by the diagonal subgroup of $SU(2)_1 \times SU(2)_R$.
Under this operation, the supercharges ${Q^i}_{\alpha}$ and $\tilde Q_{i \dot \alpha}$ that generate
4d $\CN=2$ supersymmetry transform as
\be
({\bf 2}, {\bf 1}, {\bf 2})^{\frac{1}{2}} \oplus ({\bf 1}, {\bf 2}, {\bf 2})^{-\frac{1}{2}}
\quad \xrightarrow[\text{}]{\text{topological twist}} \quad
({\bf 2}, {\bf 2})^{-\frac{1}{2}} \oplus ({\bf 3}, {\bf 1})^{\frac{1}{2}} \oplus ({\bf 1}, {\bf 1})^{\frac{1}{2}}
\label{SUSYtwist}
\ee
And the key point is that the right-hand side contains a scalar supercharge $Q$
which allows to put 4d $\CN=2$ theory on a general curved 4-manifold $M_4$.

In physics, it is now becoming increasingly clear that quantum field theories without
a traditional Lagrangian description play an important role and, possibly, even populate much of the QFT landscape.
They also offer promising opportunities in the search of new 4-manifold invariants,
especially since all Lagrangian 4d $\CN=2$ theories studied up to present day only
produce 4-manifold invariants that can be expressed via Donaldson or Seiberg-Witten invariants.
This experience further suggests that 4d $\CN=2$ strongly-coupled theories that have a dual weakly-coupled
Lagrangian description (or can be connected to one by changing marginal couplings)
are perhaps not as promising in producing new 4-manifold invariants as truly non-Lagrangian theories.

In fact, since the topological twist \eqref{SUSYtwist} relies on the supersymmetry algebra,
it applies equally well to non-Lagrangian 4d $\CN=2$ theories.
Moreover, searching for new 4-manifold invariants, it was recently pointed out \cite{Dedushenko:2017tdw}
that the Coulomb branch index could be used as a tool to identify promising candidates among 4d $\CN=2$ theories.
It clarifies, among other things, why theories that have a Lagrangian description (anywhere in the space
of marginal couplings) are less promising candidates and points to the non-Lagrangian theories, such as:

\begin{itemize}

\item Argyres-Douglas theories \cite{Argyres:1995jj} (including generalizations \cite{Xie:2012hs});

\item $\CN=3$ theories (see {\it e.g.} \cite{Aharony:2015oyb,Garcia-Etxebarria:2015wns,Aharony:2016kai});

\item theories with non-freely generated Coulomb branch chiral rings \cite{Argyres:2017tmj}.

\end{itemize}

\noindent
In particular, the latter would be an ideal candidate for producing new invariants of smooth 4-manifolds
according to the criterion of \cite{Dedushenko:2017tdw}.

Motivated by these recent developments, we wish to initiate a systematic study of the topological
twist \eqref{SUSYtwist} applied to non-Lagrangian 4d $\CN=2$ theories, including examples from the above list.
In particular, despite the lack of a Lagrangian formulation, we need to formulate a sufficiently general
and concrete proposal that will allow to define, at least in principle, 4-manifold invariants computed by
the partition function of the topologically twisted theory.
Not only that, we want it to be practically useful and efficient,
allowing to compute invariants of a given 4-manifold --- say, Horikawa surfaces --- in finite time over a cup of coffee.
How could this be possible if we do not even know the Lagrangian of the 4d theory we wish to twist?

At a conceptual level, the answer is rooted in the very fact that we know about the existence of such mysterious non-Lagrangian theories, despite the lack of their Lagrangian description. It means we know something about these theories and do have tools to analyze them. The problem, then, is to align the tools available to us with the needs of a topological twist, which does not require knowing {\it everything} about a non-Lagrangian theory in question. It only requires {\it some} information and, as long as we can get that, we are in business.

At a practical level, the method proposed below relies on two tricks (or, two ideas).
The first is that ``non-Lagrangian-ness'' is not a pathological disease, in the sense that if a given QFT lacks Lagrangian description in flat space, its dimensional reduction --- especially, topological reduction that we need for our applications --- can very well be Lagrangian.
For example, in recent years this strategy proved very successful in understanding formerly mysterious Argyres-Douglas theories,
which gain Lagrangian description upon reduction to three dimensions \cite{Benini:2010uu,Xie:2012hs}.
For our purposes, we will need to understand a similar reduction of non-Lagrangian 4d theories down to two dimensions,
with a partial topological twist along genus-$g$ Riemann surfaces.
In the case of Lagrangian theories, such topological reduction
was studied in \cite{Bershadsky:1995vm,Harvey:1995tg,Lozano:1999us,Losev:1999tu,Maldacena:2000mw,KW}
and our job will be to explore its non-Lagrangian analogue.

One might rightfully ask why --- among various possibilities --- we choose to focus on topological reduction to two dimensions.
Apart from the simple answer that compactification on a 2-manifold takes us half-way toward our main goal
(which is a compactification on a 4-manifold $M_4$) in the main text we shall see more substantial reasons
why this choice turns out to be especially convenient.
For example, a 2-form global symmetry whose gauging relates variants of 4d $\CN=3$ theories \cite{Aharony:2016kai}
becomes an ordinary symmetry upon reduction to two dimensions, {\it etc.}
Yet, topological reduction on a Riemann surface is clearly rather special and, as we pointed out a moment ago,
it only takes us half-way toward our main goal.
So, how can it help us in producing invariants of arbitrary 4-manifolds?

\begin{figure}[h]
\centering
\bigskip
\bigskip
\bigskip
$\phantom{\int_{\int}}$\\
$\phantom{Z}$\\
\begin{minipage}{0.45\textwidth}
\centering
\begin{picture}(100,100)
\put(-31,-10){\includegraphics[width=0.85\textwidth, height=0.255\textheight]{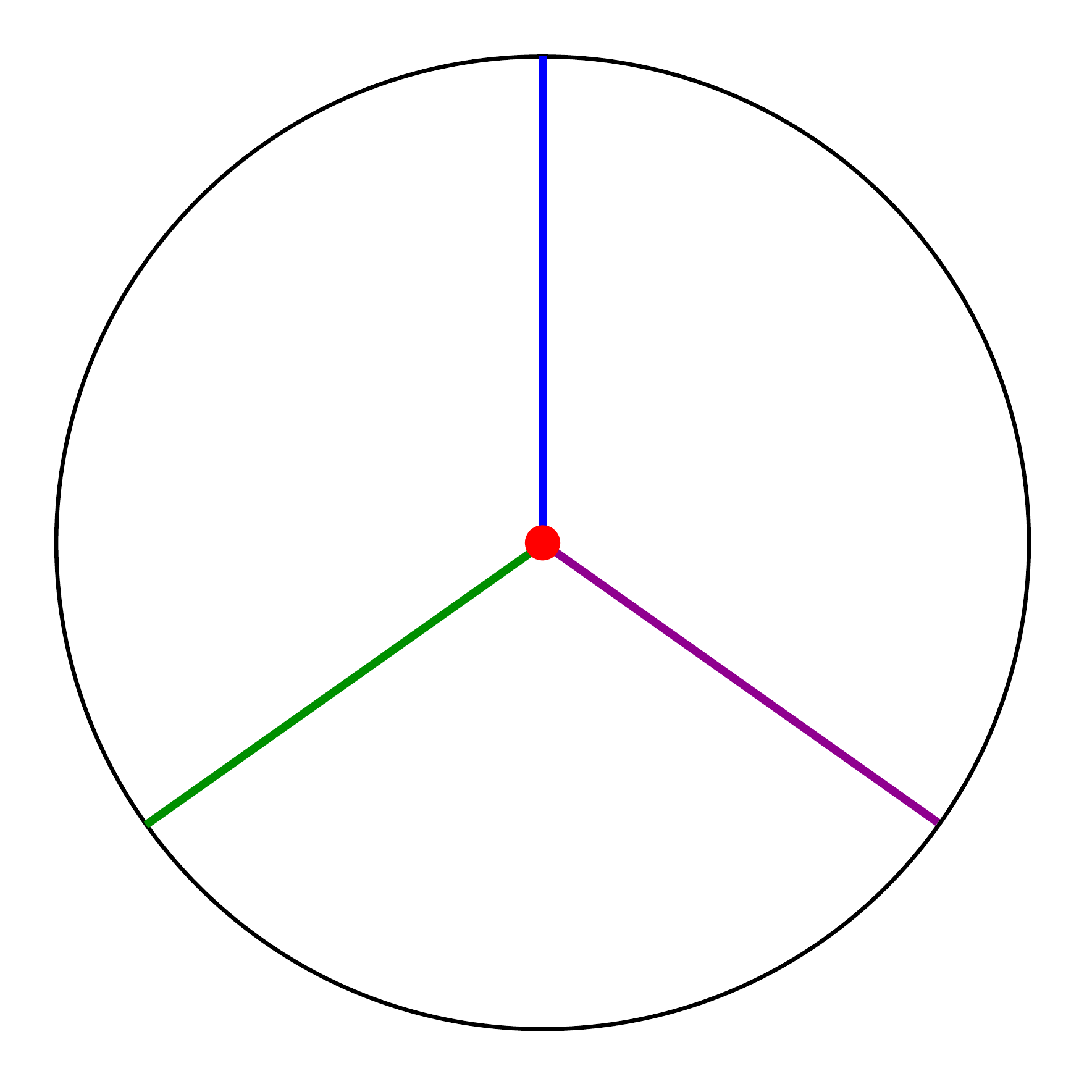}}
\put(43,54){{\Large $F_g$}}
\put(52,105){{\Large $M_3^{(\alpha)}$}}
\put(75,61){\rotatebox{-33}{{\Large $M_3^{(\beta)}$}}}
\put(-9,45){\rotatebox{33}{{\Large $M_3^{(\gamma)}$}}}
\end{picture}
\caption{Cutting a 4-manifold is like cutting a cake: with only three skillful cuts, any 4-manifold can be trisected into three basic pieces.}
\label{fig:trisectMMM}
\end{minipage}
\qquad
\begin{minipage}{0.45\textwidth}
\centering
\begin{picture}(100,100)
\put(-26,-5){\includegraphics[width=0.8\textwidth, height=0.24\textheight]{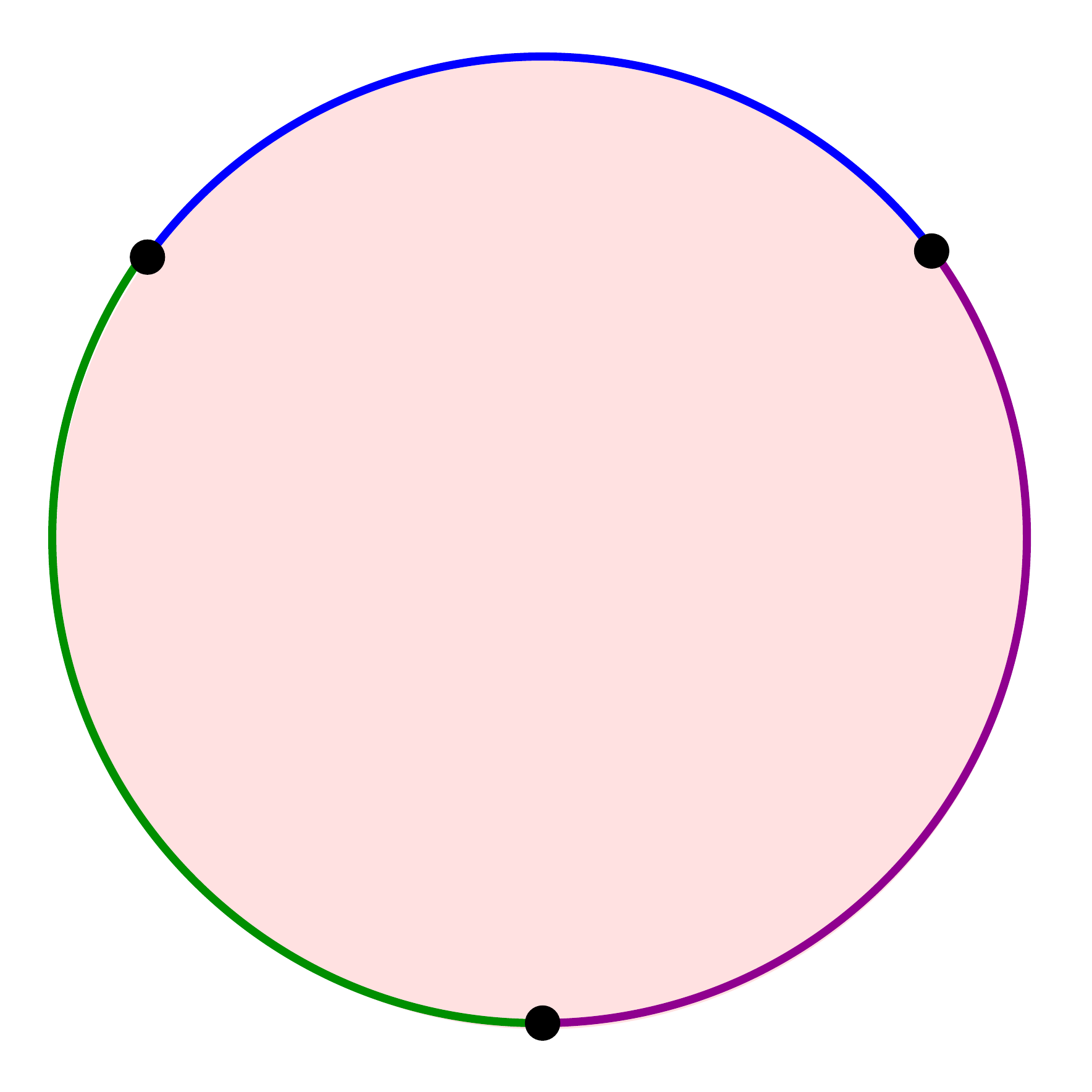}}
\put(39,147){{\huge $\CB_{\alpha}$}}
\put(118,40){{\huge $\CB_{\beta}$}}
\put(-37,40){{\huge $\CB_{\gamma}$}}
\put(23,67){{\huge $\CM (F_g)$}}
\end{picture}
\caption{Disk amplitude in the A-model of $\CM (F_g)$, with three Heegaard boundary conditions, dual to the trisection in Figure~\ref{fig:trisectMMM}.}
\label{fig:diskBBB}
\end{minipage}
\end{figure}

This is where the second important idea (trick) comes into play.
Think of amusing geometry problems like ``How many cuts does it take to divide a cake into 8 pieces?''
In the wild world of 4-manifolds, one might expect that cutting a randomly chosen smooth 4-manifold into
simple pieces requires a fairly large number of cuts. Surprisingly, this is not the case.
It takes only three cuts to decompose any 4-manifold into basic pieces~\cite{MR3590351,MR2222356}.
The trick is to cut (or, ``trisect'') along three solid handlebodies,
which share a common genus-$g$ boundary (see Figure~\ref{fig:trisectMMM}):
\be
F_g \; = \; M_3^{(\alpha)} \, \cap \, M_3^{(\beta)} \, \cap \, M_3^{(\gamma)}
\ee
As a result of this decomposition, a 4-manifold is labeled by a genus-$g$ surface $F_g$
and three $g$-tuples of simple closed curves in $F_g$ that encode the gluing data.

Thanks to this clever decomposition, we can formulate the topological twist of a 4d non-Lagrangian theory on $M_4$
in terms of 2d topological theory $\CM (F_g)$ on a disk, with three basic boundary conditions (a.k.a. branes);
generalizing the discussion in \cite{Gukov:2007ck,Gadde:2013sca} we call these ``Heegaard boundary conditions.''
Such topological branes are close cousins of the $A$-polynomial curve in $( \C^* \times \C^* ) / \Z_2$ and in many cases
can be described by equally explicit equations, as in \cite{MR3248065} or \cite{Apol}.

The advantage of this approach is that all of the individual pieces are fairly elementary,
so one might hope to understand them even in a non-Lagrangian 4d theory.
The price to pay is that gluing data is very non-trivial; in the A-model $\CM (F_g)$ this translates
into the action of the mapping class group $\text{MCG} (F_g)$ on branes.
Hence, in practice, whether one can compute 4-manifold invariants in a given non-Lagrangian theory via this method
to a large extent depends on how much can be said about its topological reduction on a genus-$g$ surface.

The formulation of 4-manifold invariants via topological 2d theories has another advantage:
in the end it only requires fairly standard tools, many of which can be made mathematically rigorous.
In fact, there are at least two examples where this formulation was already used in the mathematical literature:
one is the case of Seiberg-Witten invariants~\cite{MR2222356} and the other will be discussed in section~\ref{sec:trisec}.
The latter involves homological mirror symmetry for elliptic curves~\cite{MR1403918},
which by now is mathematically well established~\cite{MR1633036}.
Our hope is that, when sufficiently developed, the approach based on trisections and topological
disk amplitudes can give essentially a combinatorial definition of topological partition functions
on arbitrary 4-manifolds.

The paper is organized as follows. In section~\ref{sec:trisec}, we flesh out the key steps and ingredients
of the trisection approach to topological twists of non-Lagrangian theories.
Then, in sections \ref{sec:N3} and \ref{sec:AD},
we illustrate how it looks like in concrete examples of 4d non-Lagrangian theories.
To keep things simple and to avoid clutter, when we discuss concrete examples we mostly focus on `rank-1' theories;
in the case of non-Lagrangian theories, this terminology refers to the complex dimension of the Coulomb branch.
Further generalizations and calculations will appear elsewhere.


\section{A duality between trisections and disk amplitudes}
\label{sec:trisec}

Trisections of Gay and Kirby \cite{MR3590351} (and, similarly, Heegaard triples of Ozsv\'ath and Szabo \cite{MR2222356})
are basically 4-manifold versions of the more familiar Heegaard decompositions for 3-manifolds.
Much like the latter, trisections provide a simple and efficient way to build {\it any} smooth 4-manifold from elementary
pieces, such that the complexity of 4-manifolds is encoded in the gluing data.

Recall, that any closed 3-manifold $M_3$ admits a Heegaard splitting into a pair of handlebodies,
$M_3^{(1)}$ and $M_3^{(2)}$, glued along some genus-$g$ surface $F_g$,
\be
M_3 \; = \; M_3^{(1)} \; \cup_{F_g} \; M_3^{(2)}
\label{Heegaard}
\ee
While each handlebody $M_3^{(i)} \cong \natural^g (S^1 \times B^2)$ is rather simple,\footnote{We use the symbol
``$\#$'' for the connected sum and ``$\natural$'' for the boundary sum.}
the power of the construction comes from diffeomorphisms involved in gluing, which can be summarized
by the data $(F_g, \alpha, \beta)$ called the Heegaard diagram of $M_3$.
Here, $\alpha = (\alpha_1, \ldots, \alpha_g)$ is a $g$-tuple of simple closed curves in $F_g$
that bound compressing disks in $M_3^{(1)}$, and similarly $\beta$ specifies the handlebody $M_3^{(2)}$.

\begin{figure}[ht]
\centering
\includegraphics[width=3.0in]{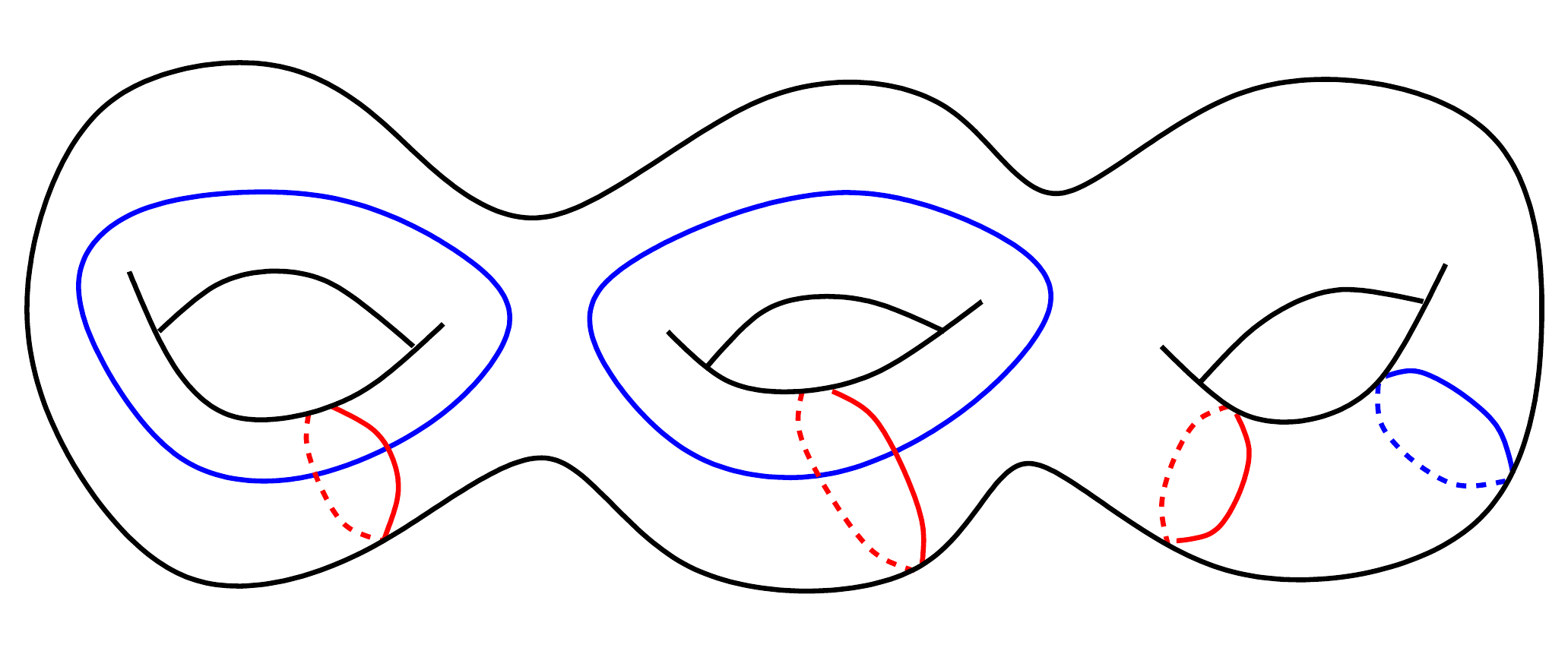}
\caption{The standard genus-$g$ Heegaard diagram for $\#^k (S^1 \times S^2)$, with $g=3$ and $k=1$.}
\label{fig:Heegaard}
\end{figure}

Curiously, 4-manifolds admit a similar decomposition into elementary pieces \cite{MR3590351}, so that the entire
construction can be summarized by a {\it trisection diagram} $(F_g, \alpha, \beta, \gamma)$,
where $F_g$ is a closed 2-dimensional surface called the {\it trisecting surface},
and each $\alpha$, $\beta$, $\gamma$ is a collection of $g$ closed curves on $F_g$ as before.
More specifically, the ingredients involved in trisecting a smooth 4-manifold $M_4$ are the following:
\begin{equation}
\begin{array}{ccl}
F_g \; \cong \; \#^g (S^1 \times S^1) & \qquad & \text{surface of genus}~g \\
M_3^{(i)} \; \cong \; \natural^g (S^1 \times B^2) & \qquad & \text{three 3-manifolds},~ i = 1, \, 2, \, 3 \\
M_4^{(i)} \; \cong \; \natural^k (S^1 \times B^3) & \qquad & \text{three 4-manifolds},~ i = 1, \, 2, \, 3 \\
\end{array}
\label{FMM}
\end{equation}
for some $0 \le k \le g$.
Here, each $M_4^{(i)}$ is diffeomorphic to $\natural^k (S^1 \times B^3)$, and
\be
\partial M_4^{(i)} \; = \; M_3^{(i)} \, \cup_{F_g} \, M_3^{(i+1)} \; \cong \; \#^k (S^1 \times S^2)
\label{HeegaardMMM}
\ee
is the genus-$g$ Heegaard splitting of $\#^k (S^1 \times S^2)$ obtained by stabilizing
the standard genus-$k$ Heegaard splitting $g-k$ times, as illustrated in Figure~\ref{fig:Heegaard}.
As usual, we can describe each $\partial M_4^{(i)} \cong \#^k (S^1 \times S^2)$
by the corresponding Heegaard diagram $(F_g, \alpha, \beta)$, $(F_g, \beta, \gamma)$, and $(F_g, \gamma, \alpha)$,
so that the entire $(g,k)$-trisection
\be
M_4 \; = \; M_4^{(1)} \, \cup \, M_4^{(2)} \, \cup \, M_4^{(3)}
\label{MMM}
\ee
can be specified by the trisection diagram $(F_g, \alpha, \beta, \gamma)$.

It is easy to read off the basic topology of a closed, connected, oriented 4-manifold $M_4$ constructed from a $(g,k)$-trisection.
Namely, $k$ is the number of 1-handles and 3-handles in $M_4$, while $g-k$ is the number of 2-handles \cite{MR3590351}.
(The number of 0-handles and 4-handles is always one.) In particular,
\be
\chi (M_4) \; = \; 2 + g - 3k
\label{chigk}
\ee
When $k=0$, each $M_4^{(i)} \cong B^4$ and $M_4$ is simply-connected.
On the other hand, if $k=g$ then $M_4$ has no 2-handles and
\be
M_4 \; \cong \; \#^k (S^1 \times S^3)
\ee
A special case of this construction with $k=g=1$ is relevant to the topological $S^1 \times S^3$ index
of \cite{Dedushenko:2017tdw}, which we can use later as a consistency check or as a way to bootstrap the A-model data.

\begin{figure}[ht]
\centering
\includegraphics[width=2.4in]{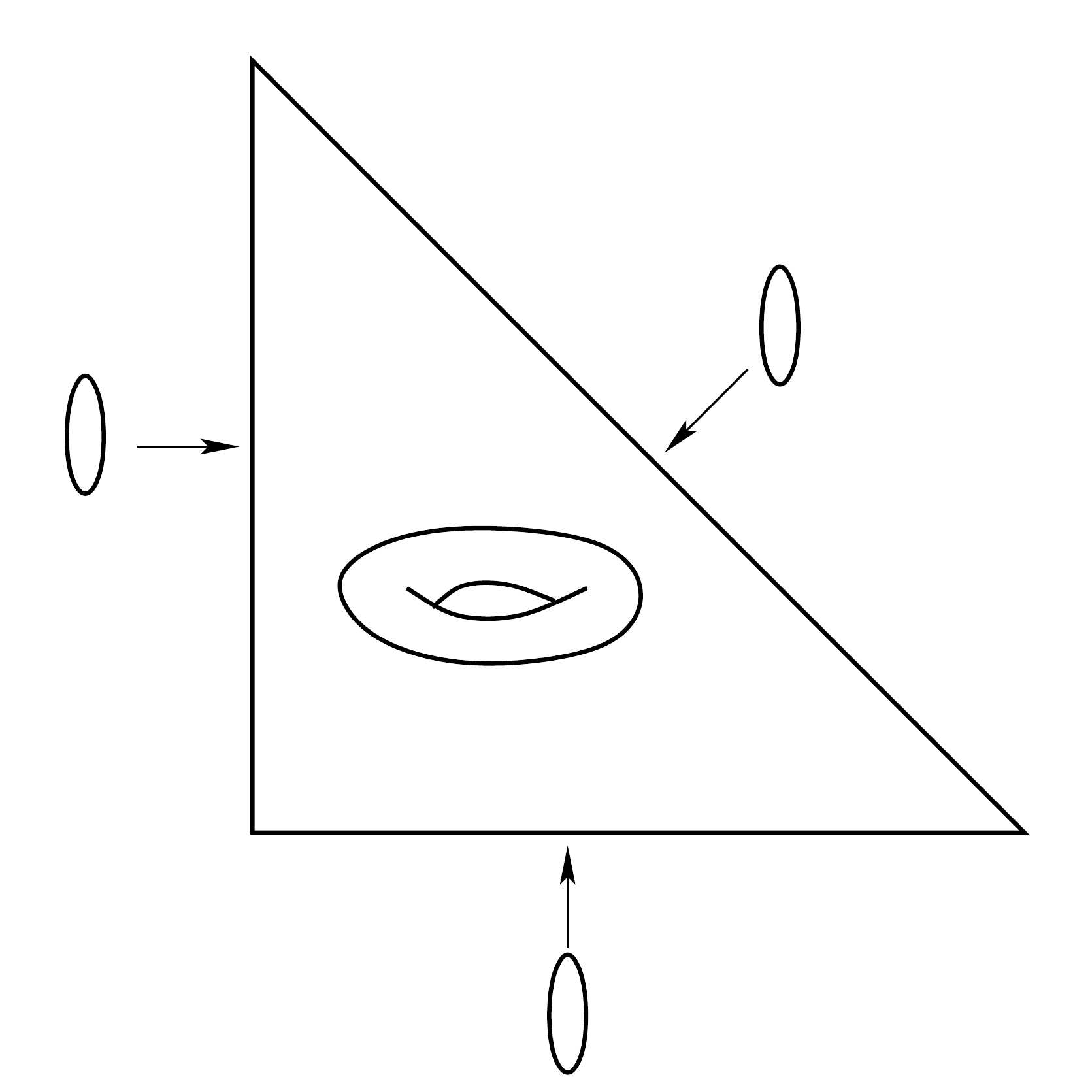}
\caption{The toric diagram for $M_4 = \cp^2$ is a simple example of a genus-1 trisection. Its trisection diagram
comprises a torus $F_g = T^2$ with three curves $\alpha$, $\beta$, and $\gamma$ in homology classes $(1,0)$, $(0,1)$, and $(1,1)$, respectively.}
\label{fig:toricCP2}
\end{figure}

Some simple examples of trisections are actually familiar in physics.
Thus, a standard toric diagram of $M_4 = \cp^2$, shown in Figure~\ref{fig:toricCP2}, is nothing but a trisection with $g=1$ and $k=0$.
This will be a useful example to keep in mind when we discuss topological reduction and topological twists of 4d $\CN=2$ theories.

Now we can describe the structure of a topological 4d theory on a trisection \eqref{MMM}.
The standard yoga of dimensional reduction tells us that geometric pieces of larger dimension
after compactification result in objects of smaller dimension, and vice versa.
In particular, if our starting point is a 4d theory, then we expect 4-dimensional pieces
in \eqref{FMM} to produce 0-dimensional objects $V(M_4^{(i)})$,
3-dimensional components $M_3^{(i)}$ to produce 1-dimensional objects $\CB (M_3^{(i)})$,
and a compactification on $F_g$ to define a 2d theory $\CM (F_g)$, all ``glued'' accordingly.
This is indeed the case, and the precise relation between $\CM (F_g)$, $\CB (M_3^{(i)})$, and $V(M_4^{(i)})$
can be deduced following \cite{Gadde:2013sca} where closely related topological reduction
on 4-manifolds and their building blocks was studied in detail.

A topological reduction of a 4d theory on a 2-manifold $F_g$ gives a 2d theory that we denote $\CM (F_g)$.
(The choice of the 4d theory is suppressed in this notation and will always be clear from the context.)
If the resulting 2d theory is a sigma-model, we use the same notation to denote its target space,
which for a Lagrangian 4d theory would be the moduli space of certain PDEs on $F_g$.
For example, if 4d theory is a $\CN=2$ (resp. $\CN=4$) super-Yang-Mills, then $\CM (F_g)$ is the moduli
space of $G$-bundles (resp. Higgs bundles) on $F_g$, {\it cf.} \cite{Bershadsky:1995vm,Harvey:1995tg,KW}.
In general, however, $\CM (F_g)$ does not need to be a sigma-model.
In fact, in general the resulting 2d theory can be a disjoint sum of several theories (``sectors'')
in which case $\CM (F_g)$ will be used to denote their entire collection.

Moving on to 3-dimensional components $M_3^{(i)}$, it is also easy to see \cite{Gukov:2007ck}
that topological reduction of a 4d theory on a 3-manifold with boundary diffeomorphic to $F_g$
defines a boundary condition --- a.k.a. ``brane'' --- in the 2d theory $\CM (F_g)$.
Moreover, the mapping class group of $F_g$ must be a symmetry of the 2d theory $\CM (F_g)$
and often is realized as a (sub)group of dualities,
\be
\text{MCG} (F_g)\ \; \rotatebox[origin=c]{-90}{$\circlearrowright$}\ \; \CM (F_g)
\label{MCGonMF}
\ee
Mathematically, it means that the mapping class group of $F_g$ is contained
in the group of self-equivalences
\be
\text{MCG} (F_g) \; \subseteq \; \text{AutEq} (D^b \text{Fuk} (\CM (F_g)))
\label{MCGinAuteq}
\ee
In particular, it acts on the boundary conditions in 2d theory $\CM (F_g)$ and gives a convenient way to describe
branes associated with 3-manifolds bounded by $F_g$. Such description of 3-manifolds in terms of branes was recently
used in 3d/3d correspondence \cite{Gadde:2013wq,Chung:2014qpa,Gukov:2016njj,MR3248065} and will be useful to us below as well.

In fact, for application to trisections of 4-manifolds, we need to understand only one basic boundary condition (brane) in $\CM (F_g)$,
namely the brane associated to a genus-$g$ solid handlebody with compressing disks bounded, say, by A-cycles of $F_g$.
If we denote this special brane by $\CB_H$, then every other brane associated with solid handlebody whose compressing
disks are bounded by an arbitrary $g$-tuple $\alpha = (\alpha_1, \ldots, \alpha_g)$ of simple closed curves in $F_g$
is given by
\be
\CB_{\alpha} \; = \; \alpha (\CB_H) \,, \qquad\qquad \alpha \in \text{MCG} (F_g)
\label{BHalpha}
\ee
where, with a small abuse of notation, we use $\alpha$ to denote the corresponding element
of the mapping class group.\footnote{Explicit examples of mapping class group action on
branes associated with 3-manifolds can be found {\it e.g.} in \cite{Gukov:2007ck,Dimofte:2011jd}.}
In other words, $\CB_{\alpha}$ is the description of the basic brane $\CB_H$ in the duality frame
of $\CM (F_g)$ labeled by $\alpha$. We refer to all such branes as ``Heegaard branes.''

Now it should be clear how to describe topological reduction of a 4d theory (Lagrangian or not)
on a 3-manifold $M_3$ given by a Heegaard splitting \eqref{Heegaard} with the Heegaard diagram $(F_g, \alpha, \beta)$.
The resulting 1d theory, {\it i.e.} supersymmetric quantum mechanics, is equivalent to 2d theory $\CM (F_g)$
on an infinite strip $\R \times I$ with two boundary conditions, $\CB_{\alpha}$ and $\CB_{\beta}$,
at the end-points of the interval $I = [a,b]$.
The space of supersymmetric ground states, {\it i.e.} the Floer homology from the 4d perspective,
is the space of ``open strings'' $\text{Hom} (\CB_{\alpha}, \CB_{\beta})$.
In our problem, for a given trisection, we have three such Heegaard splittings \eqref{HeegaardMMM}
and three Heegaard branes all of which are obtained by suitable duality transformations from the basic boundary condition $\CB_H$,
\begin{eqnarray}
\CB_{\alpha} & \; = \; & \alpha (\CB_H) \nonumber \\
\CB_{\beta} & \; = \; & \beta (\CB_H) \label{Babc} \\
\CB_{\gamma} & \; = \; & \gamma (\CB_H) \nonumber
\end{eqnarray}
In application to trisections, these are precisely the branes $\CB (M_3^{(i)})$ associated to 3-manifolds $M_3^{(i)}$;
sometimes we denote them as
\be
\CB_i \; := \; \CB (M_3^{(i)})
\label{branesBi}
\ee
when we wish to emphasize the role of 3-manifolds rather than elements of the mapping class group.

Finally, the last ingredient in \eqref{FMM} is a triple of 4-manifolds $M_4^{(i)}$ bounded by the Heegaard splitting
we just described using branes. Following considerations as in \cite{Gadde:2013sca}, we conclude that each such
4-manifold correspond to a boundary changing ``open string'' operator
\be
V_i \; := \; V (M_4^{(i)}) ~\in~ \text{Hom} (\CB_{i}, \CB_{i+1})
\label{VinHomBB}
\ee
This is a 2d world-sheet counterpart of the relation \eqref{HeegaardMMM}, obtained via topological reduction on $F_g$.
Note, just like the Heegaard branes \eqref{Babc}, our boundary changing operators $V_i$ are `canonical'
in the sense that we only need to understand what happens at the ``corner'' of the world-sheet associated with
the {\it standard} genus-$g$ Heegaard splitting of $\#^k (S^1 \times S^2)$, illustrated in Figure~\ref{fig:Heegaard}.
Indeed, each pair from the set of three $g$-tuples $\{ \alpha, \beta, \gamma \}$ can be brought to the standard form.

For example, in the case of a genus-1 trisection of $M_4 = \cp^2$ shown in Figure~\ref{fig:toricCP2},
$\alpha$ and $\beta$ define the standard Heegaard splitting of $S^3$, which bounds a 4-ball $B^4$.
After the topological reduction on $F_g$, it can be interpreted as a ``corner'' of the disk world-sheet
where boundary segments decorated by the Heegaard boundary conditions $\CB_{\alpha}$ and $\CB_{\beta}$ meet.
Therefore, in order to describe boundary changing operators for $(g,k)$-trisections with $g=1$ and $k=0$
all we need is the Floer homology of a given 4d non-Lagrangian theory on $S^3$ and a state associated to $B^4$,
\be
V (B^4) \; \in \; \CH_{4d} (S^3)
\ee
According to \cite{Dedushenko:2017tdw}, $\CH_{4d} (S^3)$ can be understood with the help of the Coulomb branch index.

Another useful example of a genus-1 trisection that makes contact with \cite{Dedushenko:2017tdw} is $M_4 = S^1 \times S^3$.
Its trisection diagram $(F_g,\alpha,\beta,\gamma)$ comprises a Riemann surface $F_g$ of genus $g=1$
together with three identical simple closed curves $\alpha \cong \beta \cong \gamma$
which, without loss of generality, we can choose to be the A-cycle of $F_g$:
\be
M_4 = S^1 \times S^3: \qquad g=1 ~~\text{and}~~
[\alpha] = [\beta] = [\gamma] = (1,0) \; \in \; H_1 (F_g, \Z)
\label{S1S3abc}
\ee
The corresponding Heegaard branes
\be
\CB_{\alpha} \; = \; \CB_{\beta} \; = \; \CB_{\gamma} \; = \; \CB_H
\label{S1S3BBBBH}
\ee
are also identical and describe what happens in 2d theory $\CM (F_g)$ when the A-cycle of $F_g$ shrinks.
In various examples of Lagrangian theories these are simple Dirichlet branes \cite{Gadde:2013sca,Chung:2014qpa},
as will also be the case in many non-Lagrangian 4d theories.
This $g=k=1$ trisection of $M_4 = S^1 \times S^3$ has $M_4^{(i)} \cong S^1 \times B^3$,
and $V_i$ introduced in \eqref{VinHomBB} are the identity operators, with $i=1,2,3$.
Therefore, the corresponding disk amplitude illustrated in Figure~\ref{fig:diskBBB}
has boundary condition $\CB_H$ along the entire boundary of the disk.
In fact, it corresponds to a familiar description of $S^3$ as a circle fibration over a disk,
where the circle shrinks at the boundary of the disk. Taking a product with another $S^1$
we obtain a fibration over the disk with the fiber $T^2 \cong S^1 \times S^1$
that degenerates to $S^1$ at the boundary of the disk.

For future reference, we summarize the dictionary between trisection ingredients \eqref{FMM}
and the corresponding elements of the two-dimensional A-model in Table~\ref{tab:dict},
supplemented with a few extra items that will be explained later in the text.

\begin{table}[htb]
\centering
\renewcommand{\arraystretch}{1.3}
\begin{tabular}{|@{\quad}c@{\quad}|@{\quad}c@{\quad}| }
\hline  {\bf Trisection data} & {\bf Topological A-model}
\\
\hline
\hline 2-manifold $F_g$ & 2d theory $\CM (F_g)$ \\
\hline mapping class group $\text{MCG} (F_g)$ & dualities of $\CM (F_g)$ \\
\hline 4d coupling constant $\tau_{\text{4d}}$ & complexified K\"ahler parameter of $\CM (F_g)$ \\
\hline S-duality in 4d theory & T-duality in $\CM (F_g)$ \\
\hline 3-manifolds $M_3^{(i)} \cong \natural^g (S^1 \times B^2)$ & ``Heegaard branes'' $\CB (M_3^{(i)})$ in $\CM (F_g)$ \\
\hline 4-manifolds $M_4^{(i)} \cong \natural^k (S^1 \times B^3)$ & boundary changing operators $V(M_4^{(i)})$ \\
\hline 4d topological partition function on  $M_4$ & disk amplitude in $\CM (F_g)$ \\
\hline 4d instantons & 2d world-sheet instantons \\
\hline
\end{tabular}
\caption{The dictionary between geometry and physics.}
\label{tab:dict}
\end{table}


\subsection{2d $\CN=(2,2)$ theories labeled by Riemann surfaces}

In this approach based on trisections and disk amplitudes, the crucial step is to identify
the two-dimensional theory $\CM (F_g)$ equipped with the action of the mapping class group \eqref{MCGonMF}.
As described earlier in this section and summarized in Table~\ref{tab:dict},
$\CM (F_g)$ is defined as a two-dimensional theory obtained by topological reduction of
the 4d (non-Lagrangian) theory in question on a genus-$g$ surface $F_g$.
And, as also noted earlier, in general $\CM (F_g)$ can be a disjoint union of several 2d theories (``sectors'')
whose entire collection we still call ``2d theory $\CM (F_g)$,'' aside from a few places where the role of
individual sectors is discussed in detail.

The partial topological twist along $F_g$ is induced by the topological twist \eqref{SUSYtwist}
on a more general 4-manifold, and can be completed to the latter once we perform
a further A-model twist of the 2d theory $\CM (F_g)$ and add Heegaard branes $\CB_i$
and boundary vertex operators $V_i$, $i=1,2,3$. They all preserve one scalar supercharge $Q$.
In order to see this more explicitly from the A-model point of view, note that the partial topological
twist along $F_g$ induced by \eqref{SUSYtwist} involves twisting by the Cartan subgroup of
the $SU(2)_R$ R-symmetry of the 4d $\CN=2$ theory:
\be
U(1)_R \; \subset \; SU(2)_R
\label{U1RCartan}
\ee
This $U(1)_R$ symmetry is always non-anomalous and for a generic metric on $F_g$ preserves
$\CN=(2,2)$ supersymmetry in the remaining two dimensions where 2d theory $\CM (F_g)$ lives.
The counting of unbroken supercharges follows directly from the supersymmetry algebra and
does not require a Lagrangian description of the 4d $\CN=2$ theory.\footnote{In particular,
the counting of unbroken supersymmetries is the same as in Lagrangian 4d $\CN=2$ theories,
whose topological reduction was extensively studied in the literature,
see {\it e.g.} \cite{Bershadsky:1995vm,Harvey:1995tg,Lozano:1999us,Losev:1999tu,Maldacena:2000mw,Gukov:2007ck,Putrov:2015jpa}.}
We describe it in detail in section~\ref{sec:N3}, where a generalization
to 4d theories with $\CN=3$ supersymmetry will also be made.

In particular, we explain in section~\ref{sec:N3} that, even when 4d theory has extended $\CN=3$ supersymmetry,
its reduction on $F_g$ with a partial topological twist by $U(1)_R \subset SU(2)_R \subset SU(3)_R$
still yields 2d theory $\CM (F_g)$ with only $\CN=(2,2)$ supersymmetry, as long as the holonomy group of $F_g$ is generic.
This has to be contrasted with the topological reduction of 4d $\CN=4$ super-Yang-Mills or
non-generic choice of $F_g = T^2$ with a flat metric, all of which lead to higher supersymmetry of
the 2d theory $\CM (F_g)$, as summarized in Table~\ref{tab:SUSY}.
While extra supersymmetry can be extremely useful for identifying 2d superconformal theory $\CM (F_g)$,
in further application to trisections and 4-manifold invariants we should resist the temptation
and focus on the A-model of $\CM (F_g)$ with respect to the $\CN=(2,2)$ subalgebra.

\begin{table}
\begin{centering}
\begin{tabular}{|c||c|c|c|}
\hline
~~~ & ~4d~$\CN=2$~ & ~4d~$\CN=3$~ & ~4d~$\CN=4$~ \tabularnewline
\hline
\hline
$\phantom{\int^{\int^\int}} g=1 \phantom{\int_{\int}}$ & $\CN=(4,4)$ & $\CN=(6,6)$ & $\CN=(8,8)$
\tabularnewline
\hline
$\phantom{\int^{\int^\int}} g \ne 1 \phantom{\int_{\int}}$ & $\CN=(2,2)$ & $\CN=(2,2)$ & $\CN = (4,4)$
\tabularnewline
\hline
\end{tabular}
\par\end{centering}
\caption{\label{tab:SUSY} 2d supersymmetry preserved by topological reduction on $F_g$.}
\end{table}

As we already stressed in the Introduction, one advantage of this approach is that all
of the basic building blocks \eqref{FMM} involved in trisection of $M_4$ admit
a concrete geometric/physics description, even when the original 4d theory is non-Lagrangian.
For instance, the intuition about 2d theory $\CM (F_g)$ and the action of its symmetry group \eqref{MCGinAuteq}
can be derived from simple examples of Lagrangian 4d theories, with and without enhanced supersymmetry.

As a simple yet instructive example, consider a 4d $\CN=2$ gauge theory with gauge group $G = U(1)$
and no matter fields. This theory is free and its topological reduction on $F_g$ can be obtained using
the standard rules of the Kaluza-Klein reduction.
Namely, the Wilson lines of a $U(1)$ gauge field along A- and B-cycles of $F_g$
give rise to $2g$ real periodic scalar fields that parametrize
(part of) the vacuum manifold of the 2d theory $\CM (F_g)$:
\be
\text{Hom} (\pi_1 (F_g), U(1)) \; \cong \; \text{Jac} (F_g) \; = \; \text{Pic}^0 (F_g)
\ee
When combined with a complex scalar $u$ of the 4d $\CN=2$ vector multiplet, they produce
a 2d $\CN=(2,2)$ sigma-model with the target space\footnote{Despite a small abuse of notations,
when 2d theory $\CM (F_g)$ is a sigma-model we find it convenient to use the same notation for its target space.}
\be
\CM (F_g) \; = \; \C_u \times \text{Jac} (F_g)
\label{U1target}
\ee
The mapping class group \eqref{MCGonMF} acts on this sigma-model in an obvious way,
and one can easily describe the Heegaard branes (see below).
Also note that the target space \eqref{U1target} is K\"ahler, as required by $\CN = (2,2)$ supersymmetry.
Although this example looks almost too trivial, it is actually relevant to all 4d $\CN=2$ non-Lagrangian
theories of rank 1, which reduce to a free $U(1)$ vector multiplet on the Coulomb branch, albeit with
a non-trivial fibration \eqref{U1target}.

An equally simple and instructive example, from which we can derive some inspiration,
is the 4d $\CN=4$ super-Yang-Mills with gauge group $G$.
Its topological reduction on $F_g$, originally studied in \cite{Bershadsky:1995vm,Harvey:1995tg},
gives a 2d $\CN=(4,4)$ sigma-model with the target space
\be
\CM (F_g) \; = \; \CM_H (G,F_g) \,,
\label{MHtarget}
\ee
the moduli space of Higgs bundles on $F_g$ (also known as the Hitchin moduli space).
Even though in this case the sigma-model has enhanced supersymmetry and a large variety
of topological twists, discussed {\it e.g.} in \cite{KW}, for purposes of the present paper
we are interested in the A-model with respect to $\omega_I$, in the conventions of \cite{Hitchin:1986vp}.
In particular, since the 4d theory is Lagrangian, the Heegaard branes $\CB_i$ can be defined as
solutions to certain PDEs on $F_g$ that can be extended to a handlebody $M_3^{(i)}$ bounded by $F_g$.
The resulting branes $\CB_i$ are holomorphic Lagrangian submanifolds in $\CM_H (G,F_g)$,
namely $(A,B,A)$ branes, that admit a convenient explicit description \cite{Gukov:2007ck}
in complex structure $J$ in terms of equations analogous to the $A$-polynomial \cite{Apol}.
Nevertheless, for applications in the present paper they are
first and foremost A-branes with respect to $\omega_I$.

In some ways, 2d theories obtained by topological reduction of non-Lagrangian 4d SCFTs are very similar.
For example, they too enjoy the action of the mapping class group \eqref{MCGonMF}, which sometimes
factors through its symplectic representation, {\it i.e.} the action of $Sp(2g,\Z) \cong Sp (H^1 (F_g))$,
just like in the case of \eqref{U1target}. In fact, as we already mentioned earlier, this simple
example is relevant to rank-1 non-Lagrangian theories, which reduce to a free $U(1)$ vector
multiplet on the Coulomb branch. There are some new elements, however.
For example, the effective theory on the Coulomb branch often involves a non-trivial
dilaton profile, {\it cf.} \cite{Nekrasov:2014xaa}:
\be
\CL_{2d} \; = \; \ldots + \frac{i}{2} \Omega (u) R \,,
\ee
which on a world-sheet of genus $\tilde g$ contributes to the partition function $e^{2\pi i (\tilde g-1) \Omega (u)}$.
This term originates from the similar gravitational couplings of the four-dimensional theory \cite{Witten:1995gf}:
\be
\CL_{4d} \; = \; \ldots + \log A(u) \, \Tr R \wedge \tilde R + \log B(u) \, \Tr R \wedge R \,,
\ee
which contribute to the measure of the path integral of the Coulomb branch theory\footnote{Here,
we allow a more general possibility that, apart from the abelian vector multiplets, the effective
theory on the Coulomb branch also contains massless hypermultiplets. These are usually
called {\it enhanced Coulomb branches} (or ECBs for short), see {\it e.g.} \cite{Argyres:2016xmc}.}
\be
Z_{\text{Coulomb}} \; = \; \int [dV] [dH] \, A(u)^{\chi} \, B(u)^{\sigma} \, e^{S_{\text{IR}} (V,H)}
\label{ZCoulomb}
\ee
By comparing the 2d and 4d gravitational couplings,
say on a 4-manifold $M_4$ which is a profuct of $F_g$ with another Riemann surface of genus $\tilde g$,
it is easy to see that the ``effective dilaton'' in the 2d theory $\CM (F_g)$ should be
\be
\Omega (u) \; = \; \frac{2}{\pi i} (g-1) \log A(u)
\label{2ddilatonA}
\ee


\subsection{A toy model}
\label{sec:toy}

As a useful metaphor for a 2d $\CN=(2,2)$ theory labeled by a Riemann surface $F_g$, let us consider
a Landau-Ginzburg model with a (twisted) superpotential $W$. To exhibit the action of the mapping class group
we construct $W$ in each duality frame separately, and then verify that the resulting Landau-Ginzburg models
describe the same IR theory.

Given a pair-of-pants decomposition of $F_g$, let us associate a 2d $\CN=(2,2)$ minimal model $A_1$ to every tube region.
Using the Landau-Ginzburg description,
\be
{\raisebox{-.3cm}{\includegraphics[width=2.0cm]{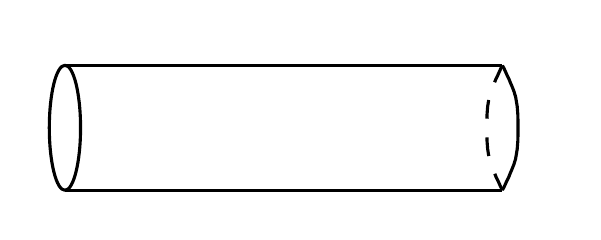}}\,} : \qquad W = x^3
\label{toytube}
\ee
we can say that each tube (cylinder) is ``colored'' by $x$ (= superfield in 2d $\CN=(2,2)$ theory) and contributes to $W$
according to \eqref{toytube}. To every pair-of-pants, where three tubes colored by $x$, $y$, and $z$ come together,
we associate a cubic interaction
\be
{\raisebox{-.6cm}{\includegraphics[width=1.6cm]{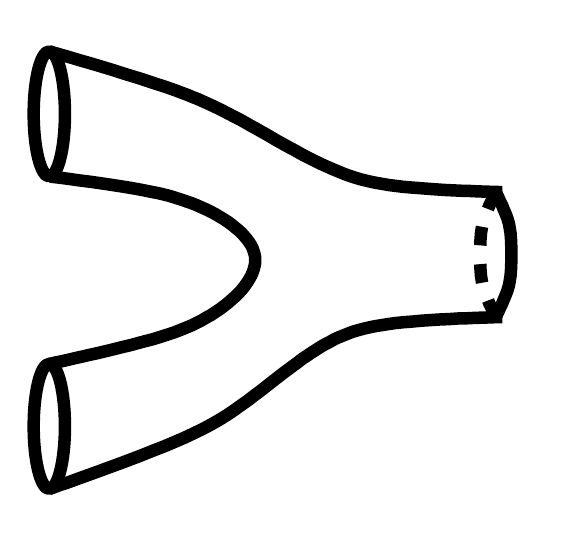}}\,} : \qquad W = xyz
\ee
These rules define an interacting 2d $\CN=(2,2)$ Landau-Ginzburg model that flows to a superconformal fixed point.
Since all terms in the superpotential $W$ are cubic, all superfields $x_i$ have the same R-charge $q_i = \frac{1}{3}$.
Therefore, the $\CN=(2,2)$ superconformal theory associated to $F_g$ in this way has the central charge
\be
\hat c \; = \; \sum_i (1 - 2 q_i) \; = \; g-1
\ee
where we used the fact that a pair-of-pants decomposition of $F_g$ has $3g-3$ tubes, so that $i=1, \ldots, 3g-3$.
Note, in particular, that the central charge and the infra-red R-charges are independent of the duality frame,
{\it i.e.} the choice of cutting $F_g$ into pairs of pants.

Moreover, it is easy to check that the chiral ring is invariant under the crossing symmetry.
For 2d $\CN=(2,2)$ theories at hand, the chiral ring is simply the Jacobi ring of the potential function $W$,
that is $\C [x_i] / \partial W$.
For example, for two different pair-of-pants decompositions of a genus-2 surface, we get
\be
{\raisebox{-0.9cm}{\includegraphics[width=2.3cm]{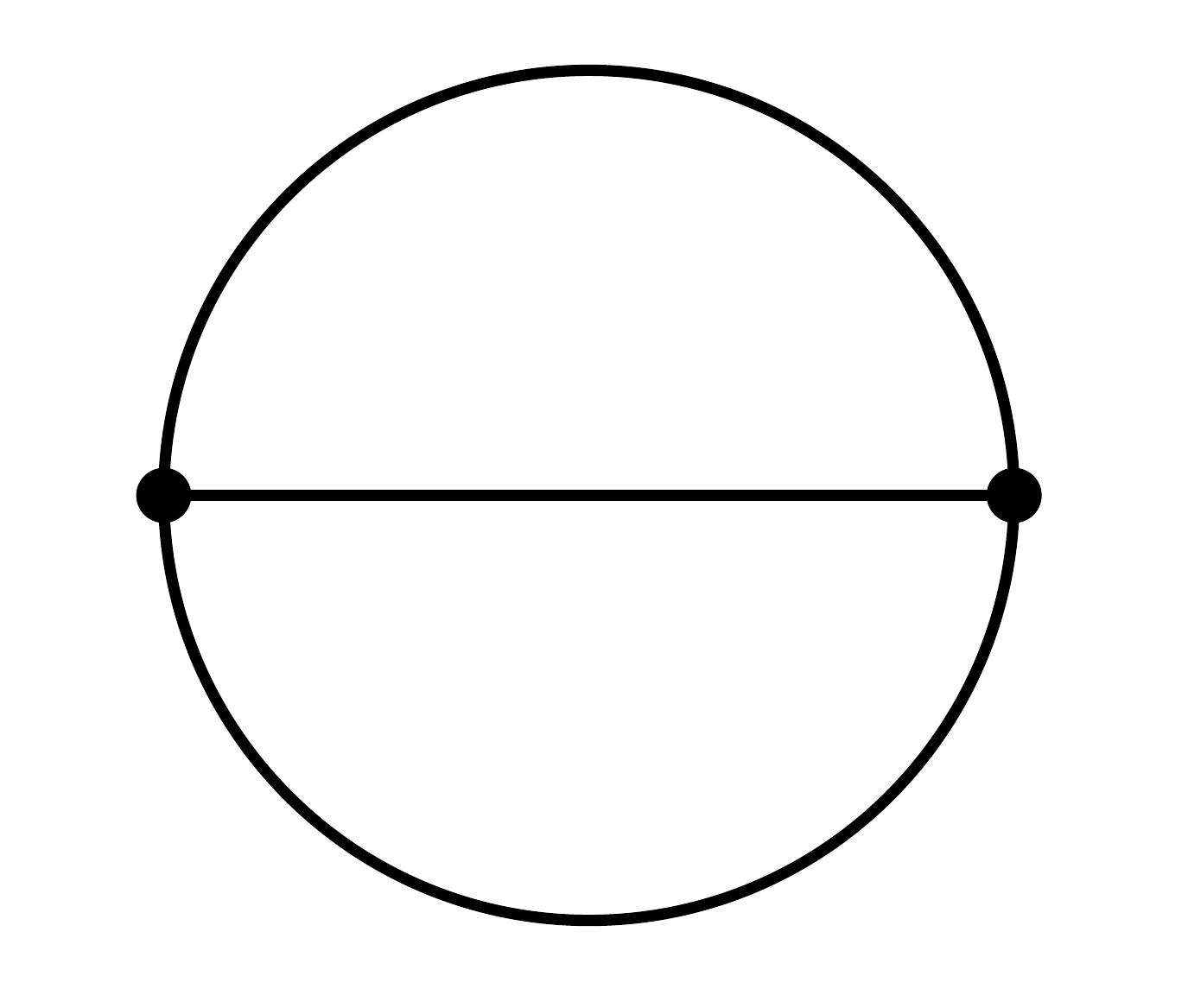}}\,} : \qquad W = x^3 + y^3 + z^3 + xyz
\ee
and
\be
{\raisebox{-.5cm}{\includegraphics[width=3.0cm]{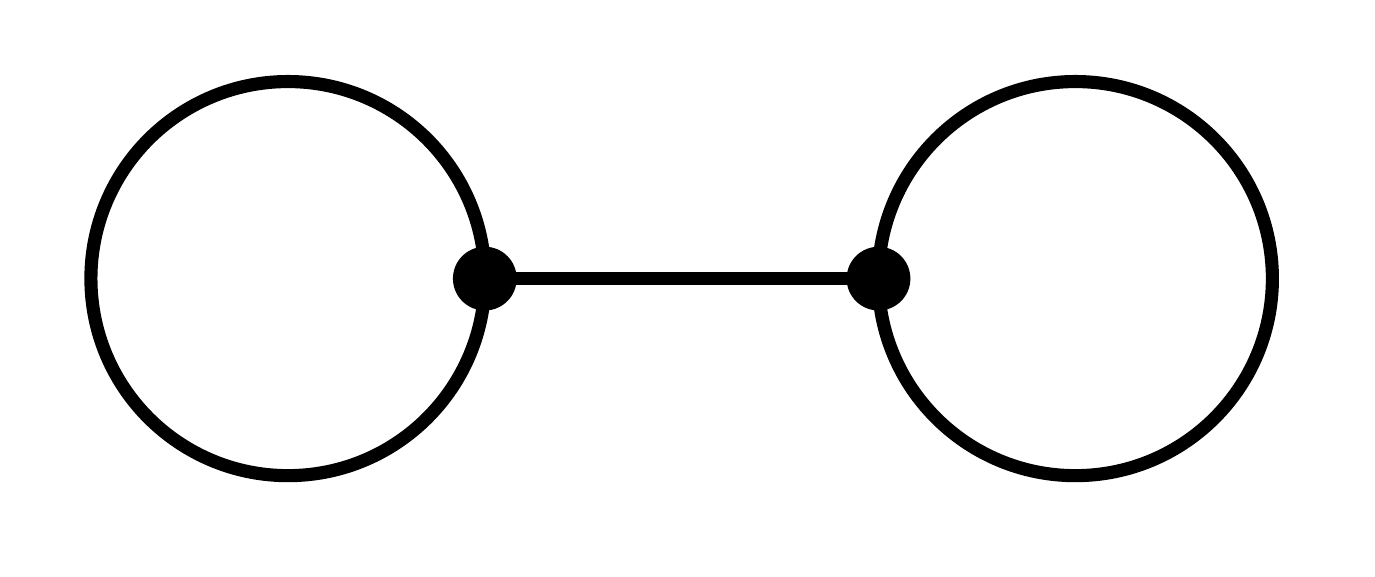}}\,} : \qquad W = x^3 + x^2 y + y^3 + yz^2 + z^3
\ee
It is easy to check that Jacobi rings for these two potentials are indeed the same and,
in the language of singularity theory, correspond to $T[3,3,3]$.
Furthermore, if $W$ is a twisted superpotential, these LG models are
equivalent to a sigma-model on the $T^2 / \Z_3$ orbifold, whose complex structure is fixed $\tau = e^{\pi i /3}$.
If, on the other hand, $W$ is treated as a superpotential, the resulting LG models are equivalent
to a mirror dual asymmetric orbifold of $T^2$ with the fixed K\"ahler modulus \cite{Lerche:1989cs,Chun:1991js}:
\be
\rho \; = \; e^{\pi i /3}
\ee
This latter case is closer to our applications that involve A-models and 4-manifolds.
In particular, it illustrates well a general feature of 4d strongly-coupled (non-Lagrangian) theories
reduced on a Riemann surface $F_g$: the resulting 2d theories are often rigid,
akin to asymmetric orbifolds \cite{Narain:1986qm} or Gepner models with gauged discrete symmetries.


\subsection{Our first Heegaard boundary conditions}

In order to develop our intution about the Heegaard branes, let us consider a genus-1 trisection
illustrated in Figure~\ref{fig:toricCP2}. As we explain momentarily, in the case of the simplest
4d $\CN=2$ Lagrangian theory, the corresponding A-model calculation comes down to the simplest
instance of mirror symmetry, where disk amplitude counting is not only easy and enjoyable
but also mathematically rigorous and well established. This gives us hope that similar steps
for non-Lagrangian theories also admit a mathematical formulation, within general framework of mirror symmetry.

Specifically, consider a 4d $\CN=2$ super-Maxwell theory with gauge group $G = U(1)$.
As we already mentioned in \eqref{U1target}, its topological reduction on a genus-$g$ surface $F_g$
gives a sigma-model with the Jacobian of $F_g$ as its target space. In the case of genus-1 trisections,
the Jacobian of $F_g$ is also a 2-torus, which for convenience we denote by $E$,
\be
E \; = \; \R^2 / (\Z + \tau \Z)
\label{Ecurve}
\ee
parametrized by the holonomies of the $U(1)$ gauge field along A- and B-cycles of $F_g = T^2$.
We denote these $U(1)$-valued holonomies by $(x,y) = (e^{i \varphi_A}, e^{i \varphi_B}) \in E$.

In order to describe the Heegaard branes $\CB_i$, we need to find conditions on $x$ and $y$
(or, equivalently, on $\varphi_A$ and $\varphi_B$) that arise from topological reduction
of 4d the theory on a handlebody $M_3^{(i)}$ bounded by $F_g = T^2$.
This condition is essentially identical to how $A$-polynomial enters the Chern-Simons theory \cite{Apol}.
Indeed, since the 4d theory in question is Lagrangian, it localizes on solutions to the anti-self-duality
equation, $F_A^+ = 0$, which upon reduction on $M_3^{(i)}$ becomes the flatness equation, $F_A = 0$.
And, since flat connections are completely characterized by their holonomies,
we conclude that flat connections on $F_g = T^2$ which can be extended to a handlebody $M_3^{(i)}$
are described by an equation of the form
\be
\CB_{\alpha} : \qquad p \varphi_A + q \varphi_B \; = \; 0 \,,
\label{Babelian}
\ee
where $[\alpha] = (p,q) \in H_1 (F_g,\Z)$ is the homology class of the cycle
that becomes trivial (contractible) in the handlebody bounded by $F_g$.
Each such equation, one for every handlebody $M_3^{(i)}$, defines an A-brane $\CB_i$, $i=1,2,3$.

Another way to see this is to observe that 2d $\CN=(2,2)$ sigma-model with the target space $E$
or $\C_u \times E$ has two $SL(2,\Z)$ symmetries\footnote{The full symmetry group also contains
a semidirect product with a few extra $\Z_2$ factors, most of which will not play a role in our
discussion, except for a $\Z_2$ symmetry that exchanges the two $SL(2,\Z)$ groups in \eqref{twoSL2Z}.
It will play an important role in section~\ref{sec:AD}.}
\be
SL(2,\Z)_{\tau} \times SL(2,\Z)_{\rho}
\label{twoSL2Z}
\ee
The former acts on the complex structure of the elliptic curve, $\tau \mapsto \frac{a \tau + b}{c \tau + d}$,
and is precisely the mapping class group action \eqref{MCGonMF} we wish to see more generally:
\be
\text{MCG} (F_g) \; \cong \; SL(2,\Z)_{\tau}
\qquad \qquad (g=1)
\label{MCGg1}
\ee
Indeed, the mapping class group of $F_g$ acts on A- and B-cycles and, therefore,
on the corresponding holonomies $(\varphi_A,\varphi_B)$.
On the other hand, the second $SL(2,\Z)$ symmetry in \eqref{twoSL2Z} acts
on the complexified K\"ahler modulus
\be
\rho = i \text{vol} (E) + \int_E B
\ee
by simultaneous T-dualities along both directions of $E$ and integer shifts of the $B$-field.
This modular group corresponds to the electric-magnetic duality of the original 4d $U(1)$ gauge theory
since under topological reduction the 4d gauge coupling is mapped to $\rho$.
This identification of the 4d couplings with complexified K\"ahler parameters of $\CM (F_g)$
is a general feature of the topological reduction \cite{Bershadsky:1995vm,Harvey:1995tg}.
And, it can serve us as a useful guide in the study of non-Lagrangian theories,
where the absence of 4d coupling constants means that 2d A-model of $\CM (F_g)$
has no large-volume limit, defined at a fixed (small) radius and large curvature.
In string theory, such targets are often dubbed ``stringy'' or ``quantum'' since classical geometry
breaks down and is replaced by its quantum version (often, an algebraic description of some sort).

\begin{figure}[ht]
\centering
\includegraphics[width=2.4in]{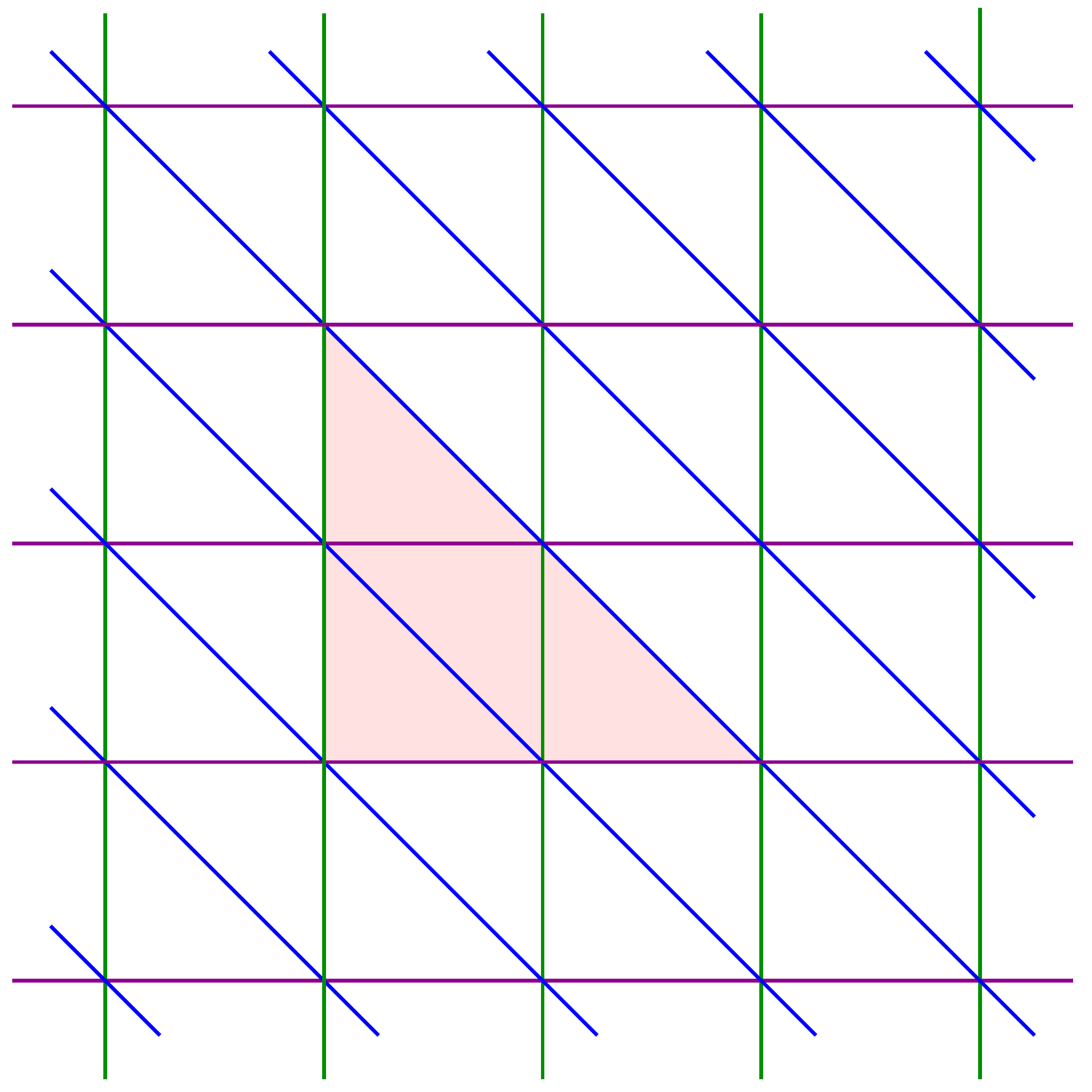}
\caption{Example of a disk instanton in 2d sigma-model associated with a genus-1 trisection of $M_4 = \cp^2$
in Figure~\ref{fig:toricCP2}. Shown here is a covering space of $\CM (F_g)$ on which each of the Heegaard branes
$\CB_{\alpha}$, $\CB_{\beta}$, $\CB_{\gamma}$ lifts to an infinite set of parallel straight lines.}
\label{fig:triangles}
\end{figure}

To summarize, the genus-1 trisection $(F_g,\alpha,\beta,\gamma)$ with $[\alpha] = (0,1)$,
$[\beta] = (0,1)$, and $[\gamma] = (1,1)$, in the case of the simplest 4d $\CN=2$ theory,
led us to the A-model of an elliptic curve $E$ with three A-branes $\CB_i$ defined by
the equations of the form \eqref{Babelian}.
Counting disks in this setup is surprisingly simple and very elegant.
It goes back to the original work of Kontsevich on the homological mirror symmetry \cite{MR1403918}
and further study of Massey products in the Fukaya category of an elliptic curve \cite{MR1633036,Brunner:2004mt,Herbst:2006nn}.
Since the target space of our A-model is not simply-connected, it is convenient to pass to
its universal cover, where each of the branes $\CB_{\alpha}$, $\CB_{\beta}$, and $\CB_{\gamma}$
lifts to an infinite set of parallel lines, as illustrated in Figure~\ref{fig:triangles}.
Counting disk instantions then becomes a simple problem of counting triangles, modulo translation,
such that each side of the triangle belongs to one of the straight lines.
Since the area of the triangles scales as $\sim (\text{size})^2$,
the result is given by a simple theta-function
\be
\vartheta (q) \; = \; \sum_{n \in \Z} q^{\frac{1}{2}n^2}
\ee
where $q = e^{2 \pi i \rho}$ and $\rho$ is the complexified K\"ahler parameter of the target space $\CM (F_g)$.
In the Fukaya category of the elliptic curve $E$ this calculation
determines the coefficient of the Massey product $m_2$.
More importantly for us here, it agrees with the partition function of $U(1)$ gauge theory on $M_4 = \cp^2$,
if we identify $\rho$ with the complexified gauge coupling constant of the 4d theory \cite{Witten:1995gf,Verlinde:1995mz}.


\subsection{From 4d chiral rings to 2d chiral rings}

The R-symmetry of the 4d $\CN=2$ theory, Lagrangian or not, is $SU(2)_R \times U(1)_r$.
As we already mentioned earlier, the partial topological twist along $F_g$ breaks $SU(2)_R$
down to its Cartan subgroup $U(1)_R$.
The abelian R-symmetries $U(1)_R$ and $U(1)_r$ then become the R-symmetries of the resulting
2d $\CN=(2,2)$ theory $\CM (F_g)$, namely the vector and axial R-symmetries,
{\it cf.} \cite{Bershadsky:1995vm,Gadde:2013ftv,Putrov:2015jpa}:
\be
R_V = 4R \qquad , \qquad R_A = 2r
\label{R2d4d}
\ee
Sometimes it will be convenient to use their linear combincations, the R-symmetries of
the left and right superconformal algebra which --- to avoid confusion with 4d R-charges
and conform to the notations frequently used in 2d literature --- we denote $q_L = 2 R - r$ and $q_R = 2 R + r$.
When we talk only about one sector (left or right) and there is no confusion,
we may omit the extra label and simply call the R-charge $q$.

With these conventions, the chiral primaries of the 2d $\CN=2$ superconformal algebra
saturate the BPS bound $h = \frac{q}{2}$,
while the anti-chiral primaries saturate the BPS bound $h = - \frac{q}{2}$.
The ring of $(\text{chiral},\text{chiral})$ operators is usually called $(c,c)$ ring or B-model ring,
and similarly the ring of $(\text{anti-chiral},\text{chiral})$ operators is the $(a,c)$ or A-model ring.

We wish to relate short (a.k.a. BPS) representations of 4d surperconformal algebra to
short representations of 2d superconformal algebra.
Note, there can be many embeddings of 2d superconformal algebra into 4d superconformal algebra.
Here, we need the one where 2d $\CN=(2,2)$ superconformal algebra is realized on a 2d space
orthogonal to the surface $F_g$, see {\it e.g.} \cite{Gadde:2013ftv}.
In particular, translating to our conventions here, we get the following relation between
the quantum numbers in the 4d $\CN=2$ theory and its 2d $\CN=(2,2)$ descendant $\CM (F_g)$:
\be
\begin{array}{l@{\qquad\qquad\qquad}l}
h \; = \; \frac{1}{2} \Delta + \frac{1}{2} j_1 + \frac{1}{2} j_2 & \bar h \; = \; \frac{1}{2} \Delta - \frac{1}{2} j_1 - \frac{1}{2} j_2 \\
q_L \; = \; 2R - r & q_R \; = \; 2 R + r
\end{array}
\label{hqLhqR}
\ee
Using these relations, we can easily see that spinless Coulomb branch operators
become elements of $(a,c)$ chiral rings in two dimensions,
whereas spinless Higgs branch operators contribute to $(c,c)$ rings, as summarized in Table~\ref{tab:N2mplets}.


\begin{table}
\begin{centering}
\begin{tabular}{|c|l|c|c|}
\hline
~Multiplet~ & ~Conditions~ & ~$(a,c)$~ & ~$(c,c)$~ \tabularnewline
\hline
\hline
$\phantom{\int^{\int^\int}} \CA^{\Delta}_{R,r (j_1, j_2)} \phantom{\int_{\int}}$ & $\Delta > 2R + 2 + j_1 + j_2 + |r+j_1 - j_2|$ & $\text{\sffamily X}$ & $\text{\sffamily X}$
\tabularnewline
\hline
$\phantom{\int^{\int^\int}} \CE_{r (0,j_2)} \phantom{\int_{\int}}$ & $\Delta = r$, \quad $R = j_1 = 0$, \quad $r > j_2 + 1$ & $\checkmark$ & $\text{\sffamily X}$
\tabularnewline
\hline
$\phantom{\int^{\int^\int}} \hat \CB_R \phantom{\int_{\int}}$ & $\Delta = 2R$, \quad $j_1 = j_2 = r = 0$ & $\text{\sffamily X}$ & $\checkmark$
\tabularnewline
\hline
$\phantom{\int^{\int^\int}} \vdots \phantom{\int_{\int}}$ & ~~~~~~~~~~~$\vdots$ & $\vdots$ & $\vdots$
\tabularnewline
\hline
\end{tabular}
\par\end{centering}
\caption{\label{tab:N2mplets} Topological reduction of 4d $\CN=2$ superconformal multiplets.}
\end{table}

One of our topics in the next subsection will be the set of observables in 4d topological theory,
{\it i.e.} operators in cohomology of the supercharge $Q$
that transforms as a scalar after the topological twist \eqref{SUSYtwist}.
By going through the list of the shortening conditions, it is easy to see that,
among 4d $\CN=2$ superconformal multiplets, the multiplets of type $\CE$ always obey this condition,
thus, providing us with a large class of topological observables.
(In fact, they obey a stronger condition, whose explicit form we won't need.)


\subsection{Observables and non-Lagrangian analogues of Donaldson polynomials}
\label{sec:structure}

One of the key messages in this paper is that 4-manifold invariants produced by the topological
twist of a non-Lagrangian 4d theory can be defined and computed in the two-dimensional A-model
on a disk, with the Heegaard boundary conditions. Schematically,
\be
Z(M_4) \; = \; Z_A \big( {\raisebox{-.2cm}{\includegraphics[width=0.6cm]{disk}}} \big)
\label{ZZ2d4d}
\ee
Moreover, on both sides we can include observables in cohomology of the scalar supercharge $Q$,
so that in general \eqref{ZZ2d4d} should be understood as a relation between topological correlators of
the 4d non-Lagrangian theory on $M_4$ and the corresponding correlators in the two-dimensional A-model of $\CM (F_g)$.
The goal of this section is to establish the dictionary between 2d and 4d topological correlators
(see also Table~\ref{tab:dict}) and to describe the structure of the resulting 4-manifold invariants.

One advantage of Lagrangian theories is that the left-hand side of \eqref{ZZ2d4d} can be expressed \cite{Witten:1988ze}
as an integral over the moduli space $\CM$ of solutions to various PDEs on $M_4$.
Consider, for example, a family of 4d theories with gauge group $G = U(1)$
and $N_f$ matter multiplets of charge $+1$. In this class of theories,
the moduli space $\CM$ has virtual dimension\footnote{We follow the conventions of \cite{Dedushenko:2017tdw},
except for the normalization of $U(1)_r$ charges, which differ by a factor of 2.
This amounts to an extra factor of 2 in all relations between $U(1)_r$ charges
and dimensions of moduli spaces or degrees of differential forms on these spaces.
Namely, to obtain the latter we now need to multiply $U(1)_r$ charges by 2.}
\be
\text{VirDim} (\CM) \; = \; \frac{N_f \lambda^2 - 2 \chi - (2 + N_f) \sigma}{4}
\label{NfSWdim}
\ee
which clearly depends on the choice of theory (the value of $N_f$), the choice of the 4-manifold $M_4$,
and the topology of the gauge bundle (the first term in \eqref{NfSWdim}).
In non-Lagrangian 4d theories, topological invariants do not reduce to counting solutions of PDEs on $M_4$,
but we still can write an analogue of \eqref{NfSWdim} which plays the same role and, in fact, does become
the dimension of a moduli space in the A-model.

Indeed, the virtual dimension \eqref{NfSWdim} is given by $U(1)_r$ anomaly on a 4-manifold $M_4$.
The latter, in turn, can be obtained from the anomaly polynomial of a general 4d $\CN=2$ theory (Lagrangian or not):
\be
A_{4d} \; = \; (a-c) \left( c_1 (\CR) p_1 (TM_4) - c_1 (\CR)^3 \right)
- (4a-2c) c_1 (\CR) c_2 (E) + \frac{k}{4} c_1 (\CR) \text{ch}_2 (F)
\ee
where $\CR$ is the $U(1)_r$ symmetry bundle and $E$ is the $SU(2)_R$ bundle.
Integrating over $M_4$ and using
\be
\int_{M_4} p_1 (TM_4) \; = \; 3 \sigma
\qquad , \qquad
\int_{M_4} c_2 (E) \; = \; \frac{2 \chi + 3 \sigma}{4}
\ee
we get
\be
\text{``VirDim $(\CM)$''}
\; = \; 2 \Delta r \; = \; \frac{k}{2} \int_{M_4} \text{ch}_2 (F) - 2 (2a-c) \chi - 3c \, \sigma
\label{rgeneral}
\ee
The first term here is due to fluxes and plays the role analogous to the first term in \eqref{NfSWdim};
we shall return to it later and first consider the ``gravitational'' part of the anomaly that involves
$\chi$ and $\sigma$, the Euler characteristic and signature of $M_4$.
Note, a single 4d $\CN=2$ vector multiplet has $c= \frac{1}{6}$ and $a = \frac{5}{24}$,
which leads to $2 \Delta r = - \frac{2\chi + 2 \sigma}{4}$, in agreement with the $N_f$-independent part of \eqref{NfSWdim}.
Similarly, the $N_f$-dependent part of \eqref{NfSWdim} is precisely what \eqref{rgeneral} gives for
a hypermultiplet, which has $c = \frac{1}{12}$ and $a = \frac{1}{24}$.

Now, we can apply \eqref{rgeneral} to any non-Lagrangian theory.
For example, the original Argyres-Douglas theory has $a = \frac{43}{120}$ and $c = \frac{11}{30}$,
{\it cf.} Table~\ref{tab:AD}. Moreover, it has no flavor symmetry, so the first term in \eqref{rgeneral} is absent.
The remaining terms give
\be
\text{``VirDim $(\CM)$''} \; = \; 2 \Delta r
\; = \; - \frac{7 \chi + 11 \sigma}{10} \; = \; \frac{4}{10} \big( \chi_h (M_4) - c (M_4) \big)
\label{rAD}
\ee
where in the last equality we used\footnote{Here, $c (M_4)$ should not be confused with 2d or 4d
conformal anomaly coefficients. To avoid confusion, when we wish to describe basic topology of $M_4$
we shall mostly use $\chi$ and $\sigma$ instead of $\chi_h$ and $c$.}
\begin{eqnarray}
\chi_h (M_4) & = & \frac{\chi (M_4) + \sigma (M_4)}{4} \label{chihc} \\
c (M_4) & = & 2 \chi (M_4) + 3 \sigma (M_4) \qquad (= K_{M_4}^2 \quad \text{when $M_4$ is a complex surface}) \nonumber
\end{eqnarray}
The expression \eqref{rAD} plays the same role as the virtual dimension of the moduli space in Donaldson or Seiberg-Witten theory:
by charge conservation, when $\Delta r \ne 0$ the partition function \eqref{ZZ2d4d}
vanishes for a generic metric on a 4-manifold $M_4$ with $b_2^+ > 1$,
unless we introduce additional observables (which will be discussed shortly).

More generally, \eqref{rgeneral} can tell us on which 4-manifolds the partition function \eqref{ZZ2d4d}
of a twisted non-Lagrangian theory can be non-zero (in the trivial flux sector and no additional observables):
\be
-1 \; \le \; \frac{\sigma}{\chi} \; \le \; 0
\label{vanishingZ}
\ee
This ``vanishing theorem'' follows from \eqref{rgeneral} combined with the $\CN=2$ version \cite{Shapere:2008zf}
of the Hofman-Maldacena bound \cite{Hofman:2008ar}:
\be
\frac{1}{2} \; \le \; \frac{a}{c} \; \le \; \frac{5}{4}
\ee
As we shall see momentarily, both bounds in \eqref{vanishingZ} can be relaxed once we
add observables to the path integral of the topological theory on $M_4$.
Specifically, the lower bound can be easily relaxed when $\chi > 0$;
in particular, this is the case for simply-connected 4-manifolds.
On the other hand, adding observables can help to overcome the upper bound when $\chi < 0$.
A similar effect can be achieved by yet another generalization: turning on fluxes for
flavor symmetries of the (non-Lagrangian) 4d theory activates the first term in \eqref{rgeneral}
and leads to an equivariant version of 4-manifiold invariants (see {\it e.g.} \cite{Dedushenko:2017tdw}
for a recent study of equivariant multi-monopole invariants on general 4-manifolds).

\begin{table}
\begin{centering}
\begin{tabular}{|c||c|c|c|c|c|}
\hline
~4d theory~ & ~$a$~ & ~$c$~ & ~$\Delta (u)$~ & ~$\dim_{\mathbb{H}} (\text{Higgs})$~ & ~Kodaira type~ \tabularnewline
\hline
\hline
$\phantom{\int^{\int^\int}} (A_1,A_2) \phantom{\int_{\int}}$ & $\frac{43}{120}$ & $\frac{11}{30}$ & $\frac{6}{5}$ & $0$ & II
\tabularnewline
\hline
$\phantom{\int^{\int^\int}} (A_1,A_3) \phantom{\int_{\int}}$ & $\frac{11}{24}$ & $\frac{1}{2}$ & $\frac{4}{3}$ & $1$ & III
\tabularnewline
\hline
\end{tabular}
\par\end{centering}
\caption{\label{tab:AD} Two simple examples of rank-1 Argyres-Douglas theories.}
\end{table}

Now let us incorporate observables which, in order to preserve topological symmetry, must be in cohomology
of the scalar supercharge $Q$, {\it cf.} \eqref{SUSYtwist}.
A large class of such observables --- that echoes the definition of Donaldson polynomials \cite{MR1066174} ---
comes from Coulomb branch operators $\CE_r$. The lowest component of such a multiplet
defines a coordinate $u$ on the Coulomb branch, of conformal dimension $\Delta (u) = r$.
Inserting such observables into the path integral of the topologically twisted theory
defines ``non-Lagrangian analogues'' of Donaldson polynomials, or topological correlators
\be
\langle u(x_1) \ldots u(x_n) \rangle
\label{uuuu}
\ee
which are non-zero (for generic metric on $M_4$) only if the total $U(1)_r$ R-charge vanishes, {\it i.e.}
\be
n \; = \; - \frac{2 (2a-c) \chi + 3c \, \sigma}{2 \Delta (u)}
\label{uuuughost}
\ee
Of course, the right-hand side should be a non-negative integer number.
When this condition is satisfied, the correlation function \eqref{uuuu}
is independent on the positions $x_i$ where we insert the observables.
For example, in the original Argyres-Douglas theory\footnote{When we write correlators like this,
we mean evaluating them in standalone Argyres-Douglas theories, not their contributions
to the topological path integral of $SU(N)$ gauge theories (the latter vanish due to
fast vanishing of the measure, at least in all examples studied so far, {\it cf.} \cite{Marino:1998tb}).}
\be
\langle u^n \rangle \; \ne \; 0 \qquad \Leftrightarrow \qquad n = - \frac{7 \chi + 11 \sigma}{24}
\label{unghostADA1A2}
\ee

Given a scalar (spin-$0$) observable $W_0 = u$ in cohomology of the topological supercharge $Q$,
one can construct topological observables of higher degree via
the standard descent procedure~\cite{Faddeev:1985iz,AlvarezGaume:1983ig,Witten:1988ze}:
\begin{eqnarray}
0 \; = \; i \{ Q, W_0 \} & \qquad , \qquad &
dW_0 \; = \; i \{ Q, W_1 \} \nonumber \\
dW_1 \; = \; i \{ Q, W_2 \} & \qquad , \qquad &
dW_2 \; = \; i \{ Q, W_3 \} \label{descentrels} \\
dW_3 \; = \; i \{ Q, W_4 \} & \qquad , \qquad &
dW_4 \; = \; 0 \nonumber
\end{eqnarray}
Constructed in this way, $W_p$ is a $p$-form on $M_4$ with $U(1)_r$ charge
\be
r (W_p) = r (W_0) - \frac{p}{2}
\ee
This follows from \eqref{descentrels} and the fact that the topological supercharge $Q$
carries $U(1)_r$ charge $r = +\frac{1}{2}$, {\it cf.} \eqref{SUSYtwist}.
Therefore, integrating $W_p$ over a $p$-cycle $\gamma$ in $M_4$ we obtain a topological observable
with $U(1)_r$ charge $\Delta (W_0) - \frac{1}{2} p$,
\be
\CO^{(\gamma)} \; := \; \int_{\gamma} W_{\dim (\gamma)}
\ee
Put differently, given a collection of cycles $\gamma_i \in H_* (M_4)$, we can write
the following correlation function (a generalization of \eqref{uuuu}):
\be
\langle \prod_i \CO^{(\gamma_i)} \rangle \; \ne \; 0 \qquad \Leftrightarrow \qquad
\sum_i \big( 2 \Delta (u) - \dim (\gamma_i) \big) =  - 2 (2a-c) \chi - 3c \, \sigma
\label{generalOOO}
\ee
Motivated by experience with Lagrangian theories, it is convenient to think of this
condition as integration of differential forms of degree $2 \Delta (u) - \dim (\gamma_i)$
over the moduli space $\CM$ of dimension \eqref{rgeneral}.
One novelty of non-Lagrangian theories, though,
is that both the ``degrees'' of the differential forms (on the left-hand side of the condition in \eqref{generalOOO})
as well as the dimension of $\CM$ (= the right-hand side of the condition in \eqref{generalOOO}) are often non-integer.
This is one of the reasons why topological twists of non-Lagrangian 4d theories, while well defined
and interesting, require new techniques and can not be described by the usual methods of cohomological field theory \cite{Witten:1988ze},
as in Donaldson or Seiberg-Witten theory.

In practice, in order to compute the topological correlation function \eqref{generalOOO}
it is convenient to write it as a correlation function in the A-model of $\CM (F_g)$ {\it a la} \eqref{ZZ2d4d}.
With a small abuse of notations --- which, however, is more intuitive and avoids clutter --- we denote
the corresponding operators in 2d theory also by $\CO^{(\gamma_i)}$ or simply $\CO_i$.
In particular, if $\gamma$ is a $p$-cycle in $F_g$ (with $p=0$, $1$, or $2$),
then $\CO^{(\gamma)}$ is a local $(a,c)$ operator in 2d $\CN=(2,2)$ theory $\CM (F_g)$.
This has a useful corollary: upon topological reduction on $F_g$,
each Coulomb branch operator contributes to the spectrum of $(a,c)$ operators ({\it cf.} Table~\ref{tab:N2mplets}):
\be
\boxed{{\text{4d Coulomb branch} \atop \text{operator } u}}
\quad \xrightarrow[~]{~\text{topological reduction on $F_g$}~} \quad
\begin{array}{l@{\;}|@{\;}c}
\multicolumn{2}{c}{~~~~~~~\text{2d}~(a,c)~\text{operators}~} \\[.1cm]
R_A & \text{multiplicity} \\\hline
2 \Delta (u) & 1 \\
2 \Delta (u) - 1 & 2g \\
2 \Delta (u) - 2 & 1
\end{array}
\label{KKaclist}
\ee
where we used the identification $R_A = 2r$.
The topological A-model correlator of such operators on a closed surface of genus $\tilde g$
is non-zero only if the total ghost number anomaly vanishes (see {\it e.g.} \cite{Hori:2003ic}):
\be
\langle \prod_i \CO_i \rangle \; \ne \; 0 \quad \Leftrightarrow \quad
\sum_i R_A (\CO_i) = 2 (1 - \tilde g) \cdot \dim_{\C} \CM (F_g) - 2 \text{deg} \big( \phi^* K_{\CM (F_g)} \big)
\ee
Even though we work with more general 2d $\CN=2$ theories, for clarity
here we use the standard conventions of a sigma-model with the target space $\CM (F_g)$.
The last term is non-zero only for non-trivial topological sectors (2d world-sheet instantons)
which are classified by
\be
H^2 (\CM (F_g), \Z)
\ee
Non-trivial elements of this homology group describe field configurations that correspond to 4d instantons,
{\it cf.} Table~\ref{tab:dict}.
In the zero-instanton sector, the 2d ghost number anomaly matches the 4d ghost number anomaly in \eqref{generalOOO},
provided we identify $R_A = 2r$ and
\be
\dim_{\C} \CM (F_g) \; = \; 4 (2a-c)(g-1)
\label{dimMFC}
\ee
Here, the subscript $\C$ can stand for either ``complex'' or ``Coulomb.''
Indeed, another convenient way to write this expression
is based on the relation between conformal anomaly coefficients of the 4d theory and the scaling dimensions
of the Coulomb branch operators conjectured in \cite{Argyres:2007tq} and further studied in \cite{Shapere:2008zf}:
\be
2a - c \; = \; \frac{1}{4} \sum_{\text{Coulomb}} (2 \Delta (u_i) - 1)
\label{AWSTrelation}
\ee
Specifically, we get
\be
\dim_{\C} \CM (F_g) \; = \; (g-1) \sum_{\text{Coulomb}} (2 \Delta (u_i) - 1)
\ee
If we interpret this as a 2d central charge $\hat c$,
then the operation of adding an extra handle to $F_g$,
that is increasing the genus $g \to g + 1$,
contributes to the 2d central charge
\be
\hat c \, (\text{handle}) \; = \; 2 \Delta (u) - 1
\label{chandleguess}
\ee
For example, a single $U(1)$ vector multiplet in four dimensions has $a = \frac{5}{24}$, $c = \frac{1}{6}$,
and $\Delta (u) = 1$, so that $\hat c \, (\text{handle}) = 1$, in agreement with \eqref{U1target}.
Similarly, when the gauge group is $G$ we get $4 (2a-c) (g-1) = (g-1) \dim G = \dim \text{Bun}_G$.
On the other hand, in two basic rank-1 Argyres-Douglas theories of Table~\ref{tab:AD}
we have $\hat c \, (\text{handle}) = \frac{7}{5}$ and $\hat c \, (\text{handle}) = \frac{5}{3}$, respectively.
Curiously, these are precisely the central charges of $\CN=2$ super-Liouville theories
with levels $k=5$ and $k=3$.

Also note, that matching the $U(1)_r$ anomaly of the UV theory \eqref{rgeneral}
with that of the effective IR theory on the Coulomb branch \eqref{ZCoulomb} determines
the $U(1)_r$ charges of the measure factors $A(u)$ and $B(u)$, {\it cf.} \cite{Argyres:2016xmc,Shapere:2008zf}:
\begin{eqnarray}
r_A & = & \frac{1}{4} n_v - (2a-c) \\
r_B & = & \frac{1}{4} n_v + \frac{1}{8} n_h - \frac{3}{2} c \nonumber
\end{eqnarray}
where $n_v$ and $n_h$ is the number of massless vector and hypermultiplets on the Coulomb branch.
(Unless we deal with an enhanced Coulomb branch, $n_h = 0$.)
In particular, using \eqref{AWSTrelation}, in the case of rank-1 theories ($n_v=1$)
this gives
\be
r_A \; = \; \frac{1 - \Delta (u)}{2}
\ee
and, therefore, according to \eqref{2ddilatonA}:
\be
\Omega (u) \; = \; \frac{1}{\pi i} (1-g) \left(1 - \frac{1}{\Delta (u)} \right) \log u
\ee
\\

In the traditional Donaldson-Witten theory, correlation functions of topological observables \eqref{generalOOO}
can be conveniently packaged into a generating function that,
for a large class of 4-manifolds --- namely, for manifolds of Kronheimer-Mrowka simple type  --- has a rather simple structure:
\begin{multline}
\Big\langle \exp \left( p \cdot u + \sum_i S_i \cdot \CO^{(D_i)} \right) \Big\rangle
\; = \;
\\
\; = \;
\sum_{v \, \in \, \text{vacua}} \;
\exp \left(a_v \chi + b_v \sigma + p u_v + \frac{\eta_v}{2} \sum_{i,j} S_i S_j \, \# (D_i \cap D_j) \right) \, Z_{\text{vortex}}
\phantom{~~~~~~~}
\label{KMsimpletype}
\end{multline}
Here, to avoid clutter, we only included the 0-observable $u$ and 2-observables $\CO^{(D_i)}$
associated with two-dimensional surfaces $D_i \subset M_4$.
The resulting generating function is a function of (formal) variables $p$ and $S_i$,
with some theory-dependent constants $a_v$, $b_v$, $u_v$, and $\eta_v$, indexed by $v$ that runs over a finite set.
Suggestively, this finite set is called the set of ``vacua'' since this is how \eqref{KMsimpletype}
can be interpreted when $M_4$ admits a K\"ahler metric \cite{Witten:1994ev}
and the topological twist requires only $U(1)_R$ symmetry \eqref{U1RCartan}
of the $\CN=1$ supersymmetry (sub)algebra \cite{Johansen:1994aw}.
In this case, the sum over $v$ in \eqref{KMsimpletype} is indeed a sum over isolated massive vacua
of the $\CN=1$ deformation of the theory consistent with topological invariance.

In this interpretation, $u_v := \langle u \rangle_v$ is simply the expectation value of $u$
in a vacuum labeled by $v$. And the factor $Z_{\text{vortex}}$ is a contribution of
the ``vortex strings'' supported on components of the divisor representing the canonical class of $M_4$;
it plays an important role for 2-observables but does not affect correlators of 0-observables which,
thanks to topological invariance, can be placed away from the vortex strings (see \cite{Witten:1994ev} for more details).
Only part of this elegant structure generalizes to topological twists of non-Lagrangian theories,
and the other part is replaced by a different structure, which we describe next.

First, we could not expect \eqref{KMsimpletype} to hold in non-Lagrangian 4d theories considered here
simply because all such theories are superconformal.
Indeed, even in Lagrangian theories, such as $\CN=2$ SQCD with $N_f$ matter fields
in the fundamental representation of the $SU(2)$ gauge group, the instanton contribution
to the $U(1)_r$ ghost number anomaly is multiplied by $4-N_f$, see {\it e.g.} \cite{Losev:1997tp}.
In particular, it vanishes when $N_f = 4$, {\it i.e.} when the theory is superconformal.
Then, the exponentials on the right-hand side of \eqref{KMsimpletype} that package Donaldson
polynomials of all degrees ``truncate'' to a single polynomial,
whose degree is set by the virtual dimension \eqref{rgeneral}.

This is precisely what happens in all our examples of non-Lagrangian theories;
even if there are instanton-like non-perturbative effects in such theories,
they do not affect the ghost number anomaly \eqref{rgeneral}.
In other words, instead of exponentials on the right-hand side of \eqref{KMsimpletype},
in non-Lagrangian theories we have only one polynomial in $p$ and $S_i$, whose degree is set by \eqref{generalOOO}.
Of course, if the non-Lagrangian theory in question has a flavor symmetry, we can consider
an equivariant version of the invariants, summed over flux sectors for the flavor symmetry group.
Such generalization will formally look similar to the traditional Donaldson-Witten theory \eqref{KMsimpletype},
as long as the cofficient $k$ of the first term in \eqref{rgeneral} is non-zero.

We can, however, hope to find the structure \eqref{KMsimpletype} for individual correlators even in non-Lagrangian theories.
Thus, optimistically, we could expect
that, for a certain class of 4-manifolds, the partition function \eqref{ZZ2d4d} without any observables
can be written as a sum
\be
Z(M_4) \; = \; \sum_{v \, \in \, \text{vacua}} \; e^{a_v \chi + b_v \sigma}
\label{Zsmpltp}
\ee
such that the constants $a_v$ and $b_v$ depend on the choice of 4d theory and its vacuum, $v$,
but not on the 4-manifold $M_4$.
And, similarly, we could expect a correlation function of 0-observables \eqref{uuuu}
to have the following structure:
\be
\langle u^n \rangle \; = \;
\sum_{v \, \in \, \text{vacua}} \; (u_v)^n \, e^{a_v \chi + b_v \sigma}
\label{uuusmpltp}
\ee
Of course, these invariants can be non-zero only if the conditions \eqref{rgeneral} and \eqref{uuuughost} are satisfied.
As we shall see in examples, however, the structure \eqref{Zsmpltp} and \eqref{uuusmpltp} is a bit too optimistic
and, even for simple K\"ahler 4-manifolds (and without 2-observables), the correct structure is similar to the
generating function of Seiberg-Witten invariants, now in non-Lagrangian 4d TQFT:
\be
\langle u^n \rangle \; = \;
\sum_{v \, \in \, \text{vacua}} \; (u_v)^n \, e^{a_v \chi + b_v \sigma}
\sum_{\lambda \, \in \, \text{Basic}} Z_{\text{vortex}} (v,\lambda)
\label{uuusmpltpvortex}
\ee
As in Seiberg-Witten theory, we call $\lambda \in H^2 (M_4, \Z)$ which contribute to this sum ``basic classes''
and, by analogy with \eqref{KMsimpletype}, interpret their contribution as vortex strings supported on components
of the divisor representing $\hat \lambda \in H_2 (M_4)$.
These contributions are important even when $n=0$, {\it i.e.} for the computation of the partition function \eqref{ZZ2d4d}
without any observables on $M_4$.

As we explain below, the structure \eqref{uuusmpltpvortex} is indeed rather natural and,
in some cases, can even be interpreted as a sum over Coulomb branch and Higgs branch vacua
of the A-model {\it a la} \cite{Vafa:1990mu,Melnikov:2005tk}.
For example, the simplest instance of this phenomenon is when $M_4 = F_g \times \tilde F_{\tilde g}$
is a product of $F_g$ with another surface of genus $\tilde g$.
(Note, this is also an example where $M_4$ admits a K\"ahler metric.)
Then, under favorable conditions that will be summarized in a moment,
a topological correlator in 4d non-Lagrangian theory
is equal to the corresponding genus-$\tilde g$ correlator in the A-model of $\CM (F_g)$:
\be
\langle \CO_1 \ldots \CO_n \rangle_{\text{A-model}} \; = \;
\sum_{\text{vacua} : \;  d \tilde \CW_{\text{eff}} = 0} \, \CO_1 (v) \ldots \CO_n (v)
\left( e^{2\pi i \Omega (u_v)} \, \det \text{Hess} \, \tilde \CW_{\text{eff}} \right)^{{\tilde g}-1}
\label{AmodelOOO}
\ee
or, yet another way, to a genus-$g$ correlator in the A-model of $\CM (\tilde F_{\tilde g})$.
This happens when the A-model has a phase where all vacua are realized as Coulomb branch vacua which,
in turn, are the critical points of the effective twisted superpotential $\tilde \CW_{\text{eff}} (u)$.

As a special important example of the relation between \eqref{uuusmpltpvortex} and \eqref{AmodelOOO},
directly related to trisections, let us consider $M_4 = F_g \times S^2$.
Furthermore, let us for a moment focus on a class of 4d non-Lagrangian theories whose $\CN=1$ deformations
have the same number of vacua on $S^2$ as in flat space. Then, first reducing on a 2-sphere,
we should expect a direct match between the vacua in \eqref{uuusmpltpvortex} and critical points in \eqref{AmodelOOO}.
In other words, for such 4d theories there exists a function $\tilde \CW_{\text{eff}} (u)$
that depends only on 4d theory and not on the 4-manifold $M_4$ (or genus $g$, for the case at hand),
such that the sum in \eqref{uuusmpltpvortex} is precisely the sum over critical points of $\tilde \CW_{\text{eff}} (u)$,
and $u_v$ are the corresponding values of $u$:
\be
u_v : \qquad \exp \left( \frac{\partial \tilde \CW_{\text{eff}}}{\partial u} \right) \; = \; 1
\label{Wexpcrit}
\ee
Moreover, in such non-Lagrangian theories
\be
a_v \; = \; - \frac{\pi i}{2} \Omega (u_v) - \frac{1}{4} \log \left( \det \text{Hess} \, \tilde \CW_{\text{eff}} \right) \vert_{u=u_v}
\ee
This is essentially another manifestation of the relation between 4d and 2d chiral rings
(see section \ref{sec:AD} for a further discussion in the context of Argyres-Douglas theories).
Generalizations that do not rely on the above assumptions require understanding $\CN=1$ deformations of a given
non-Lagrangian theory that, on the one hand, are consistent with topological invariance and,
on the other hand, lead to a massive theory in the infra-red.

Note, our present discussion of the Coulomb branch localization in $\CM (F_g)$ suggests a potentially
interesting application to the Bethe/gauge correspondence \cite{Nekrasov:2009uh} and questions like this:
Is there a quantum integrable system that corresponds, say, to the Argyres-Douglas theory $(A_1,A_3)$ and genus $g=2$?
If the answer is yes, it would be interesting to proceed further and identify integrable systems
labeled by a choice of non-Lagrangian 4d theory and $g$ (= genus of $F_g$).
Perhaps one way to tackle these questions is to study a special case of the 6d fivebrane theory
compactified on a 4-manifold \cite{Dedushenko:2017tdw,Gadde:2013sca,Putrov:2015jpa}
which, for our purposes here, should be taken to be a product of two Riemann surfaces,
one of which is singular.

It should also be noted that, while it can be very useful for applications to K\"ahler 4-manifolds,
the Coulomb branch localization in the 2d theory $\CM (F_g)$ captures only part of the physics of the superconformal theory $\CM (F_g)$.
In particular, it obscures the action of the mapping class group \eqref{MCGonMF} and the analysis
of the Heegaard branes that we need for applications to general trisections.
(Note that, in the wild world of 4-manifolds, K\"ahler ones are rather special.)
Plus, of course, the Coulomb branch localization in general may not be sufficient for
computing the partition function \eqref{ZZ2d4d}, as it happens {\it e.g.} in \cite{Melnikov:2005tk}.

Another obvious challenge involves incorporating 2-observables; it requires a detailed understanding
of the factor $Z_{\text{vortex}}$ in \eqref{uuusmpltpvortex} that accounts for vortex strings,
which is highly non-trivial even in Lagrangian theories with non-abelian gauge groups.
In the case of $(A_1,A_3)$ Argyres-Douglas theory, vortex strings were studied in \cite{Tong:2006pa},
where it was proposed that their world-sheet theory is described by 2d $\CN=(2,2)$ minimal model $A_1$,
{\it i.e.} the IR fixed point of the Landau-Ginzburg theory with the cubic twisted superpotential, {\it cf.} \eqref{toytube}:
\be
\tilde \CW \; = \; x^3
\label{WA1twisted}
\ee
Assuming these are the right vortex strings for our applications here, this proposal
suggests an intriguing generalization of \eqref{KMsimpletype} where the divisor representing
the canonical class of a K\"ahler 4-manifold $M_4$ carries a suitable number of copies of the $A_1$ model.
In particular, for the topological reduction on $F_g$, it implies that ``adding a handle'' to the surface $F_g$,
{\it i.e.} increasing its genus by 1, adds to the 2d theory $\CM (F_g)$
some fixed number of $A_1$ supersymmetric minimal models \eqref{WA1twisted}.
We present some tests of this intriguing scenario in section~\ref{sec:AD}.

To the best of our knowledge, not much is known about vortex strings in the Argyres-Douglas theory $(A_1,A_2)$
and there is no proposal analogous to \cite{Tong:2006pa}.\footnote{Looking at surface operators --- which are
basically non-dynamical vortices --- may offer some help. Thus, sending the simplest surface operator
(represented by a single M2-brane in the M-theory construction) to probe
the Argyres-Douglas singularity $P =x^N \pm 2 \Lambda^{N}$ \cite{Argyres:1995jj}
returns $\tilde \CW' = - \log \frac{P + \sqrt{P^2 - 4 \Lambda^{2N}}}{2}$, see {\it e.g.} \cite{Gaiotto:2013sma}.
Expanding this expression and the corresponding chiral ring relation $\exp (\tilde \CW') = 1$
near $x=0$ gives the leading behavior $\sim x^{N/2}$ that agrees with \eqref{WA1twisted} for $N=2$
and suggests $\tilde \CW \sim x^{\frac{N}{2}+1}$ for general $N$.
One should be careful with such interpretation, though, since the full system has $N$ vacua.
Also to keep in mind is that, while surface operators can be regarded as a non-dynamical limit
of vortex strings with respect to parameters which luckily drop out after the topological twist,
the limiting procedure should be treated with great care since we are dealing with massless theories.}

\subsection*{Refinement}

The truncation of \eqref{KMsimpletype} to a single polynomial of degree \eqref{generalOOO} discussed above has a flip side:
the same superconformal symmetry that was responsible for this effect guarantees a non-anomalous $U(1)_r$ symmetry.
Mathematically, it means that homology groups {\it a la} Floer, $\CH (M_3)$, assigned by 4d non-Lagrangian TQFT
to 3-manifolds carry a $\Z$ grading by $U(1)_r$ symmetry, analogous to a $\Z_8$ grading in the standard Donaldson-Floer theory.
The situation is similar for homological invariants of 3-manifolds that categorify Witten-Reshetikhin-Turaev invariants~\cite{Gukov:2016gkn}.

In particular, this $U(1)_r$ symmetry can be extremely useful in ``regularizing'' otherwise divergent partition
functions on $M_4 = S^1 \times M_3$ by replacing the trace over an infinite-dimensional space $\CH (M_3)$
with its graded version,
\be
\dim_{{\frak t}} \CH (M_3) \; : = \; \sum_r {\frak t}^r \dim \CH_r (M_3)
\label{trefinement}
\ee
It is well-defined as long as each graded component is finite-dimensional, $\dim \CH_r (M_3) < \infty$.
We adopt this refinement of \eqref{ZZ2d4d} when we work with $M_4 = S^1 \times M_3$ or $M_4 = T^2 \times F_g$.

Even for more general 4-manifolds, the ingredients (``constants'') in \eqref{uuusmpltpvortex} can be replaced by functions of ${\frak t}$.
One advantage of doing this is that it gives an opportunity to study various limits, as in \cite{Gukov:2016gkn},
{\it e.g.} the unrefined limit ${\frak t} \to 1$ or the limit ${\frak t} \to 0$
which has the effect of lifting the Coulomb branch in our rank-1 examples.


\section{$\CN=3$ theories}
\label{sec:N3}

A nice class of 4d non-Lagrangian theories, that only recently came in the spotlight,
consists of superconformal theories with $\CN=3$ supersymmetry.
The constraints imposed by supersymmetry put such theories right in-between
SCFTs with $\CN=2$ and $\CN=4$ supersymmetry, making them convenient examples
in various problems, including the study of topological twists and trisections.

For example, much like their $\CN=4$ cousins, $\CN=3$ theories
have no global symmetries except the R-symmetry $SU(3)_R \times U(1)_{{\tilde r}}$,
and the conformal anomaly coefficients in such theories are not independent \cite{Aharony:2015oyb}:
\be
a \; = \; c
\ee
In particular, the F-theory construction \cite{Garcia-Etxebarria:2015wns,Aharony:2016kai}
that involves $N$ D3-branes probing a non-perturbative $\Z_k$ orbifold
gives a large family of $\CN=3$ theories labeled by $N \in \Z_{+}$, $k = 2, 3, 4, 6$,
and $\ell \vert k$, with central charges
\be
a \; = \; c \; = \; \frac{1}{4} k N^2 + \frac{1}{4} (2 \ell - k - 1) N
\label{N3ac}
\ee
The special case $k=2$ involves the ordinary type IIB orientifolds,
whereas other values of $k=3,4,6$ correspond to more esoteric ``S-folds.''
Rank-1 theories, that we like to use as our examples in this paper,
have central charges given by \eqref{N3ac} with $N=1$
and have only one Coulomb branch operator with \cite{Nishinaka:2016hbw}:
\be
\Delta (u) \; = \; \ell
\ee
Theories with $\ell = 1$ and $2$ turn out to have enhanced $\CN=4$ supersymmetry,
so that the ``minimal'' pure $\CN=3$ theory has $N=1$ and $k = \ell =3$.

Now, let us take a look at the topological twist of these theories,
in particular, its implementation based on trisections of 4-manifolds.
In the case of rank-1 $\CN=3$ theories on 4-manifolds with $b_2^+ > 1$, from \eqref{uuuughost}
we obtain the following analogue of the condition \eqref{NfSWdim} on `basic classes':
\be
\langle u^n \rangle \; \ne \; 0 \qquad \Leftrightarrow \qquad
n = - \frac{(2 \chi + 3 \sigma)(2 \ell - 1)}{8 \ell}
\label{N3ucorrelators}
\ee
where $n$ has to be integer, of course.
This `selection rule' can be easily generalized to higher-rank $\CN=3$ theories
and more general topological observables, {\it cf.} \eqref{generalOOO}.
Note, since $\CN=3$ theories have no global symmetries aside from R-symmetry,
there is no equivariant version of the corresponding 4-manifold invariants
and there is no way to activate the first term in \eqref{rgeneral}.

\subsection{$\CN=3$ version of the Hitchin moduli space}

In order to compute the topological invariants \eqref{N3ucorrelators} in practice
by using the trisection approach, we need to know the 2d theory $\CM (F_g)$
obtained by topological reduction on a surface of genus $g$.
By analogy with the earlier example \eqref{MHtarget} and for reasons that will become clear shortly,
$\CM (F_g)$ in this section can be called ``$\CN=3$ version of the Hitchin moduli space.''
Apart from the surface $F_g$, it is labeled by the same data as the $\CN=3$ theory in question;
for example, in the case of $\CN=3$ theories constructed via S-folds, $\CM (F_g)$ is labeled by
$N$, $k$, and $\ell$.

What does this $\CN=3$ version of the Hitchin moduli space look like?
The first interesting feature of $\CM (F_g)$, already mentioned in Table~\ref{tab:SUSY},
is that larger supersymmetry in four dimensions does not necessarily lead to larger supersymmetry in two dimensions.
Thus, $\CM (F_g)$ preserves only $\CN=(2,2)$ supersymmetry when $g \ne 1$.
In other words, if $\CM (F_g)$ were described by classical geometry,
it should be thought of as a K\"ahler rather than hyper-K\"ahler manifold.
As promised in section~\ref{sec:trisec}, we demonstrate this by writing
explicitly the supercharges of the 4d $\CN=3$ supersymmetry algebra
and their behavior under the partial topological twist along $F_g$.
The result of this simple exercise is presented in Table~\ref{tab:N3Q}.
(Our conventions are consistent with those in \cite{Nishinaka:2016hbw,Putrov:2015jpa}.)
Only 4 out of 12 supercharges transform as scalars on $F_g$,
the same four that are present in $\CN=2$ subalgebra.

\begin{table}
\begin{centering}
\begin{tabular}{|c|c|c|c|c|c|c|}
\hline
~~~ & ~$j_1$~ & ~$j_2$~ & ~$R$~ & ~$r$~ & ~$F$~ & ~$R + j_1 - j_2$~
\tabularnewline
\hline
\hline
${Q^1}_-$ & $- \frac{1}{2}$ & $0$ & $\frac{1}{2}$ & $\frac{1}{2}$ & $0$ & $0$
\tabularnewline
${Q^1}_+$ & $\frac{1}{2}$ & $0$ & $\frac{1}{2}$ & $\frac{1}{2}$ & $0$ & $1$
\tabularnewline
${Q^2}_-$ & $- \frac{1}{2}$ & $0$ & $- \frac{1}{2}$ & $\frac{1}{2}$ & $0$ & $-1$
\tabularnewline
${Q^2}_+$ & $\frac{1}{2}$ & $0$ & $- \frac{1}{2}$ & $\frac{1}{2}$ & $0$ & $0$
\tabularnewline
\hline
$\tilde Q_{2 \dot -}$ & $0$ & $- \frac{1}{2}$ & $\frac{1}{2}$ & $- \frac{1}{2}$ & $0$ & $1$
\tabularnewline
$\tilde Q_{2 \dot +}$ & $0$ & $\frac{1}{2}$ & $\frac{1}{2}$ & $- \frac{1}{2}$ & $0$ & $0$
\tabularnewline
$\tilde Q_{1 \dot -}$ & $0$ & $- \frac{1}{2}$ & $- \frac{1}{2}$ & $- \frac{1}{2}$ & $0$ & $0$
\tabularnewline
$\tilde Q_{1 \dot +}$ & $0$ & $\frac{1}{2}$ & $- \frac{1}{2}$ & $- \frac{1}{2}$ & $0$ & $-1$
\tabularnewline
\hline
${Q^3}_+$ & $\frac{1}{2}$ & $0$ & $0$ & $- \frac{1}{2}$ & $1$ & $\frac{1}{2}$
\tabularnewline
${Q^3}_-$ & $- \frac{1}{2}$ & $0$ & $0$ & $- \frac{1}{2}$ & $1$ & $- \frac{1}{2}$
\tabularnewline
$\tilde Q_{3 \dot +}$ & $0$ & $\frac{1}{2}$ & $0$ & $\frac{1}{2}$ & $-1$ & $-\frac{1}{2}$
\tabularnewline
$\tilde Q_{3 \dot -}$ & $0$ & $- \frac{1}{2}$ & $0$ & $\frac{1}{2}$ & $-1$ & $\frac{1}{2}$
\tabularnewline
\hline
\end{tabular}
\par\end{centering}
\caption{\label{tab:N3Q} Supercharges of the 4d $\CN=3$ supersymmetry. Upon partial topological twist,
the spin $j_1 - j_2$ along $F_g$ is replaced by a linear combination $R + j_1 - j_2$ summarized in the last column.
It clearly illustrates that topological reduction of a 4d $\CN=3$ theory on $F_g$ preserves only 2d $\CN=(2,2)$ supersymmetry,
just like topological reduction of a 4d $\CN=2$ theory.}
\end{table}

In order to describe $\CM (F_g)$ in more detail,
recall that $\CN=3$ theories of García-Etxebarria and Regalado \cite{Garcia-Etxebarria:2015wns}
are constructed by taking the ``quotient'' of the $\CN=4$ super-Yang-Mills with gauge group $G = U(N)$
by a $\Z_k$ symmetry that combines the R-symmetry and $SL(2,\Z)$ duality.
The latter acts on 4d gauge coupling in the usual way,
\be
\tau_{\text{4d}} \; \mapsto \; \frac{a \tau_{\text{4d}} + b}{c \tau_{\text{4d}} + d}
\ee
with the appropriate order-$k$ elements listed in Table~\ref{tab:N3tau}.
Topological reduction of the 4d $\CN=4$ super-Yang-Mills on $F_g$ gives 2d $\CN=(4,4)$ sigma-model
whose target space is $\CM_H (G,F_g)$ and the complexified K\"ahler parameter $\rho$ is given by $\tau_{\text{4d}}$.
In particular, $SL(2,\Z)$ duality of the 4d theory acts on $\rho$,
\be
SL(2,\Z)_{\rho} : \qquad \rho \; \mapsto \; \frac{a \rho + b}{c \rho + d}
\label{SL2Zrho}
\ee
and its standard generators,
the $S = \begin{pmatrix} 0 & -1 \\ 1 & 0 \end{pmatrix}$
and $T = \begin{pmatrix} 1 & 1 \\ 0 & 1 \end{pmatrix}$ elements,
act on the Hitchin moduli space $\CM_H (G,F_g)$
as mirror symmetry (Langlands duality) and $B$-field transform, respectively \cite{Bershadsky:1995vm,Harvey:1995tg,KW}.

Therefore, upon topological reduction on a surface $F_g$, a rather unusual ``quotient''
of the $\CN=4$ super-Yang-Mills becomes a slightly more familiar quotient of
the 2d sigma-model with target space $\CM_H (G,F_g)$.
It is still somewhat peculiar even in two dimensions, because part of the $\Z_k$ action
involves mirror symmetry (T-duality along SYZ fibers of the Hitchin fibration).
Indeed, the part of $\Z_k$ action that involves R-symmetry is a simple geometric symmetry;
it acts on eigenvalues of the Higgs field by multiplication with $k$-th roots of unity,
\be
z_j \; \mapsto \; e^{2\pi i /k} \, z_j
\label{ZkHiggsfield}
\ee
However, the part of the $\Z_k$ quotient that involves $SL(2,\Z)$ action \eqref{SL2Zrho} is more interesting.
In string theory literature, such spaces go by the name {\it asymmetric orbifolds} or {\it T-folds}.

In our present case, the asymmetric orbifold combines the ordinary, geometric $\Z_k$ quotient
on the Hitchin base ${\bf B}$ with the order-$k$ Fourier-Mukai transform on the Hitchin fibers
(which, in turn, is a combination of $S = \{$mirror symmetry$\}$ and $T = \{ B$-field transform$\}$).
For example, when $k = 4$, the action on the Hitchin fibers is literally by mirror symmetry.
On the other hand, in order to understand what happens on the Hitchin base, it is convenient to
think of $\CM (F_g)$ as a locus of singularities in the moduli space of $U(kN)$ Higgs bundles,
\be
\CM (F_g) \quad \xhookrightarrow{~~~} \quad \CM_H (U(kN),F_g)
\label{MFMH}
\ee
much like the moduli space of $Sp(2N)$ Higgs bundles arises as a locus of singularities in
the moduli space of $SU(2N)$ Higgs bundles, or as the moduli space of unramified Higgs bundles
is a locus of singularities in the moduli space of Higgs bundles with ramification.
Indeed, the characteristic polynomial of $U(kN)$ Higgs bundles
\be
\det (x - \Phi) \; = \;
x^{kN} + u_1 x^{kN-1} + \ldots + u_{kN} \; = \; 0
\label{kNcurve}
\ee
defines the spectral curve in the total space of the canonical bundle $K$.
The coefficients $u_n \in H^0 (F_g, K^n)$ are holomorphic sections of $K^n$;
they define a map from $\CM_H (U(kN),F_g)$ to the base of the Hitchin fibration,
\be
{\bf B} \; = \; \bigoplus_{n=1}^{kN} H^0 (F_g , K^n)
\ee
whose generic fibers are complex tori of dimension $g + (k^2 N^2 -1)(g-1)$
that can be identified with the Jacobian of the spectral curve \eqref{kNcurve}.

\begin{table}
\begin{centering}
\begin{tabular}{c|ccc}
~~~ & ~$k=3$~ & ~$k=4$~ & ~$k=6$~
\tabularnewline
\hline
$\tau_{\text{4d}}$ & $\phantom{\oint^{\oint}_{\oint}} e^{\pi i /3} \phantom{\oint^{\oint}_{\oint}}$ & $\phantom{\oint^{\oint}_{\oint}} i \phantom{\oint^{\oint}_{\oint}}$ & $\phantom{\oint^{\oint}_{\oint}} e^{\pi i /3} \phantom{\oint^{\oint}_{\oint}}$
\tabularnewline
duality & $\begin{pmatrix} 0 & -1 \\ 1 & -1 \end{pmatrix}$ & $\begin{pmatrix} 0 & -1 \\ 1 & 0 \end{pmatrix}$ & $\begin{pmatrix} 1 & -1 \\ 1 & 0 \end{pmatrix}$
\tabularnewline
\end{tabular}
\par\end{centering}
\caption{\label{tab:N3tau} Order-$k$ elements of the $SL(2,\Z)$ duality group
and the corresponding values of $\tau_{\text{4d}}$ fixed by the $\Z_k$ action.}
\end{table}

Now consider such $U(kN)$ Higgs bundles that the eigenvalues of the Higgs field
come in groups of $k$, related by the $\Z_k$ action \eqref{ZkHiggsfield}.
Since the coefficients $u_n$ are symmetric polynomials of $z_i$'s,
such configurations define a subspace of ${\bf B}$,
\be
{\bf B}_k \; = \; \bigoplus_{j=1}^{N} H^0 (F_g , K^{jk})
\label{Bkbase}
\ee
where the characteristic polynomial has the form
\be
\det (x - \Phi) \; = \;
x^{kN} + u_k x^{kN-k} + \ldots + u_{k(N-1)} x^k + u_{kN}
\ee
In other words, $u_n =0$ unless $n$ is an integer multiple of $k$.
Using
\be
\dim H^0 (F_g, K^n) \; = \; (2n-1) (g-1)
\label{HKdim}
\ee
we find
\be
\dim_{\C} {\bf B}_k \; = \; N (kN + k - 1) (g-1)
\ee
Note, when $k=2$ this agrees with the dimension of the Hitchin base for $Sp(2N)$ and $SO(2N+1)$ Higgs bundles.
For us, though, the most interesting cases are $k=3$, $4$ and $6$, which generalize
the moduli space of Higgs bundles to systems labeled by Shephard-Todd complex reflection groups
instead of Lie groups $G$.

The list of invariant polynomials $\{ u_k, \ldots, u_{kN} \}$ correctly accounts
for all Coulomb branch parameters of 4d $\CN=3$ theories when $\ell = k$.
For other variants of $\CN=3$ theories labeled by $\ell \ne k$, the correct list
includes the generalized Pfaffian operator $(z_1 z_2 \ldots z_N)^{\ell}$
and looks like $\{ u_k , u_{2k} , \ldots , u_{Nk-k} , u_{N\ell} \}$ \cite{Aharony:2016kai}.
Therefore, following the same reasoning as above, we conclude that after topological
reduction on $F_g$ the base of the $\CN=3$ version of the Hitchin system is, {\it cf.} \eqref{Bkbase}:
\be
{\bf B}_{k, \ell} \; = \; H^0 (F_g, K^{N \ell}) \oplus \bigoplus_{j=1}^{N-1} H^0 (F_g , K^{jk})
\label{Bklbase}
\ee
Again, note that for $k=2$ this gives the correct base of the Hitchin fibration for $SO(2N)$ Higgs bundles.
Using \eqref{HKdim}, we find the dimension of the vector space \eqref{Bklbase}:
\be
\dim_{\C} {\bf B}_{k,\ell} \; = \; N (kN + 2\ell - k - 1) (g-1)
\ee
which agrees with the general expression \eqref{dimMFC} and \eqref{N3ac}.

If all this sounds a bit too abstract, as usual a low genus and low rank example should help.
Consider the case of $g=1$ and $N=1$, that is $G=U(1)$ in the 4d $\CN=4$ theory before the $\Z_k$ quotient.
Then, the topological reduction of 4d $\CN=3$ theory on $F_g \cong T^2$ gives the following
asymmetric orbifold:
\be
\CM (F_g) \; = \; \frac{\C^3 \times E}{\Z_k}
\label{MFN3N1g1}
\ee
where $\Z_k$ acts on $E$ by an order-$k$ element of $SL(2,\Z)_{\rho}$
and on $\C^3$ via $(u,v,w) \mapsto (e^{2\pi i /k} u, e^{- 2\pi i /k} v, e^{2\pi i /k} w)$.
As before, this action fixes the complexified K\"ahler modulus $\rho$.
Note that $SL(2,\Z)_{\tau}$ still acts,
and has to be identified with the action of the mapping class group \eqref{MCGonMF}.
Up to exchange of $SL(2,\Z)_{\rho}$ and $SL(2,\Z)_{\tau}$,
the target space \eqref{MFN3N1g1} of our 2d theory $\CM (F_g)$ is basically
the target space of the F-theory construction in \cite{Garcia-Etxebarria:2015wns}.

In order to see the origin of \eqref{MFN3N1g1} and to describe the Heegaard branes,
recall that topological reduction of 4d abelian $\CN=4$ super-Maxwell theory on $F_g = T^2$
gives a 2d sigma-model with the target space
\be
\C^3 \, \times \, E \quad \cong \quad \mathbb{H} \, \times \, \CM_H (U(1), F_g)
\label{N8target}
\ee
where $E = T^2$ is parametrized by holonomies of the $U(1)$ gauge field,
and three copies of $\C$ are parametrized by complex scalars in three chiral
multiplets that appear in $\CN=1$ decomposition of the $\CN=4$ supermultiplet.
Two of these chirals can be combined into a full $\CN=2$ hypermultiplet;
as will be discussed further in section \ref{sec:AD},
since they carry R-charge $R=1$ they do not contribute to the Kaluza-Klein spectrum
of the higher-genus topological reduction, defined by coupling to the background \eqref{U1RCartan}.
This is the reason why, in genus-$1$ case with no topological twist, the resulting
target space \eqref{N8target} is larger compared to what one might expect from the higher-genus analogues \eqref{MHtarget}.

In order to compute 4-manifold invariants in the topologically twisted $\CN=3$ theory,
we need to compute disk amplitudes \eqref{ZZ2d4d} that, apart from $\CM (F_g)$ itself,
also involve Heegaard branes in 2d theory $\CM (F_g)$ and the action of the mapping class group \eqref{MCGonMF}.
As explained around \eqref{BHalpha}, it suffices to know only the basic boundary condition $\CB_H \; = \; \CB (\natural^g (S^1 \times B^2))$
since the entire set of Heegaard branes can be generated from it by the action of $\text{MCG} (F_g)$.

As far as the mapping class group action is concerned, the $\CN=3$ version of the Hitchin moduli space
introduced here is very similar to the ordinary Hitchin moduli space; in particular,
the action of $\text{MCG} (F_g)$ on branes is a straightforward generalization of \cite{Gukov:2007ck,Dimofte:2011jd}.
We leave the exciting project of studying branes $\CB_H$ in the self-mirror 2d $\CN=(2,2)$ theories
based on $\CN=3$ version of the Hitchin moduli space to future work.


\section{Argyres-Douglas theories}
\label{sec:AD}

Another interesting (and, perhaps, the oldest) class of non-Lagrangian 4d $\CN=2$ theories goes back to
the original work of Argyres and Douglas \cite{Argyres:1995jj}, who discovered a strongly interacting SCFT
in the moduli space of 4d $\CN=2$ super-Yang-Mills theory with gauge group $SU(3)$.
Nowadays, this theory goes by many different names
\be
H_0 = (A_1,A_2) = I_{3,2} : \qquad y^2 = x^3 + u
\label{A1A2curve}
\ee
and has been generalized in various directions. In the present section we wish to study
topological reduction of this, still growing, class of non-Lagrangian 4d $\CN=2$ theories
and discuss its application to trisections of 4-manifolds.

As in the rest of this paper, for concreteness we shall focus on rank-1 theories,
namely on the two simplest examples summarized in Table~\ref{tab:AD}.
The first example is the original theory of Argyres and Douglas with
the Seiberg-Witten curve \eqref{A1A2curve}, and the second example
is its close cousin that can be found in the moduli space of $SU(4)$ super-Yang-Mills:
\be
H_1 = (A_1,A_3) = (A_1,D_3) : \qquad y^2 = x^3 + ux
\label{A1A3curve}
\ee
In our discussion below, we shall refer to these two theories as $(A_1,A_2)$ and $(A_1,A_3)$, respectively.
The former has no Higgs branch and no flavor symmetry, whereas the latter has
a one-dimensional Higgs branch that can be identified with a 1-instanton moduli space for $SU(2)$.

\subsection{Elliptic genus of $\CM (F_g)$}

In order to identify the 2d $\CN=(2,2)$ theory $\CM (F_g)$ obtained by topological reduction of
a (generalized) Argyres-Douglas theory on a surface $F_g$ of genus $g$, it is convenient
to study its elliptic genus. The latter is equal to the $T^2 \times F_g$ partition function
of the four-dimensional theory, partially twisted along $F_g$.

Recall, that the 2d index (a.k.a. equivariant elliptic genus) of a $(0,2)$ Fermi multiplet
is given by \cite{Gadde:2013wq,Gadde:2013ftv,Benini:2013nda}:
\be
\CI_{\text{Fermi}} (x;q) \; = \; \theta (x;q)
\ee
where\footnote{In \cite{Honda:2015yha} and \cite{Benini:2016hjo} the same expression is written
as a ratio of the Dedekind eta-function $\eta (q) = q^{\frac{1}{24}} \prod_{n=1}^{\infty} (1 - q^n)$
and the Jacobi theta-function
\be
\theta_1 (x;q) \; = \; -i q^{\frac{1}{8}} x^{\frac{1}{2}}
\prod_{n=1}^{\infty} (1-q^n) (1-xq^n) (1 - x^{-1} q^{n-1}) \,.
\ee}
\be
\theta (x;q) \; = \; q^{\frac{1}{12}} x^{- \frac{1}{2}} \prod_{n=0}^{\infty} (1 - x q^n) (1 - x^{-1} q^{n+1})
\label{mytheta}
\ee
Similarly, the equivariant elliptic genus of a 2d $(0,2)$ chiral multiplet is
\be
\CI_{\text{chiral}} (x;q) \; = \; \frac{1}{\theta (x;q)}
\ee
Combining the two we can obtain the index of a 2d $\CN=(2,2)$ chiral multiplet, {\it etc.}

\begin{figure}[ht]
\centering
\includegraphics[trim={0 0.5in 0 0.5in},clip,width=4.5in]{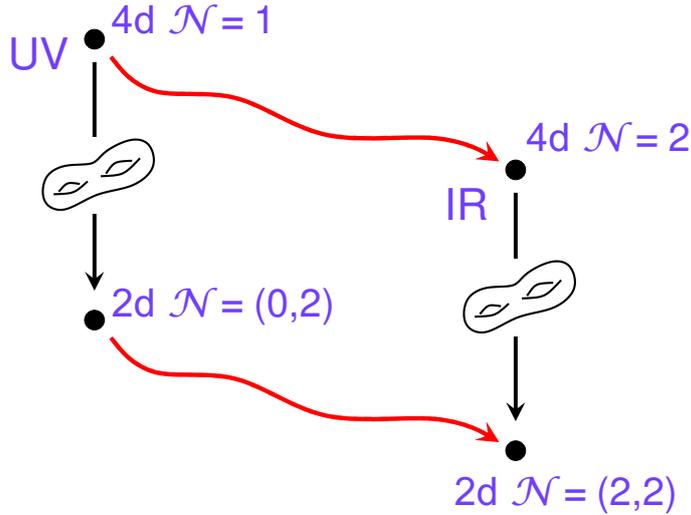}
\caption{Topological reduction of the endpoints of the four-dimensional
RG flow from a UV theory with $\CN=1$ supersymmetry to an IR theory with $\CN=2$ supersymmetry.
When the diagram commutes, 2d $\CN=(2,2)$ theory $\CM (F_g)$ can be identified by studying 2d RG flow.}
\label{fig:2d4dRGflows}
\end{figure}

For a 2d $\CN=(2,2)$ theory $\CM (F_g)$, defined via topological reduction
of a non-Lagrangian 4d $\CN=2$ theory on a surface $F_g$ of genus $g$,
the elliptic genus is equal to the partition function of the four-dimensional theory on $T^2 \times F_g$.
Moreover, these 4d and 2d partition functions are invariant under RG flow
and require only $\CN=1$ and $\CN=(0,2)$ supersymmetry, respectively.
This can be extremely handy for those 4d non-Lagrangian theories that can be realized as end-points of
RG flows from Lagrangian 4d $\CN=1$ theories, as {\it e.g.} $E_6$ SCFT \cite{Gadde:2015xta}
and many Argyres-Douglas theories \cite{Maruyoshi:2016tqk,Maruyoshi:2016aim,Agarwal:2016pjo}.

Thus, both of our main examples, \eqref{A1A2curve} and \eqref{A1A3curve},
can be realized as IR fixed points of RG flows in certain variants
of 4d $\CN=1$ adjoint SQCD with $SU(2)$ gauge group and $N_f = 1$.
\begin{table}[htb]
\be
\begin{array}{l@{\;}|@{\;}cccc}
\multicolumn{5}{c}{(A_1,A_2)} \\[.1cm]
& q & q' & \phi & u \\\hline
SU(2)_{\text{gauge}} & \Box & \Box & \text{adj} & 1 \\
U(1)_R & 1 & 1 & 0 & 0 \\
U(1)_{r} & \frac{2}{5} & -\frac{1}{5} & \frac{1}{5} & \frac{6}{5}
\end{array}
\qquad\qquad
\begin{array}{l@{\;}|@{\;}cccc}
\multicolumn{5}{c}{(A_1,A_3)} \\[.1cm]
& q & \tilde q & \phi & u \\\hline
SU(2)_{\text{gauge}} & \Box & \Box & \text{adj} & 1 \\
U(1)_{R} & 1 & 1 & 0 & 0 \\
U(1)_{r} & - \frac{1}{6} & - \frac{1}{6} & \frac{1}{3} & \frac{4}{3}
\end{array}
\notag \ee
\caption{Field content of 4d $\CN=1$ Lagrangian theories that flow to $(A_1,A_2)$ and $(A_1,A_3)$
Argyres-Douglas theories.}
\label{tab:ADLagr}
\end{table}
%
Note, the operator $\Tr \phi^2$ decouples from the IR fixed point
in the ordinary $\CN=1$ adjoint SQCD with $N_c =2$ and $N_f = 1$,
and the same is true for the variants in Table~\ref{tab:ADLagr}
coupled to an extra singlet $\CN=1$ chiral multiplet $u$
via a superpotential.\footnote{The field content of
$\CN=1$ theories in \cite{Maruyoshi:2016tqk,Maruyoshi:2016aim},
originally obtained via nilpotent Higgsing of 4d $\CN=2$ SQCD,
contains a few other gauge singlets, all of which decouple from the IR fixed point,
much like $\Tr \phi^2$. The only singlet matter superfields that
do not decouple in the IR are $M_5$ in the case of $(A_1,A_2)$
and $M_3$ in the case of $(A_1,A_3)$ theory.
Since they become Coulomb branch operators of the Argyres-Douglas
fixed points --- and to avoid confusion with $n$-manifolds denoted $M_n$ --- we call them $u$ here.}
Specifically, the theory that flows to the Argyres-Douglas fixed point $(A_1,A_3)$
is basically a deformation of the $\CN=1$ adjoint SQCD
with $SU(2)$ gauge group and $N_f = 1$ by a superpotential
\be
W \; = \; u q \tilde q
\label{SQCDWuqq}
\ee
where $u$ is a gauge singlet.
Similarly, the theory that flows to the Argyres-Douglas fixed point $(A_1,A_2)$
is a 4d $\CN=1$ gauge theory with gauge group $SU(2)$, one adjoint and two fundamental
chiral multiplets, coupled to a gauge singlet $u$ via the superpotential
\be
W \; = \; \phi q q + u \phi q' q'
\ee
Note, the Coulomb branch parameter $u$ plays the role of the mass in $\CN=1$ adjoint SQCD
with the superpotential \eqref{SQCDWuqq}. In particular, when $u \ne 0$ the quarks are massive
and the system is effectively a pure $\CN=2$ super-Yang-Mills with gauge group $SU(2)$,
whereas at the origin of the Coulomb branch ($u=0$) we have $\CN=1$ adjoint SQCD with $N_f = 1$.

In general, a calculation of the partition function on $T^2 \times F_g$ with a topological twist along $F_g$
requires a choice of non-anomalous $U(1)_R$ symmetry under which all matter fields have integer charges.
Different choices of R-symmetry lead to different 2d theories, many of which have been explored in the recent
literature \cite{Honda:2015yha,Putrov:2015jpa,Gadde:2015wta,Benini:2016hjo,Amariti:2017cyd}.
In our applications to 4-manifolds, the choice of the $U(1)_R$ symmetry is uniquely fixed \eqref{U1RCartan}:
it must be the Cartan subgroup of the $SU(2)_R$ symmetry at the IR fixed point with 4d $\CN=2$ supersymmetry,
\be
U(1)_R \; \subset \; SU(2)_R
\ee

With this choice of the R-symmetry flux through $F_g$, we now need to calculate the supersymmetric
partition function of the 4d theory on $T^2 \times F_g$. For theories with $\CN=1$ supersymmetry (or higher)
this can be conveniently done by using localization techniques, applied to a Lagrangian description of
a UV theory that flows to the desired (non-Lagrangian) fixed point. As usual in such calculations,
the result is a sum over gauge fluxes and an integral over the holonomies in the Cartan part of the gauge group,
with the integrand given by ratios of various 1-loop determinants.

For example, in the case of 4d $\CN=1$ theories that flow to $(A_1,A_2)$ and $(A_1, A_3)$ Argyres-Douglas theories
the localization of $T^2 \times F_g$ partition function leads to expressions of the form
\be
\frac{1}{2} \, \sum_{{\frak m} \in \Z} \, \oint_C \, \frac{dz}{2\pi i z} \, Z_{\text{1-loop}} (z, {\frak m};x, {\frak n})
\label{4dindex-general}
\ee
where the factor $\frac{1}{2}$ comes from the Weyl group of $SU(2)_{\text{gauge}}$.
Moreover, 4d $\CN=1$ chiral multiplets contribute to the integrand $Z_{\text{1-loop}}$
as follows\footnote{There is a slight clash
of notations here since $R$ used to denote the quantum number for $\CN=2$ R-symmetry,
whereas here it is used for the choice of R-symmetry in $\CN=1$ theory. In our applications to topological twists,
however, this choice will always be the $U(1)_R$ symmetry of the IR $\CN=2$ fixed point, meaning that we can
use the same notation without any reservations.}
\be
\prod_{\text{chirals}}
\left( \frac{1}{\theta (z^{\text{gauge}} x^{\text{flavor}}; q)} \right)^{(1-g)(1-R) + {\frak m} \cdot (\text{gauge}) + {\frak n} \cdot (\text{flavor})}
\label{4dindex-chiral}
\ee

Now let us consider the topological reduction of a 4d $\CN=1$ vector multiplet.
When $g=0$, the gauge field has no Kaluza-Klein modes and its topological reduction on $F_g$
gives a 2d $(0,2)$ vector multiplet (with the same gauge group).
Consequently, its contribution to the $T^2 \times F_g$ index of the 4d $\CN=1$ theory
is identical to the elliptic genus of the 2d $(0,2)$ vector multiplet.

When $g>0$, the Kaluza-Klein modes of the gauge field, {\it i.e.} holonomies along $2g$ generators of $H_1 (F_g, \Z)$,
give rise to $g$ two-dimensional $(0,2)$ chiral multiplets for each root of the gauge group $G$.
Therefore, compared to the genus-0 case, the topological reduction of a 4d $\CN=1$ vector multiplet
leads to the following extra factor in the integrand $Z_{\text{1-loop}}$ of the $T^2 \times F_g$ index:
\be
\left(
\eta (q)^{-2 \text{rank} (G)} \prod_{\alpha \in \text{Ad}(G)} \frac{1}{\theta (z^{\alpha}; q)} \right)^{g}
\ee
where $\eta (q)^{-2g \cdot \text{rank} (G)}$ comes from the Cartan generators \cite{Benini:2016hjo}.

As a simple example and a consistency check, let us consider a free 4d $\CN=2$ vector multiplet with $G=U(1)$,
viewed as a pair of a 4d $\CN=1$ vector and a 4d $\CN=1$ chiral multiplet $\Phi$.
On the one hand, from the discussion around \eqref{U1target}
we already know that topological reduction on $F_g$ should give us
a 2d $\CN=(2,2)$ vector multiplet together with $g$ chiral multiplets that parametrize $\text{Jac} (F_g)$.
On the other hand, 4d $\CN=1$ vector multiplet gives a 2d $\CN=(0,2)$ vector and $g$ copies of $(0,2)$ chiral multiplet.
They combine with the products of the topologically reduced 4d $\CN=1$ chiral multiplet $\Phi$ to produce the correct result.

Indeed, the 4d $\CN=1$ chiral multiplet $\Phi$ carries R-charge $R=0$ with respect to $U(1)_R$ that becomes part of
the $SU(2)_R$ R-symmetry when the theory is viewed as a 4d $\CN=2$ vector multiplet.
Therefore, its topological reduction on $F_g$, controlled by the sign of $(1-g)(1-R)$,
produces a 2d $\CN=(0,2)$ chiral multiplet when $g=0$ and then, as we increase the genus $g > 0$,
starts producing Fermi multiplets, whose ``net number'' is $g-1$.
Here, the ``net number'' is what the index \eqref{4dindex-chiral} can see, that is the difference
between the number of Fermi and $(0,2)$ chiral multiplets.
However, this simple example teaches us a good lesson, namely to keep in the Kaluza-Klein spectrum
{\it both} Fermi and $(0,2)$ chiral multiplets, even though one of their pairs does not contribute
to the index (since it can be gapped out by adding a mass term).

To summarize, in this example --- which can be easily generalized to non-abelian theory with
arbitrary background fluxes --- we get the following rules:
\be
\boxed{\phantom{\int} \text{4d } \CN=1 \text{ vector} \phantom{\int}}
\quad \xrightarrow[~]{~\text{topological reduction on $F_g$}~} \quad
\begin{cases}
\text{2d } \CN = (0,2) \text{ vector} & \\
+~g~ \CN = (0,2) \text{ chirals} &
\end{cases}
\ee
and
\be
\boxed{{\text{4d } \CN=1 \text{ chiral} \atop \text{with } R=0}}
\quad \xrightarrow[~]{~\text{topological reduction on $F_g$}~} \quad
\begin{cases}
\text{2d } \CN = (0,2) \text{ chiral} & \\
+~g~ \text{Fermi} &
\end{cases}
\ee

Now, let us apply this to 4d $\CN=1$ theories that flow to 4d $\CN=2$ Argyres-Douglas SCFTs
\eqref{A1A2curve} and \eqref{A1A3curve}.
In both cases,
the $T^2 \times F_g$ index is given by \eqref{4dindex-general} with the integrand of the form:
%
\begin{multline}
Z_{\text{1-loop}} (z,{\frak m}) \; = \;
\overbrace{ \eta (q)^{2(1-g)} \theta (z^2;q)^{1-g} \theta (z^{-2};q)^{1-g} }^{SU(2) \text{ gauge}}
\overbrace{ \theta (1;q)^{1-g} }^{\Tr \phi^2}
\times
\\
\times
\underbrace{ \frac{1}{\theta (z^2;q)^{2{\frak m} +1-g} \theta (1;q)^{1-g} \theta (z^{-2};q)^{-2{\frak m} +1-g}} }_{\phi}
\underbrace{ \frac{1}{\theta (z;q)^{2{\frak m}} \theta (z^{-1};q)^{-2{\frak m}}} }_{q's}
\underbrace{ \frac{1}{\theta (x_u;q)^{1-g}} }_{u}
\label{A1A3integrand}
\end{multline}
where we describe the origin of each factor and, to avoid clutter, omit flavor fugacities of all
chiral multiplets except $u$. (They can be easily restored by analogy with the contribution of $u$.)
Also, as in the computation of the superconformal index \cite{Maruyoshi:2016tqk},
in this expression we removed the contribution of the gauge singlet operators
that hit the unitarity bound and decouple along the RG flow.

By inspecting the resulting expressions for $Z_{\text{1-loop}} (z,{\frak m})$ we see that in the case ${\frak m} = 0$
the integral computation is basically identical to that of a 2d $\CN=(2,2)$ vector multiplet with gauge group $SU(2)$
coupled to $g$ adjoint $\CN=(2,2)$ chirals. When $g=0$, this theory flows to a theory of a free chiral $\Tr \phi^2$ \cite{Aharony:2016jki}.
However, in our case, interactions with other fields force this chiral to hit the unitarity bound and decouple from the IR fixed point.
So, the resulting elliptic genus is essentially that of a Coulomb branch operator $u$.

Note, both 4d $\CN=1$ theories in Table~\ref{tab:ADLagr} have non-negative spectrum of $U(1)_R$ charges.
In such situation, as pointed out in \cite{Gadde:2015wta}, only ${\frak m}=0$ flux sector contributes when $g=0$.
When $g>0$, according to \eqref{4dindex-chiral},
the R-charges of chiral matter multiplets are effectively replaced by\footnote{Note,
that we consider the reduction with zero flavor fluxes.}
\be
R \quad \to \quad R + g(1-R)
\ee
This fact has a curious consequence: in a theory where all matter multiplets have $U(1)_R$ charges
in the range between 0 and 1 (with values 0 and 1 included), only the flux sector ${\frak m} = 0$ contributes
to the $T^2 \times F_g$ index for any $g \ge 0$.
In particular, this is true of our two examples in Table~\ref{tab:ADLagr}.

The expression \eqref{A1A3integrand} appears to give
a deceivingly simple answer in the genus-1 case, $g=1$, which of course is expected from
the fact that 2d theory $\CM (F_g)$ has enhanced $\CN = (4,4)$ supersymmetry
in this case, see Table~\ref{tab:SUSY}.
On the other hand, precisely in this case we can easily identify the 2d theory $\CM (F_g)$
with a sigma-model whose target space is the moduli space of Higgs bundles with
wild ramification.\footnote{In string theory, this can be easily seen by realizing
Argyres-Douglas theory as a 6d fivebrane theory on a Riemann surface with wild
ramification and then exchanging the order of compactification on this surface with $F_g = T^2$.}
In the case of rank-1 theories $(A_1,A_2)$ and $(A_1,A_3)$, this target space is
a hyper-K\"ahler elliptic fibration with singular fibers of Kodaira type II and III,
respectively, as summarized in Table~\ref{tab:AD} (see also \cite{Fredrickson:2017yka}
for a detailed discussion of the geometry of these spaces).
The spectral curves are precisely \eqref{A1A2curve} and \eqref{A1A3curve}.

\begin{figure}[ht]
\centering
\includegraphics[width=2.3in]{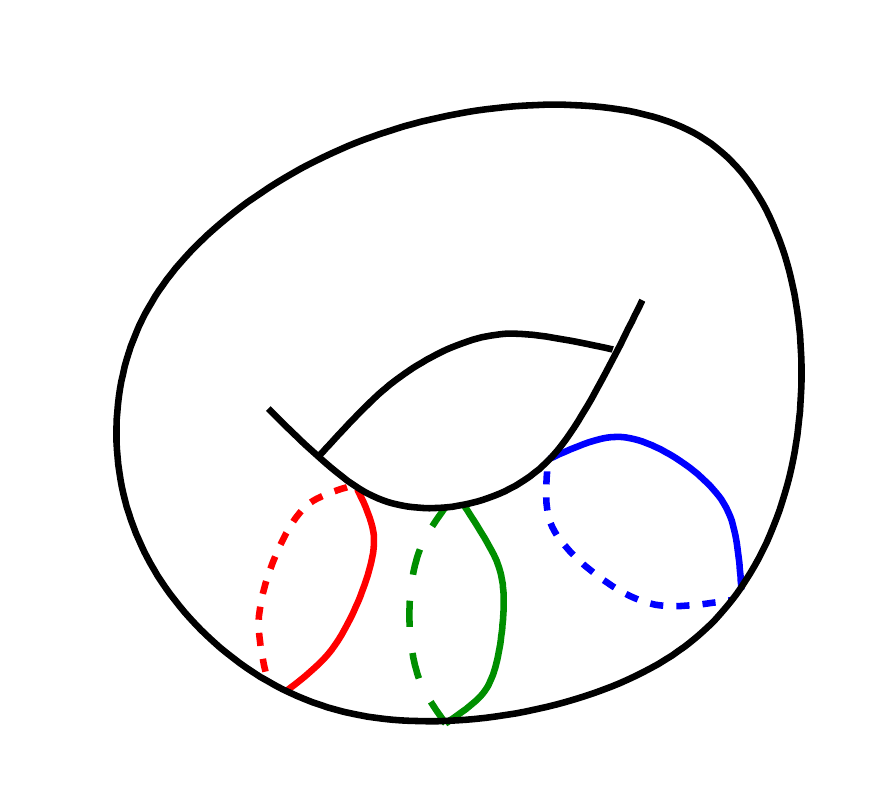}
\caption{Trisection diagram for $M_4 = S^1 \times S^3$.}
\label{fig:S1S3}
\end{figure}


\subsection{Heegaard branes}

The two singular elliptic fibrations that we encounter in the case of
Argyres-Douglas theories $(A_1,A_2)$ and $(A_1,A_3)$ correspond, respectively,
to a cuspidal rational curve and two rational curves tangent at a point.
They carry Euler number 2 and 3.
In the former case, the surface defined by the elliptic fibration is smooth,
whereas in the latter case it has a singularity of type $A_1$.

Away from singular fibers at $u=0$, these geometries
look like our simple examples in \eqref{Ecurve} and \eqref{N8target}.
In particular, there are two $SL(2,\Z)$ symmetries \eqref{twoSL2Z},
one of which acts on the complexified K\"ahler structure $\rho$
of the elliptic fiber and plays the role of mirror symmetry / Langlands duality.
The second group $SL(2,\Z)_{\tau}$ acts on the complex structure of the elliptic fiber
and in all of our previous examples was identified with the mapping class group of $F_g$.
If we make the same identification now, however, we run into a problem since $SL(2,\Z)_{\tau}$ is not really
a symmetry of $\CM (F_g)$ due to the monodromy around $u=0$. Then, what plays the role
of $\text{MCG} (F_g) \cong SL(2,\Z)$?

The only other candidate is $SL(2,\Z)_{\rho}$ and we claim that this is indeed the right
symmetry to be identified with $\text{MCG} (F_g)$. Indeed, suppose we start with
a 6d conformal theory of a single or several fivebranes,
first compactified on a surface $\Sigma$ and then further on a surface $F_g$:
\be
\begin{CD}
\text{6d fivebrane} \\
@VV\text{on}~\Sigma\ \; \rotatebox[origin=c]{90}{$\circlearrowleft$}\ \; \text{e/m 4d duality}V \\
\text{4d theory} \\
@VV\text{on}~F_g\ \; \rotatebox[origin=c]{90}{$\circlearrowleft$}\ \; \text{MCG} (F_g)V \\
\text{2d theory}
\end{CD}
\label{6d4d2d}
\ee
Mapping class groups of both surfaces, $\Sigma$ and $F_g$, are realized as symmetries (dualities)
in the two-dimensional theory that we call $\CM (F_g)$ throughout the paper.
However, they are realized a bit differently.
With this order of compactification, $\text{MCG} (\Sigma)$ is the electric-magnetic duality of
the 4d theory and becomes mirror symmetry (Langlands duality) in 2d theory $\CM (F_g)$.
In our rank-1 examples, it acts on the complexified K\"ahler parameter $\rho$ of the elliptic fiber,
whereas $\text{MCG} (F_g)$ acts on its complex structure.

If we reverse the order of compactification, the role of $SL(2,\Z)_{\tau}$ and $SL(2,\Z)_{\rho}$ is also reversed.
Indeed, this is what happens when, in our examples with $F_g = T^2$, we identify $\CM (F_g)$ with
the elliptic fibration that describes the moduli space of ramified Higgs bundles on $\Sigma$, rather than on $F_g$:
\be
\CM (F_g) \; \cong \; \CM_H (G,\Sigma)
\label{MFHS}
\ee
Indeed, if $\Sigma$ is a surface with wild ramification point(s) that leads
to the desired Argyres-Douglas theory in four dimensions \cite{Xie:2012hs},
we can switch the order and first compactify the 6d theory on $F_g = T^2$.
This gives 4d $\CN=4$ super-Yang-Mills with electric-magnetic duality group $\text{MCG} (F_g) = SL(2,\Z)$.
Upon further topological reduction on $\Sigma$, we get 2d sigma-model with the target space \eqref{MFHS}
on which $\text{MCG} (F_g) = SL(2,\Z)$ acts via fiberwise T-duality (mirror symmetry)
and $B$-field transform \cite{Bershadsky:1995vm,Harvey:1995tg,KW}.

A similar subtlety appears when we consider branes in $\CM (F_g)$ associated with 3-manifolds bounded by $F_g$.
If, instead, we were interested in 3-manifolds bounded by $\Sigma$ (or, mapping tori of $\Sigma$),
then the corresponding branes would be of type $(A,B,A)$ with respect to the three standard
K\"ahler structures on \eqref{MFHS}, see {\it e.g.} \cite{Gukov:2007ck,MR3248065,Chung:2014qpa}.
For example, for $\Sigma = T^2$ bounding a knot complement $M_3$ the corresponding $(A,B,A)$ brane would be
supported on the zero-locus of a simple polynomial equation \cite{Apol}, holomorphic in complex structure $J$.
However, since we are interested in 3-manifolds bounded by $F_g$, not $\Sigma$,
the corresponding branes in \eqref{MFHS} are a bit different;
in particular, they are of type $(B,A,A)$ with respect to the K\"ahler structures on $\CM_H (G,\Sigma)$.

\begin{figure}[h]
\centering
\begin{minipage}{0.45\textwidth}
\centering
\includegraphics[width=0.6\textwidth, height=0.19\textheight]{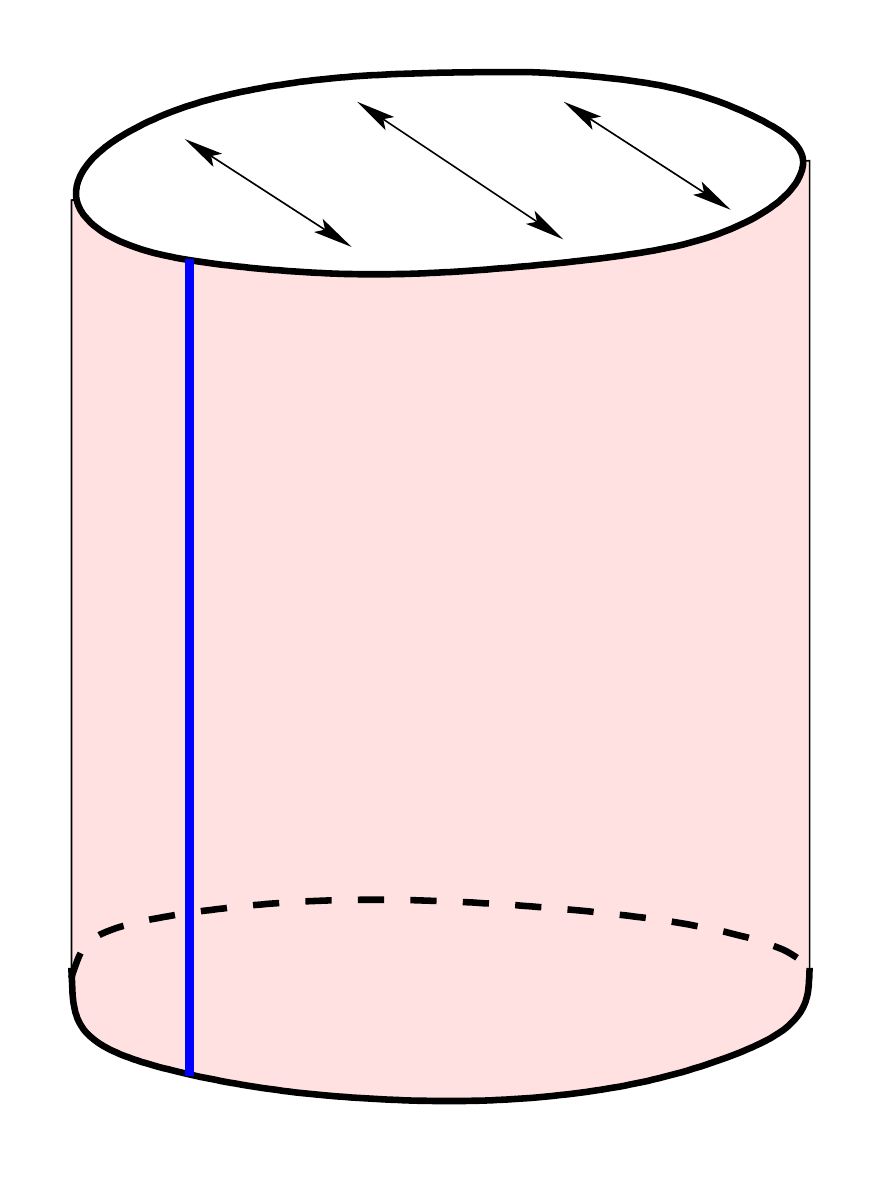}
\end{minipage}
$\simeq$
\begin{minipage}{0.45\textwidth}
\centering
\includegraphics[width=0.6\textwidth, height=0.19\textheight]{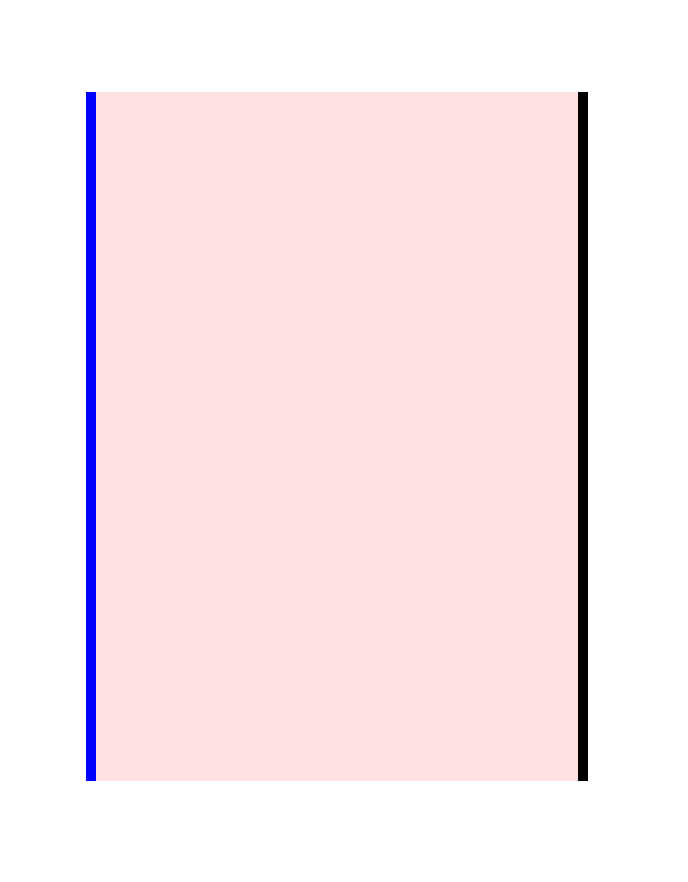}
\end{minipage}
\caption{A folding trick that leads to A-model interpretation of the equivariant Verlinde formula.
The Hilbert space of a 2d theory $\CM$ on a circle with a line defect (interface) $\CB$ is equivalent to
a 2d theory $\CM \times \bar \CM$ on a strip with the ``diagonal'' (totally reflective) boundary condition
at one end and $\CB$ at the other.}
\label{fig:foldingtrick}
\end{figure}

Since $\text{MCG} (F_g) = SL(2,\Z)$ acts on \eqref{MFHS} via mirror symmetry and $B$-field transform,
and since in our conventions (which follow \cite{Hitchin:1986vp,KW,Gukov:2010sw}) both operations are of type $(B,A,A)$,
so are the branes associated with mapping tori of $F_g$ or, in fact,
with more general 3-manifolds either fibered by $F_g$ or bounded by $F_g$.
For concreteness, let $M_3$ be a mapping torus of $F_g = T^2$ associated\footnote{More general
choices of the $SL(2,\Z)$ elements should lead to various generalizations of the equivariant
Verlinde formula and make contact with \cite{Ganor:2014pha}.} with the element
$\begin{pmatrix} 1 & k \\ 0 & 1 \end{pmatrix} \in SL(2,\Z) = \text{MCG} (F_g)$.
In the 2d theory $\CM (F_g)$, it is described by putting the theory on a circle
and requiring that it undergoes a transformation by a duality as we go around the circle.
Since the standard generator $\begin{pmatrix} 1 & 1 \\ 0 & 1 \end{pmatrix} \in SL(2,\Z)$
acts on the category of $(B,A,A)$ branes in \eqref{MFHS}
by tensoring with a line bundle $\CL$, such that $c_1 (\CL) = \omega_I$,
we learn that the space of supersymmetric states (= $Q$-cohomology) of
the 2d theory $\CM (F_g)$ on a circle with such a duality twist is given by
the space of open strings between two branes on $\CM \times \bar \CM$,
\be
\text{Hom} (\CL^{\otimes k}, \Delta_{\CM}) \; \cong \; H^0 (\CM, \CL^{\otimes k})
\label{HomLkM}
\ee
where we used the ``folding trick'' in Figure~\ref{fig:foldingtrick}
and a more compact notation $\CM = \CM (F_g) \cong \CM_H (G,\Sigma)$,
such that $\Delta_{\CM}$ is the (anti-)diagonal:
\be
\Delta_{\CM} \; \xhookrightarrow{~~} \; \CM \times \bar \CM
\ee
The space \eqref{HomLkM} is infinite-dimensional, but because both branes
are invariant under the $U(1)_{\beta} \equiv U(1)_r$ action on $\CM_H (G,\Sigma)$,
we can consider its graded dimension \cite{Fredrickson:2017yka}:
\begin{multline}
\dim_{\frak t} H^0 (\CM, \CL^{\otimes k}) \; = \;
\int_{\CM} \text{ch} \big(\CL^{\otimes k}, \beta \big) \wedge \text{Td} (\CM, \beta)
\; = \;
\\
\; = \;
\begin{cases}
\frac{1}{(1-{\frak t}^{\frac{2}{5}}) (1-{\frak t}^{\frac{3}{5}})}
+ \frac{{\frak t}^{\frac{k}{5}}}{(1-{\frak t}^{\frac{6}{5}}) (1-{\frak t}^{-\frac{1}{5}})}, & \text{for } (A_1,A_2) \\
 & \\
\frac{1}{(1-{\frak t}^{\frac{1}{3}}) (1-{\frak t}^{\frac{2}{3}})}
+ \frac{{\frak t}^{\frac{\lambda}{3}} + {\frak t}^{\frac{k-\lambda}{3}}}{(1-{\frak t}^{\frac{4}{3}}) (1-{\frak t}^{-\frac{1}{3}})}, & \text{for } (A_1,A_3)
\end{cases}
\phantom{~~~~~~~}
\label{ADCoulindex}
\end{multline}
where, as usual, $\frak t = e^{- \beta}$.

Now, following \cite{Gukov:2010sw}, we can apply mirror symmetry (Langlands duality)
to both branes in \eqref{HomLkM}, considering them as branes on $\CM \cong \Delta_{\CM}$.
Indeed, \eqref{HomLkM} can be understood as the space of open strings between two branes on $\CM$,
one of which is a $(B,A,A)$ brane that we denote $\CL^{\otimes k}$
and the other is a rank-1 $(B,B,B)$ brane supported on all of $\CM$,
that is $\CO_{\CM}$ in any of the complex structures on $\CM$.
(In fact, this was the point of view in \cite{Fredrickson:2017yka}.)
Under mirror symmetry (Langlands duality), the $(B,A,A)$ brane $\CL^{\otimes k}$
maps to a hyper-holomorphic higher-rank bundle, $(B,B,B)$ brane, that was studied in \cite{Gukov:2010sw}.

And, the original rank-1 $(B,B,B)$ brane $\CO_{\CM}$ maps to a $(B,A,A)$ brane
supported on a holomorphic Lagrangian submanifold that
in a case of tame or no ramification would be called the ``Hitchin section.''
We propose to identify this $(B,A,A)$ brane on \eqref{MFHS} with the basic boundary condition $\CB_H$
that determines all other Heegaard branes \eqref{BHalpha} -- \eqref{Babc}
in this class of examples with $F_g = T^2$:
\be
\CB_H ~=~ \text{mirror of the } (B,B,B) \text{ brane } \CO_{\CM}
\label{HBAA}
\ee
Note how conveniently the subscript $H$ can stand for either ``Heegaard'' or ``Hitchin.''

As a test of the proposal \eqref{HBAA}, we can calculate the partition function \eqref{ZZ2d4d}
of the topologically twisted Argyres-Douglas theory on $M_4 = S^1 \times S^3$.
The trisection diagram for this 4-manifold was already described in \eqref{S1S3abc}
and, for convenience, is reproduced here, in Figure~\ref{fig:S1S3}.
In particular, all three Heegaard branes are identical \eqref{S1S3BBBBH} and equal to the basic brane $\CB_H$.
Moreover, all boundary changing operators $V_i$ are copies of the identity operator,
and so the disk amplitude that gives the 4-manifold invariant \eqref{ZZ2d4d}
has the boundary condition \eqref{HBAA} along the entire boundary of the disk.
Equivalently, we can think of it as having two boundary components, both decorated with $\CB_H$,
in which case the problem is reduced to the familiar space of open strings going from $\CB_H$ to itself.
Either way, the key point is that we end up with open string zero-modes parametrizing $\CB_H$,
which is precisely the Coulomb branch of the 4d theory\footnote{Mathematically,
the A-model version of this calculation boils down to
$HF^* (\Delta_{\CM}, \Delta_{\CM}) \cong HH^* (\text{Fuk} (\CM))$,
and from the B-model viewpoint it involves a version of the Hochschild-Kostant-Rosenberg
isomorphism $\text{Ext}^*_{\CM \times \CM} (\CO_{\Delta}, \CO_{\Delta}) \cong H^* (\CM, \Lambda^* T_{\CM})$.}
\be
\text{Hom} (\CB_H, \CB_H) \; \cong \; \C [u]
\ee
Hence, their ``counting'' gives precisely the Coulomb branch index of the 4d theory
which, according to the conjecture of \cite{Dedushenko:2017tdw}, is equal to
the topological partition function on $M_4 = S^1 \times S^3$,
and for the Argyres-Douglas theories $(A_1,A_2)$ and $(A_1,A_3)$ is given by \eqref{ADCoulindex} with $k=1$.

In general, it seems that mirror symmetry for branes on singular elliptic fibrations of
different Kodaira types has not been studied systematically in the literature,
aside from certain cases like $I_0^*$ or $I_2$ considered in \cite{Gukov:2010sw,Frenkel:2007tx}.
Based on these examples, and as a small step toward future development of this subject,
we conjecture that $(B,A,A)$ branes supported on rational components of a singular
elliptic fiber of a given Kodaira type are mirror to fractional $(B,B,B)$ branes
localized at the singularity of the mirror elliptic fibration, of the same Kodaira type.
For example, in the case of Kodaira type III relevant to the Argyres-Douglas theory $(A_1,A_3)$,
two $(B,A,A)$ branes supported on two rational curves (tangent at a point) conjecturally
are mirror to two fractional $(B,B,B)$ branes localized at the $A_1$ singularity
of the mirror elliptic fibration, also of Kodaira type III.
It would be interesting to test this conjecture and to explore the role these
special branes may play in disk amplitudes dual to genus-1 trisections.

As explained in section~\ref{sec:trisec}, once all building blocks are in place,
for a given 4d theory and genus $g$, we can start composing them in various ways to
compute topological partition functions \eqref{ZZ2d4d} of arbitrary genus-$g$ trisections.
For example, aside from the simple genus-1 trisection $M_4 = S^1 \times S^3$ considered above,
we can now use \eqref{MFHS} and \eqref{HBAA} to study topological twist
of Argyres-Douglas theories on other genus-1 trisections, such as $M_4 = \cp^2$.
The trisection diagram $(F_g, \alpha, \beta, \gamma)$ for $M_4 = \cp^2$ comprises a genus-1
surface $F_g = T^2$ with three closed curves in homology classes $[\alpha] = (1,0)$,
$[\beta] = (0,1)$, and $[\gamma] = (1,1)$, {\it cf.} Figure~\ref{fig:toricCP2}.

The corresponding Heegaard branes \eqref{Babc} can all be obtained from
the basic boundary condition \eqref{HBAA} by repeated action of the order-3 element
$\gamma$ of the duality group $SL(2,\Z) = \text{MCG} (F_g)$
that we already encountered in Table~\ref{tab:N3tau} in a slightly different context:
\begin{eqnarray}
\CB_{\alpha} & \; = \; & \CB_H \nonumber \\
\CB_{\beta} & \; = \; & \gamma^2 (\CB_H)  \\
\CB_{\gamma} & \; = \; & \gamma (\CB_H) \nonumber
\end{eqnarray}
Here, $\gamma = \begin{pmatrix} 0 & -1 \\ 1 & -1 \end{pmatrix}$
acts by a particular composition of mirror symmetry and $B$-field transform,
which allows to describe the Heegaard branes $\CB_{\alpha}$, $\CB_{\beta}$, and $\CB_{\gamma}$
explicitly, as in \cite{Gukov:2010sw}.
Note, if not for $b_2^+ = 1$ which makes things more interesting,
we could conclude that \eqref{ZZ2d4d} should vanish without computing it directly.
Indeed, in the case of the Argyres-Douglas theory $(A_1,A_2)$,
the condition \eqref{unghostADA1A2} gives $n = - \frac{4}{3}$ for $M_4 = \cp^2$.
Since the Argyres-Douglas theory $(A_1,A_2)$ has no global flavor symmetries,
we can not take advantage of the first term in \eqref{rgeneral} to change this result.
On the other hand, in the case of the Argyres-Douglas theory $(A_1,A_3)$ on $M_4 = \cp^2$,
the general condition \eqref{uuuughost} gives $n = - \frac{3}{2}$,
which can be compensated by the flux for the global $SU(2)$ symmetry
to produce potentially non-zero topological correlators $\langle u \ldots u \rangle_{\text{flux}}$.

Now, let us consider the case of general $g$, for which $\CM (F_g)$ may not have a nice geometric description as in the $g=1$ case.


\subsection{Higher-genus trisections}

In order to determine 2d $\CN = (2,2)$ theory $\CM (F_g)$, it is helpful to compute various protected quantities,
especially those which only require $\CN=1$ supersymmetry, so that we can use various Lagrangian deformations
of a non-Lagrangian theory in question.

One such protected quantity is the elliptic genus of $\CM (F_g)$ that equals the partition function of the 4d theory on $T^2 \times F_g$.
While very powerful in principle, computationally it is rather bulky; in particular, this was one of the reasons why
we chose to suppress fugacities of some chiral multiplets in writing \eqref{A1A3integrand}.
A simpler version of the elliptic genus is the Witten index, which is also a partition function on $T^2$
(resp. $T^2 \times F_g$ in 4d theory), but the corresponding expressions are much simpler and easier to manage
since theta-functions \eqref{mytheta} are replaced by rational functions, $\theta (x;q) \xrightarrow{~~~} x^{1/2} - x^{-1/2}$.

Furthermore, from the viewpoint of the 4d theory, this version of the $T^2 \times F_g$ partition function is more
natural because it does not involve extra variable $q$ and directly corresponds to the path integral of
the topological theory on $M_4 = T^2 \times F_g$. Since it is also a particular instance of $M_4 = S^1 \times M_3$,
we need to be careful and regularize a potentially divergent trace over $\CH (M_3)$
by introducing a fugacity ${\frak t}$ for the $U(1)_r$ charge, as in \eqref{trefinement}.
The result can be interpreted as a graded trace over the Floer-type homology of $M_3 = S^1 \times F_g$ computed by
a non-Lagrangian theory of our choice, see {\it e.g.} \cite{Dedushenko:2017tdw} and \cite{Gukov:2016gkn}
for various examples of this calculation that we shall follow here.

Specifically, in the case of $(A_1, A_3)$ Argyres-Douglas theory --- which, curiously,
turns out to be simpler than $(A_1,A_2)$ theory (for reasons that will become clear shortly) ---
we obtain the following result:
\be
Z (T^2 \times F_g) \; = \;
\begin{cases}
- \frac{2 {\frak t}^{5/6}}{(1 - {\frak t}^{1/3}) (1 - {\frak t}^{4/3})} \,, & g=0 \\
2 \,, & g=1 \\
8 {\frak t}^{-1/2} (1 + {\frak t}^{1/3} + {\frak t}^{2/3} + {\frak t}) \,, & g=2 \\
8 {\frak t}^{-4/3} (1 + 6 \, {\frak t}^{1/3} + {\frak t}^{2/3}) (1 + {\frak t}^{1/3} + {\frak t}^{2/3} + {\frak t})^2 \,, & g=3 \\
\quad \vdots &
\end{cases}
\label{A1A3Fgindex}
\ee
As explained above, this result has several equivalent interpretations, as
\begin{itemize}

\item
the Witten index of $\CM (F_g)$,

\item
the invariant of a 4-manifold $M_4 = T^2 \times F_g$,

\item
the character of the ``non-Lagrangian Floer homology'' $\CH (M_3)$ for $M_3 = S^1 \times F_g$,

\item
the genus-$g$ A-model partition function of the 2d theory $\CM (T^2)$.

\end{itemize}

Let us briefly outline the derivation of \eqref{A1A3Fgindex};
further details or conventions can be found in \cite{Gukov:2016gkn} or \cite{Dedushenko:2017tdw}.
If we start with 4d $\CN=1$ theory that flows to $(A_1, A_3)$ fixed point,
via compactification on $S^1$ we can reduce the problem to a standard calculation
in 3d $\CN=2$ theory whose field content is essentially identical to that in Table~\ref{tab:ADLagr}.
The $S^1 \times F_g$ partition function of this 3d $\CN=2$ theory is still given by
the contour integral \eqref{4dindex-general}, though with a much simpler integrand, {\it cf.} \eqref{A1A3integrand}:
\begin{multline}
Z_{\text{1-loop}} \; = \;
\overbrace{ \left( \frac{1}{z^2} - 2 + z^2 \right)^{1-g} }^{SU(2) \text{ gauge}}
\overbrace{ \left( \frac{z^{1/2} x_q^{1/2}}{1 - z x_q} \right)^{2 {\frak m}}
\left( \frac{z^{-1/2} x_q^{1/2}}{1 - z^{-1} x_q} \right)^{- 2 {\frak m}} }^{q \text{ and } \tilde q}
\overbrace{ \left( \frac{x_u^{1/2}}{1 - x_u} \right)^{1-g} }^{u}
\times
\\
\times
\underbrace{ \left( \frac{1 - x_{\phi}^2}{ x_{\phi} } \right)^{1 - g} }_{\Tr \phi^2}
\underbrace{ \left( \frac{z x_{\phi}^{1/2}}{1 - z^2 x_{\phi}} \right)^{2 {\frak m} + 1 - g}
\left( \frac{x_{\phi}^{1/2}}{1 - x_{\phi}} \right)^{1 - g}
\left( \frac{z^{-1} x_{\phi}^{1/2}}{1 - z^{-2} x_{\phi}} \right)^{ - 2 {\frak m} + 1 - g} }_{\phi}
\label{A1A3T2Fgintegrand}
\end{multline}
In particular, here we introduced fugacities for all chiral multiplets (using, hopefully, self-explanatory notations).
To reproduce \eqref{A1A3Fgindex}, in the end we need to set $x = {\frak t}^r$, where $r$ is the $U(1)_r$
R-charge.\footnote{Note, the values of the $U(1)_r$ R-charge listed in Table~\ref{tab:ADLagr} represent
linear combinations of charges in the UV theory that correspond to the $U(1)_r$ symmetry at the IR fixed point.
Some of these values are negative, meaning that only gauge-invariant composite fields with positive total
$U(1)_r$ R-charge are in the spectrum of the IR theory.}
For this reason, we use the same fugacity $x_q$ for the chirals $q$ and $\tilde q$,
because their $U(1)_r$ R-charges are equal, {\it cf.} Table~\ref{tab:ADLagr}.
This is one way in which $(A_1, A_3)$ Argyres-Douglas theory is simpler than $(A_1,A_2)$.

By summing over residues, the contour integral that gives \eqref{A1A3Fgindex} can be expressed as a sum,
{\it cf.} \eqref{AmodelOOO},
\be
Z (T^2 \times F_g) \; = \; \frac{1}{2} \sum_{\text{vacua} : \;  d \tilde \CW = 0} \,
Z_{\text{1-loop}} \vert_{{\frak m} = 0} \, \left( \tilde \CW'' \right)^{g-1}
\label{ADindexW}
\ee
over critical points of the twisted superpotential $\tilde \CW (z)$:
\be
1 \; = \; \exp \left( \frac{\partial \tilde \CW}{\partial \log z} \right) \; = \;
\left( \frac{z^2 - x_{\phi}}{1 - z^2 x_{\phi}} \cdot \frac{z - x_q}{1 - z x_q} \right)^2
\label{BAEA1A3}
\ee
where $\tilde \CW''$ denotes $\frac{\partial^2 \tilde \CW}{(\partial \log z)^2}$ and
$\tilde \CW' = \frac{\partial \tilde \CW}{\partial \log z} = \frac{ \partial \log Z_{\text{1-loop}} }{ \partial {\frak m} }$.
Because the chiral fields $q$ and $\tilde q$ have $R=1$ ({\it cf.} Table~\ref{tab:ADLagr}),
they do not contribute to $e^{2\pi i (g-1) \Omega} = Z_{\text{1-loop}} \vert_{{\frak m} = 0}$ which, therefore, is identical to
what one finds in 3d $\CN=2$ adjoint SQCD that computes the equivariant Verlinde formula \cite{Gukov:2015sna}.
The contribution of $q$ and $\tilde q$ is present, however, in $\tilde \CW''$ and, of course,
in the equation \eqref{BAEA1A3}. When $g=1$, from the explicit form of \eqref{A1A3T2Fgintegrand}
it is clear that each summand in \eqref{ADindexW} is equal to 1 in this case,
so that the total sum is simply half the number of solutions to \eqref{BAEA1A3}.

We can also summarize our result for \eqref{ADindexW}--\eqref{BAEA1A3}
by writing it in the form \eqref{AmodelOOO}--\eqref{Wexpcrit}, with
\be
2\pi i \Omega \; = \;
\log \frac{(1 - z^2 x_{\phi}) (1 - z^{-2} x_{\phi})}{(1-z^{-1})^2 (1+z)^2}
+ \log \frac{1 - x_u}{\sqrt{x_u x_{\phi}} (1 + x_{\phi})}
\label{OmA1A3g1}
\ee
and
\be
\tilde \CW' \; = \;
2 \log \left( \frac{z^2 - x_{\phi}}{1 - z^2 x_{\phi}} \right)
+ 2 \log \left( \frac{z - x_q}{1 - z x_q} \right)
\label{WA1A3g1}
\ee

Note, even though $\tilde \CW$ is a rather non-trivial function, the equation \eqref{BAEA1A3} for its critical
points is a simple polynomial equation.
Depending on the context, the polynomial equation $\exp \left( \frac{\partial \tilde \CW}{\partial \log z} \right) = 1$
has a number of interesting interpretations. For example, in the context of Bethe/gauge correspondence,
it is the Bethe ansatz equation for the dual integrable system \cite{Nekrasov:2009uh,Nekrasov:2014xaa}.
In the context of 3d/3d correspondence, it is the $A$-polynomial equation or its generalization~\cite{Apol,Gadde:2013wq,Chung:2014qpa}.

In our present case of a topologically twisted Argyres-Douglas theory $(A_1, A_3)$ on $M_4 = T^2 \times F_g$,
the polynomial equation \eqref{BAEA1A3} has a total of 6 solutions, two of which (namely, $z = \pm 1$)
have to be discarded because they correspond to the points on the maximal torus of $SU(2)$ fixed by the Weyl group \cite{Benini:2016hjo}.
The total sum \eqref{ADindexW} over the remaining four solutions gives the final answer \eqref{A1A3Fgindex}.
In particular, using the explicit form of \eqref{OmA1A3g1} and \eqref{WA1A3g1}, it is easy to verify
that four admissible solutions consist of two pairs which produce the same values
of the ``handle gluing operator'' $H_v := e^{2\pi i \Omega} \, \tilde \CW''$, namely
\be
H_{\pm} \; = \;
\pm 2 \, (1 \pm {\frak t}^{1/6})^2 {\frak t}^{-2/3} \,
\left( 1 + {\frak t}^{1/3} + {\frak t}^{2/3} + {\frak t} \right)
\label{vortexindexA1A3}
\ee
where we already specialized the fugacities of the chiral multiplets to the corresponding values ${\frak t}^r$
according to Table~\ref{tab:ADLagr}.

Since the sum in \eqref{ADindexW} runs over four admissible solutions to \eqref{BAEA1A3},
organized in two pairs with handle gluing operators $H_+$ and $H_-$, respectively, we can conveniently
cancel the multiplicity factor of 2 within each group with the factor $\frac{1}{2}$ in \eqref{ADindexW}
to write it as
\be
Z (T^2 \times F_g) \; = \; (H_+)^{g-1} + (H_-)^{g-1}
\label{HsumA1A3}
\ee
Then, following our discussion in section \ref{sec:structure}, we can interpret \eqref{vortexindexA1A3}
as the refined Witten index of ``vortex strings'' supported\footnote{Recall that all components of
the divisor representing the canonical class of $M_4 = T^2 \times F_g$ are disjoint copies of $T^2$.} on $T^2 \subset M_4$.
We see that there are two types of vortex strings --- suggesting that the deformed $(A_1, A_3)$ Argyres-Douglas theory
has two vacua --- one of which has vanishing Witten index in the unrefined limit ${\frak t} \to 1$:
\be
Z_{\text{vortex}} (\pm , T^2) \; = \;
\pm 2 \, (1 \pm {\frak t}^{1/6})^2 {\frak t}^{-2/3} \,
\left( 1 + {\frak t}^{1/3} + {\frak t}^{2/3} + {\frak t} \right)
\ee

Even though our prime interest is in 4-manifolds, for now let us use the interpretation of \eqref{A1A3Fgindex}
as the Witten index of 2d theory $\CM (F_g)$.
For $g=0$ it has a simple interpretation: the two factors in the denominator describe a chiral ring
freely generated by two operators with $\frac{R_A}{2} = r = \frac{1}{3}$ and $\frac{4}{3}$.
These are precisely the two $(a,c)$ operators we found in \eqref{KKaclist}, with the correct quantum numbers.

Similarly, \eqref{A1A3Fgindex} gives a clear picture of what happens in higher genus:
the topological reduction of the Argyres-Douglas theory $(A_1, A_3)$ on a surface of genus $g$
looks like a non-compact Gepner model {\it a la} \cite{Mukhi:1993zb,Ghoshal:1995wm,Ooguri:1995wj,Giveon:1999zm,Giveon:1999px,Eguchi:2000tc,Lerche:2000uy,Eguchi:2004yi,Eguchi:2004ik,Ashok:2007ui,Ashok:2012qy} whose number of minimal model / Liouville factors grows with $g$.
In particular, adding an extra handle to $F_g$ ({\it i.e.} increasing $g$ by $+1$)
each time contributes to $\CM (F_g)$ a 2d $\CN=(2,2)$ theory with a finite $(a,c)$ ring
represented by a factor $1 + {\frak t}^{1/3} + {\frak t}^{2/3} + {\frak t}$,
along with some other changes.

For $g=1$, \eqref{A1A3Fgindex} gives the Witten index of the Argyres-Douglas theory $(A_1, A_3)$.
In particular, it indicates that generic $\CN=1$ SUSY-preserving deformations of the theory
have at least two supersymmetric vacua. Of course, as in any index calculation, the actual number
can be larger, but the extra vacua must always come in pairs.
If the minimal number 2 is realized, as suggested by \eqref{HsumA1A3},
then the topological partition function of the Argyres-Douglas theory $(A_1, A_3)$
on general K\"ahler 4-manifolds is given by
\be
Z(M_4) \; = \; e^{a_+ \chi \, + \, b_+ \sigma}
\sum_{\lambda} Z_{\text{vortex}} (+ , \lambda)
+
e^{a_- \chi \, + \, b_- \sigma}
\sum_{\lambda} Z_{\text{vortex}} (- , \lambda)
\label{A1A3Kahler}
\ee
with some universal constants $a_{\pm}$ and $b_{\pm}$ that do not depend on $M_4$.
Even though 4d theories considered here are non-Lagrangian, they are still local, which means
that contributions $Z_{\text{vortex}} (\pm , \lambda)$ depend only on topology of the components
representing $\lambda$ and their normal bundle in $M_4$.
Furthermore, simple examples here suggest that, in the case of K\"ahler manifolds,
$\lambda$ which contribute to \eqref{A1A3Kahler} coincide with Seiberg-Witten basic classes
(but their contribution is very different, of course); needless to say, this observation requires more extensive checks.
We can determine the constants $a_{\pm}$ and $b_{\pm}$ by computing, say, the topological $S^2$
partition function of $\CM (F_g)$.
Alternatively, deriving inspiration from \cite{Witten:1994ev}, one might be tempted to conjecture that
$e^{a_- \chi + b_- \sigma} = i^{\frac{1}{2} \text{VirDim} (\CM)} e^{a_+ \chi + b_+ \sigma}$,
where ``$\text{VirDim} (\CM)$'' was defined in \eqref{rgeneral}:
\be
e^{a_- \chi \, + \, b_- \sigma} \; = \; \exp \left( - \frac{\pi i}{2} (2a-c) \chi - \frac{3 \pi i}{4} c \, \sigma \right)
\cdot e^{a_+ \chi \, + \, b_+ \sigma}
\ee
and, similarly, $u_- = - u_+$ in the two vacua of the deformed Argyres-Douglas theory $(A_1, A_3)$.


\section{Discussion}

We proposed a duality between disk amplitudes and 4-manifolds, or invariants of 4-manifolds, to be more precise.
The 2d ``closed string'' sector $\CM (F_g)$ in the interior of the disk depends on the choice
of 4d (non-Lagrangian) theory and the genus $g$ of the surface on which it is compactified.
Given this data, the construction of 4-manifold invariants is essentially combinatorial;
it is determined by a fixed set of branes in $\CM (F_g)$ that we call Heegaard branes
and a particular set of boundary changing operators $V_i$.
Composing these elements like LEGO produces invariants of 4-manifolds represented by genus-$g$ trisections.

The next obvious goal is to explore this duality in a variety of interesting non-Lagrangian theories
and interesting trisections, especially where smooth structures play an important role.
A relatively simple classic example involves $M_4$ defined as a degree-$d$ hypersurface in $\cp^3$.
The first few cases in low degree are simple familiar 4-manifolds: $d=1$ gives $\cp^2$,
$d=2$ gives $S^2 \times S^2$, $d=3$ gives $M_4$ diffeomorphic to $\cp^2 \; \# \; 6 \bar{\cp}^2$,
$d=4$ gives $K3$, and $d \ge 5$ give surfaces of general type.
For odd values of $d \ge 5$, such hypersurfaces have the same intersection form as
\be
M_4' \; = \; m \cp^2 \; \# \; n \bar{\cp}^2
\ee
with
\be
m \; = \; \frac{d^3 - 6 d^2 + 11 d - 3}{3}
\qquad,\qquad
n \; = \; \frac{2 d^3 - 6 d^2 + 7 d - 3}{3}
\ee
but different Seiberg-Witten invariants, so that $M_4$ and $M_4'$ are homeomorphic but not diffeomorphic.
Using \eqref{generalOOO}, it is easy to see that both rank-1 Argyres-Douglas theories $(A_1,A_2)$ and $(A_1,A_3)$
can potentially produce non-zero invariants for this class of non-diffeomorphic pairs, see Table~\ref{tab:degd}
where we describe the `interesting range' of values of the degree $d$.
This class of examples provides a good opportunity to explore the structure \eqref{uuusmpltpvortex} (on the $M_4$ side)
and to understand the behavior of non-Lagrangian 4-manifold invariants under connected sum (on the $M_4'$ side).

\begin{table}
\begin{centering}
\begin{tabular}{|c||c|c|c|c|c|c|c|c|c|}
\hline
~degree~ & $d=1$ & $d=2$ & $d=3$ & $d=4$ & $d=5$ & $d=6$ & $d=7$ & $d=8$ & $d=9$ \tabularnewline
\hline
\hline
$\phantom{\int^{\int^\int}} (A_1,A_2) \phantom{\int_{\int}}$ & $- \frac{16}{5}$ & $-\frac{14}{5}$ & $-\frac{4}{5}$ & $\frac{4}{5}$ & $0$ & $-\frac{26}{5}$ & $-\frac{84}{5}$ & $-\frac{184}{5}$ & $-\frac{336}{5}$
\tabularnewline
\hline
$\phantom{\int^{\int^\int}} (A_1,A_3) \phantom{\int_{\int}}$ & $-4$ & $-\frac{10}{3}$ & $0$ & $4$ & $\frac{20}{3}$ & $6$ & $0$ & $-\frac{40}{3}$ & $-36$
\tabularnewline
\hline
\end{tabular}
\par\end{centering}
\caption{\label{tab:degd} The virtual dimension ``VirDim$(\CM)$'' or, equivalently, ``ghost number'' anomaly \eqref{rgeneral}
in rank-1 Argyres-Douglas theories on a 4-manifold $M_4$ defined by a degree-$d$ hypersurface in~$\cp^3$.}
\end{table}

As was stressed a number of times throughout the paper, the computation of the 4-manifold invariants \eqref{ZZ2d4d}
via trisections requires a rather detailed understanding of the 2d theory $\CM (F_g)$.
In general, it is natural to expect that 2d SCFTs $\CM (F_g)$ can be constructed from basic building blocks,
as in the toy model of section~\ref{sec:toy}, so that assembling the pieces
together corresponds to gauging continuous or (in simple cases) discrete symmetries, {\it e.g.}
\be
\CM (F_g) \; = \;
\Big( \CM_0 \otimes \underbrace{ \CM_{\text{handle}} \otimes \ldots \otimes \CM_{\text{handle}} }_{g~\text{copies}} \Big) \, \Big/ \, H
\label{MFghandles}
\ee
Furthermore, our preliminary analysis of the chiral ring, elliptic genus, and other protected quantities
suggests that, at least in a certain class of examples, these basic ingredients involve
$\CN=2$ minimal models, Liouville theory, or the cigar supercoset model $\frac{SL(2)}{U(1)}$
related to Liouville theory by supersymmetric FZZ duality \cite{Hori:2001ax}.
If we denote by $Q$ the background charge of the Liouville theory
and by $k$ the supersymmetric level of the dual non-compact Kazama-Suzuki coset $\frac{SL(2)}{U(1)}$,
then such a dual pair has central charge
\be
\hat c \; = \; 1 + Q^2 \; = \; 1 + \frac{2}{k}
\label{cSLcoset}
\ee
and the spectrum of discrete primaries that looks like $q = \frac{1}{k}, \frac{2}{k}, \ldots, \frac{k}{k}$.
Here, the lowest state with R-charge $q = \frac{1}{k}$ is ``almost normalizable'' in the terminology of \cite{Ashok:2007ui};
if we identify it with the lowest value of the R-charge $q = \frac{R_A}{2}$
that appears in the list \eqref{KKaclist} of $(a,c)$ primaries, we obtain
\be
\frac{1}{k} \; = \; \Delta (u) - 1
\ee
Curiously, $\Delta (u) - 1$ is indeed an inverse integer for rank-1 Argyres-Douglas theories
and leads to values\footnote{Namely, it gives $k=5$ for the $(A_1,A_2)$ theory, and $k=3$ for the $(A_1,A_3)$ theory.}
of $k$, such that the central charge \eqref{cSLcoset} matches
the corresponding value of $\hat c \, (\text{handle})$ discussed below \eqref{chandleguess}.
It is not clear how to square this with \eqref{WA1twisted},
which suggests that $\CM_{\text{handle}}$ is made of $\CN=2$ minimal models;
however, both that and the present discussion strongly suggest that the 2d theory \eqref{MFghandles}
is a non-compact Gepner model similar to world-sheet theories of
\cite{Mukhi:1993zb,Ghoshal:1995wm,Ooguri:1995wj,Giveon:1999zm,Giveon:1999px,Eguchi:2000tc,Lerche:2000uy,Eguchi:2004yi,Eguchi:2004ik,Ashok:2007ui,Ashok:2012qy},
as also corroborated by the structure of the elliptic genus and other indices.

Another potentially fruitful direction is to study the topological twist in the higher-dimensional string
constructions of non-Lagrangian theories which, among other things, provide a relation to vertex operator algebras \cite{Dedushenko:2017tdw}.
In the correspondence between 4-manifolds and vertex operator algebras (VOAs),
the topological correlators \eqref{generalOOO} on $M_4$ are identified with chiral correlators on a Riemann surface $\Sigma$
which is determined by a choice of 4d $\CN=2$ theory and which we already encountered in \eqref{6d4d2d}.
In particular, Argyres-Douglas theories are obtained by inserting irregular vertex operators on $\Sigma$,
and the results of this paper could be viewed as a step toward ``bootstrapping'' these operators;
once the right operators are identified, the calculation of 4-manifold invariants is reduced to a much
simpler problem of computing 2d correlators.


\acknowledgments{It is pleasure to thank
M.~Atiyah,
M.~Dedushenko,
T.~Dumitrescu,
D.~Gay,
A.~Giveon,
N.~Hitchin,
R.~Kirby,
D.~Kutasov,
D.~L\"ust,
M.~Mari\~no,
J.~Meier,
I.~Melnikov,
P.~Putrov,
L.~Schaposnik,
E.~Sharpe,
S.~Shatashvili,
E.~Silverstein,
J.~Song,
J.~Troost,
and K.~Ye
for useful discussions and comments.
This work is supported in part by the U.S. Department of Energy, Office of Science,
Office of High Energy Physics, under Award Number DE-SC0011632.}

%


\newpage

\bibliographystyle{JHEP}
\bibliography{classH}

\providecommand{\href}[2]{#2}\begingroup\raggedright\begin{thebibliography}{10}

\bibitem{Witten:1988ze}
E.~Witten, \emph{{Topological Quantum Field Theory}},
  \href{http://dx.doi.org/10.1007/BF01223371}{\emph{Commun.Math.Phys.} {\bf
  117} (1988) 353}.

\bibitem{Dedushenko:2017tdw}
M.~Dedushenko, S.~Gukov and P.~Putrov, \emph{{Vertex algebras and 4-manifold
  invariants}},  \href{https://arxiv.org/abs/1705.01645}{{\tt 1705.01645}}.

\bibitem{Argyres:1995jj}
P.~C. Argyres and M.~R. Douglas, \emph{{New phenomena in SU(3) supersymmetric
  gauge theory}},
  \href{http://dx.doi.org/10.1016/0550-3213(95)00281-V}{\emph{Nucl. Phys.} {\bf
  B448} (1995) 93--126}, [\href{https://arxiv.org/abs/hep-th/9505062}{{\tt
  hep-th/9505062}}].

\bibitem{Xie:2012hs}
D.~Xie, \emph{{General Argyres-Douglas Theory}},
  \href{http://dx.doi.org/10.1007/JHEP01(2013)100}{\emph{JHEP} {\bf 01} (2013)
  100}, [\href{https://arxiv.org/abs/1204.2270}{{\tt 1204.2270}}].

\bibitem{Aharony:2015oyb}
O.~Aharony and M.~Evtikhiev, \emph{{On four dimensional N = 3 superconformal
  theories}}, \href{http://dx.doi.org/10.1007/JHEP04(2016)040}{\emph{JHEP} {\bf
  04} (2016) 040}, [\href{https://arxiv.org/abs/1512.03524}{{\tt 1512.03524}}].

\bibitem{Garcia-Etxebarria:2015wns}
I.~García-Etxebarria and D.~Regalado, \emph{{$ \mathcal{N}=3 $ four
  dimensional field theories}},
  \href{http://dx.doi.org/10.1007/JHEP03(2016)083}{\emph{JHEP} {\bf 03} (2016)
  083}, [\href{https://arxiv.org/abs/1512.06434}{{\tt 1512.06434}}].

\bibitem{Aharony:2016kai}
O.~Aharony and Y.~Tachikawa, \emph{{S-folds and 4d N=3 superconformal field
  theories}}, \href{http://dx.doi.org/10.1007/JHEP06(2016)044}{\emph{JHEP} {\bf
  06} (2016) 044}, [\href{https://arxiv.org/abs/1602.08638}{{\tt 1602.08638}}].

\bibitem{Argyres:2017tmj}
P.~C. Argyres, Y.~Lu and M.~Martone, \emph{{Seiberg-Witten geometries for
  Coulomb branch chiral rings which are not freely generated}},
  \href{https://arxiv.org/abs/1704.05110}{{\tt 1704.05110}}.

\bibitem{Benini:2010uu}
F.~Benini, Y.~Tachikawa and D.~Xie, \emph{{Mirrors of 3d Sicilian theories}},
  \href{http://dx.doi.org/10.1007/JHEP09(2010)063}{\emph{JHEP} {\bf 09} (2010)
  063}, [\href{https://arxiv.org/abs/1007.0992}{{\tt 1007.0992}}].

\bibitem{Bershadsky:1995vm}
M.~Bershadsky, A.~Johansen, V.~Sadov and C.~Vafa, \emph{{Topological reduction
  of 4-d SYM to 2-d sigma models}},
  \href{http://dx.doi.org/10.1016/0550-3213(95)00242-K}{\emph{Nucl. Phys.} {\bf
  B448} (1995) 166--186}, [\href{https://arxiv.org/abs/hep-th/9501096}{{\tt
  hep-th/9501096}}].

\bibitem{Harvey:1995tg}
J.~A. Harvey, G.~W. Moore and A.~Strominger, \emph{{Reducing S duality to T
  duality}}, \href{http://dx.doi.org/10.1103/PhysRevD.52.7161}{\emph{Phys.
  Rev.} {\bf D52} (1995) 7161--7167},
  [\href{https://arxiv.org/abs/hep-th/9501022}{{\tt hep-th/9501022}}].

\bibitem{Lozano:1999us}
C.~Lozano and M.~Marino, \emph{{Donaldson invariants of product ruled surfaces
  and two-dimensional gauge theories}},
  \href{http://dx.doi.org/10.1007/s002200100442}{\emph{Commun. Math. Phys.}
  {\bf 220} (2001) 231--261}, [\href{https://arxiv.org/abs/hep-th/9907165}{{\tt
  hep-th/9907165}}].

\bibitem{Losev:1999tu}
A.~Losev, N.~Nekrasov and S.~L. Shatashvili, \emph{{Freckled instantons in
  two-dimensions and four-dimensions}},
  \href{http://dx.doi.org/10.1088/0264-9381/17/5/327}{\emph{Class. Quant.
  Grav.} {\bf 17} (2000) 1181--1187},
  [\href{https://arxiv.org/abs/hep-th/9911099}{{\tt hep-th/9911099}}].

\bibitem{Maldacena:2000mw}
J.~M. Maldacena and C.~Nunez, \emph{{Supergravity description of field theories
  on curved manifolds and a no go theorem}},
  \href{http://dx.doi.org/10.1142/S0217751X01003935,
  10.1142/S0217751X01003937}{\emph{Int. J. Mod. Phys.} {\bf A16} (2001)
  822--855}, [\href{https://arxiv.org/abs/hep-th/0007018}{{\tt
  hep-th/0007018}}].

\bibitem{KW}
A.~Kapustin and E.~Witten, \emph{{Electric-Magnetic Duality And The Geometric
  Langlands Program}}, {\emph{Commun.Num.Theor.Phys.} {\bf 1} (2007) 1--236},
  [\href{https://arxiv.org/abs/hep-th/0604151}{{\tt hep-th/0604151}}].

\bibitem{MR3590351}
D.~Gay and R.~Kirby, \emph{Trisecting 4-manifolds},
  \href{http://dx.doi.org/10.2140/gt.2016.20.3097}{\emph{Geom. Topol.} {\bf 20}
  (2016) 3097--3132}.

\bibitem{MR2222356}
P.~Ozsv\'ath and Z.~Szab\'o, \emph{Holomorphic triangles and invariants for
  smooth four-manifolds},
  \href{http://dx.doi.org/10.1016/j.aim.2005.03.014}{\emph{Adv. Math.} {\bf
  202} (2006) 326--400}.

\bibitem{Gukov:2007ck}
S.~Gukov, \emph{{Gauge theory and knot homologies}},
  \href{http://dx.doi.org/10.1002/prop.200610385}{\emph{Fortsch.Phys.} {\bf 55}
  (2007) 473--490}, [\href{https://arxiv.org/abs/0706.2369}{{\tt 0706.2369}}].

\bibitem{Gadde:2013sca}
A.~Gadde, S.~Gukov and P.~Putrov, \emph{Fivebranes and 4-manifolds},  in
  \emph{Arbeitstagung {B}onn 2013}, vol.~319 of \emph{Progr. Math.},
  pp.~155--245.
\newblock Birkh\"auser/Springer, Cham, 2016.
\newblock \href{https://arxiv.org/abs/1306.4320}{{\tt 1306.4320}}.

\bibitem{MR3248065}
D.~Baraglia and L.~P. Schaposnik, \emph{Higgs bundles and {$(A,B,A)$}-branes},
  \href{http://dx.doi.org/10.1007/s00220-014-2053-6}{\emph{Comm. Math. Phys.}
  {\bf 331} (2014) 1271--1300}.

\bibitem{Apol}
S.~Gukov, \emph{Three-dimensional quantum gravity, chern-simons theory, and the
  {$A$}-polynomial}, {\emph{Commun. Math. Phys.} {\bf 255} (2005) 577--627},
  [\href{https://arxiv.org/abs/hep-th/0306165}{{\tt hep-th/0306165}}].

\bibitem{MR1403918}
M.~Kontsevich, \emph{Homological algebra of mirror symmetry},  in
  \emph{Proceedings of the {I}nternational {C}ongress of {M}athematicians,
  {V}ol.\ 1, 2 ({Z}\"urich, 1994)}, pp.~120--139, Birkh\"auser, Basel, 1995.

\bibitem{MR1633036}
A.~Polishchuk and E.~Zaslow, \emph{Categorical mirror symmetry: the elliptic
  curve}, \href{http://dx.doi.org/10.4310/ATMP.1998.v2.n2.a9}{\emph{Adv. Theor.
  Math. Phys.} {\bf 2} (1998) 443--470}.

\bibitem{Gadde:2013wq}
A.~Gadde, S.~Gukov and P.~Putrov, \emph{{Walls, Lines, and Spectral Dualities
  in 3d Gauge Theories}},
  \href{http://dx.doi.org/10.1007/JHEP05(2014)047}{\emph{JHEP} {\bf 05} (2014)
  047}, [\href{https://arxiv.org/abs/1302.0015}{{\tt 1302.0015}}].

\bibitem{Chung:2014qpa}
H.-J. Chung, T.~Dimofte, S.~Gukov and P.~Sulkowski, \emph{{3d-3d Correspondence
  Revisited}},  \href{https://arxiv.org/abs/1405.3663}{{\tt 1405.3663}}.

\bibitem{Gukov:2016njj}
S.~Gukov, M.~Marino and P.~Putrov, \emph{{Resurgence in complex Chern-Simons
  theory}},  \href{https://arxiv.org/abs/1605.07615}{{\tt 1605.07615}}.

\bibitem{Dimofte:2011jd}
T.~Dimofte and S.~Gukov, \emph{{Chern-Simons Theory and S-duality}},
  \href{http://dx.doi.org/10.1007/JHEP05(2013)109}{\emph{JHEP} {\bf 05} (2013)
  109}, [\href{https://arxiv.org/abs/1106.4550}{{\tt 1106.4550}}].

\bibitem{Putrov:2015jpa}
P.~Putrov, J.~Song and W.~Yan, \emph{{(0,4) dualities}},
  \href{http://dx.doi.org/10.1007/JHEP03(2016)185}{\emph{JHEP} {\bf 03} (2016)
  185}, [\href{https://arxiv.org/abs/1505.07110}{{\tt 1505.07110}}].

\bibitem{Hitchin:1986vp}
N.~J. Hitchin, \emph{{The Self-duality equations on a Riemann surface}},
  \href{http://dx.doi.org/10.1112/plms/s3-55.1.59}{\emph{Proc.Lond.Math.Soc.}
  {\bf 55} (1987) 59--131}.

\bibitem{Nekrasov:2014xaa}
N.~A. Nekrasov and S.~L. Shatashvili, \emph{{Bethe/Gauge correspondence on
  curved spaces}}, \href{http://dx.doi.org/10.1007/JHEP01(2015)100}{\emph{JHEP}
  {\bf 01} (2015) 100}, [\href{https://arxiv.org/abs/1405.6046}{{\tt
  1405.6046}}].

\bibitem{Witten:1995gf}
E.~Witten, \emph{{On S duality in Abelian gauge theory}},
  \href{http://dx.doi.org/10.1007/BF01671570}{\emph{Selecta Math.} {\bf 1}
  (1995) 383}, [\href{https://arxiv.org/abs/hep-th/9505186}{{\tt
  hep-th/9505186}}].

\bibitem{Argyres:2016xmc}
P.~Argyres, M.~Lotito, Y.~Lü and M.~Martone, \emph{{Geometric constraints on
  the space of N=2 SCFTs III: enhanced Coulomb branches and central charges}},
  \href{https://arxiv.org/abs/1609.04404}{{\tt 1609.04404}}.

\bibitem{Lerche:1989cs}
W.~Lerche, D.~Lust and N.~P. Warner, \emph{{Duality Symmetries in $N=2$
  Landau-ginzburg Models}},
  \href{http://dx.doi.org/10.1016/0370-2693(89)90686-2}{\emph{Phys. Lett.} {\bf
  B231} (1989) 417--424}.

\bibitem{Chun:1991js}
E.~J. Chun, J.~Lauer and H.~P. Nilles, \emph{{Equivalence of Z(N) orbifolds and
  Landau-Ginzburg models}},
  \href{http://dx.doi.org/10.1142/S0217751X9200096X}{\emph{Int. J. Mod. Phys.}
  {\bf A7} (1992) 2175--2192}.

\bibitem{Narain:1986qm}
K.~S. Narain, M.~H. Sarmadi and C.~Vafa, \emph{{Asymmetric Orbifolds}},
  \href{http://dx.doi.org/10.1016/0550-3213(87)90228-8}{\emph{Nucl. Phys.} {\bf
  B288} (1987) 551}.

\bibitem{Brunner:2004mt}
I.~Brunner, M.~Herbst, W.~Lerche and J.~Walcher, \emph{{Matrix factorizations
  and mirror symmetry: The Cubic curve}},
  \href{http://dx.doi.org/10.1088/1126-6708/2006/11/006}{\emph{JHEP} {\bf 11}
  (2006) 006}, [\href{https://arxiv.org/abs/hep-th/0408243}{{\tt
  hep-th/0408243}}].

\bibitem{Herbst:2006nn}
M.~Herbst, W.~Lerche and D.~Nemeschansky, \emph{{Instanton geometry and quantum
  A-infinity structure on the elliptic curve}},
  \href{https://arxiv.org/abs/hep-th/0603085}{{\tt hep-th/0603085}}.

\bibitem{Verlinde:1995mz}
E.~P. Verlinde, \emph{{Global aspects of electric - magnetic duality}},
  \href{http://dx.doi.org/10.1016/0550-3213(95)00431-Q}{\emph{Nucl. Phys.} {\bf
  B455} (1995) 211--228}, [\href{https://arxiv.org/abs/hep-th/9506011}{{\tt
  hep-th/9506011}}].

\bibitem{Gadde:2013ftv}
A.~Gadde and S.~Gukov, \emph{{2d Index and Surface operators}},
  \href{http://dx.doi.org/10.1007/JHEP03(2014)080}{\emph{JHEP} {\bf 03} (2014)
  080}, [\href{https://arxiv.org/abs/1305.0266}{{\tt 1305.0266}}].

\bibitem{Shapere:2008zf}
A.~D. Shapere and Y.~Tachikawa, \emph{{Central charges of N=2 superconformal
  field theories in four dimensions}},
  \href{http://dx.doi.org/10.1088/1126-6708/2008/09/109}{\emph{JHEP} {\bf 09}
  (2008) 109}, [\href{https://arxiv.org/abs/0804.1957}{{\tt 0804.1957}}].

\bibitem{Hofman:2008ar}
D.~M. Hofman and J.~Maldacena, \emph{{Conformal collider physics: Energy and
  charge correlations}},
  \href{http://dx.doi.org/10.1088/1126-6708/2008/05/012}{\emph{JHEP} {\bf 05}
  (2008) 012}, [\href{https://arxiv.org/abs/0803.1467}{{\tt 0803.1467}}].

\bibitem{MR1066174}
S.~K. Donaldson, \emph{Polynomial invariants for smooth four-manifolds},
  \href{http://dx.doi.org/10.1016/0040-9383(90)90001-Z}{\emph{Topology} {\bf
  29} (1990) 257--315}.

\bibitem{Marino:1998tb}
M.~Marino, G.~W. Moore and G.~Peradze, \emph{{Superconformal invariance and the
  geography of four manifolds}},
  \href{http://dx.doi.org/10.1007/s002200050694}{\emph{Commun. Math. Phys.}
  {\bf 205} (1999) 691--735}, [\href{https://arxiv.org/abs/hep-th/9812055}{{\tt
  hep-th/9812055}}].

\bibitem{Faddeev:1985iz}
L.~D. Faddeev and S.~L. Shatashvili, \emph{{Algebraic and Hamiltonian Methods
  in the Theory of Nonabelian Anomalies}},
  \href{http://dx.doi.org/10.1007/BF01018976}{\emph{Theor. Math. Phys.} {\bf
  60} (1985) 770--778}.

\bibitem{AlvarezGaume:1983ig}
L.~Alvarez-Gaume and E.~Witten, \emph{{Gravitational Anomalies}},
  \href{http://dx.doi.org/10.1016/0550-3213(84)90066-X}{\emph{Nucl. Phys.} {\bf
  B234} (1984) 269}.

\bibitem{Hori:2003ic}
K.~Hori, S.~Katz, A.~Klemm, R.~Pandharipande, R.~Thomas, C.~Vafa et~al.,
  \emph{{Mirror symmetry}}, vol.~1 of \emph{Clay mathematics monographs}.
\newblock AMS, Providence, USA, 2003.

\bibitem{Argyres:2007tq}
P.~C. Argyres and J.~R. Wittig, \emph{{Infinite coupling duals of N=2 gauge
  theories and new rank 1 superconformal field theories}},
  \href{http://dx.doi.org/10.1088/1126-6708/2008/01/074}{\emph{JHEP} {\bf 01}
  (2008) 074}, [\href{https://arxiv.org/abs/0712.2028}{{\tt 0712.2028}}].

\bibitem{Witten:1994ev}
E.~Witten, \emph{{Supersymmetric Yang-Mills theory on a four manifold}},
  \href{http://dx.doi.org/10.1063/1.530745}{\emph{J. Math. Phys.} {\bf 35}
  (1994) 5101--5135}, [\href{https://arxiv.org/abs/hep-th/9403195}{{\tt
  hep-th/9403195}}].

\bibitem{Johansen:1994aw}
A.~Johansen, \emph{{Twisting of $N=1$ SUSY gauge theories and heterotic
  topological theories}},
  \href{http://dx.doi.org/10.1142/S0217751X9500200X}{\emph{Int. J. Mod. Phys.}
  {\bf A10} (1995) 4325--4358},
  [\href{https://arxiv.org/abs/hep-th/9403017}{{\tt hep-th/9403017}}].

\bibitem{Losev:1997tp}
A.~Losev, N.~Nekrasov and S.~L. Shatashvili, \emph{{Issues in topological gauge
  theory}}, \href{http://dx.doi.org/10.1016/S0550-3213(98)00628-2}{\emph{Nucl.
  Phys.} {\bf B534} (1998) 549--611},
  [\href{https://arxiv.org/abs/hep-th/9711108}{{\tt hep-th/9711108}}].

\bibitem{Vafa:1990mu}
C.~Vafa, \emph{{Topological Landau-Ginzburg models}},
  \href{http://dx.doi.org/10.1142/S0217732391000324}{\emph{Mod. Phys. Lett.}
  {\bf A6} (1991) 337--346}.

\bibitem{Melnikov:2005tk}
I.~V. Melnikov and M.~R. Plesser, \emph{{A-model correlators from the Coulomb
  branch}},  \href{https://arxiv.org/abs/hep-th/0507187}{{\tt hep-th/0507187}}.

\bibitem{Nekrasov:2009uh}
N.~A. Nekrasov and S.~L. Shatashvili, \emph{{Supersymmetric vacua and Bethe
  ansatz}},
  \href{http://dx.doi.org/10.1016/j.nuclphysbps.2009.07.047}{\emph{Nucl.Phys.P%
roc.Suppl.} {\bf 192-193} (2009) 91--112},
  [\href{https://arxiv.org/abs/0901.4744}{{\tt 0901.4744}}].

\bibitem{Tong:2006pa}
D.~Tong, \emph{{Superconformal vortex strings}},
  \href{http://dx.doi.org/10.1088/1126-6708/2006/12/051}{\emph{JHEP} {\bf 12}
  (2006) 051}, [\href{https://arxiv.org/abs/hep-th/0610214}{{\tt
  hep-th/0610214}}].

\bibitem{Gaiotto:2013sma}
D.~Gaiotto, S.~Gukov and N.~Seiberg, \emph{{Surface Defects and Resolvents}},
  \href{https://arxiv.org/abs/1307.2578}{{\tt 1307.2578}}.

\bibitem{Gukov:2016gkn}
S.~Gukov, P.~Putrov and C.~Vafa, \emph{{Fivebranes and 3-manifold homology}},
  \href{https://arxiv.org/abs/1602.05302}{{\tt 1602.05302}}.

\bibitem{Nishinaka:2016hbw}
T.~Nishinaka and Y.~Tachikawa, \emph{{On 4d rank-one $ \mathcal{N}=3 $
  superconformal field theories}},
  \href{http://dx.doi.org/10.1007/JHEP09(2016)116}{\emph{JHEP} {\bf 09} (2016)
  116}, [\href{https://arxiv.org/abs/1602.01503}{{\tt 1602.01503}}].

\bibitem{Benini:2013nda}
F.~Benini, R.~Eager, K.~Hori and Y.~Tachikawa, \emph{{Elliptic genera of
  two-dimensional N=2 gauge theories with rank-one gauge groups}},
  \href{https://arxiv.org/abs/1305.0533}{{\tt 1305.0533}}.

\bibitem{Honda:2015yha}
M.~Honda and Y.~Yoshida, \emph{{Supersymmetric index on $T^2 \times S^2$ and
  elliptic genus}},  \href{https://arxiv.org/abs/1504.04355}{{\tt 1504.04355}}.

\bibitem{Benini:2016hjo}
F.~Benini and A.~Zaffaroni, \emph{{Supersymmetric partition functions on
  Riemann surfaces}},  \href{https://arxiv.org/abs/1605.06120}{{\tt
  1605.06120}}.

\bibitem{Gadde:2015xta}
A.~Gadde, S.~S. Razamat and B.~Willett, \emph{{"Lagrangian" for a
  Non-Lagrangian Field Theory with $\mathcal N=2$ Supersymmetry}},
  \href{http://dx.doi.org/10.1103/PhysRevLett.115.171604}{\emph{Phys. Rev.
  Lett.} {\bf 115} (2015) 171604},
  [\href{https://arxiv.org/abs/1505.05834}{{\tt 1505.05834}}].

\bibitem{Maruyoshi:2016tqk}
K.~Maruyoshi and J.~Song, \emph{{Enhancement of Supersymmetry via
  Renormalization Group Flow and the Superconformal Index}},
  \href{http://dx.doi.org/10.1103/PhysRevLett.118.151602}{\emph{Phys. Rev.
  Lett.} {\bf 118} (2017) 151602},
  [\href{https://arxiv.org/abs/1606.05632}{{\tt 1606.05632}}].

\bibitem{Maruyoshi:2016aim}
K.~Maruyoshi and J.~Song, \emph{{$ \mathcal{N}=1 $ deformations and RG flows of
  $ \mathcal{N}=2 $ SCFTs}},
  \href{http://dx.doi.org/10.1007/JHEP02(2017)075}{\emph{JHEP} {\bf 02} (2017)
  075}, [\href{https://arxiv.org/abs/1607.04281}{{\tt 1607.04281}}].

\bibitem{Agarwal:2016pjo}
P.~Agarwal, K.~Maruyoshi and J.~Song, \emph{{$ \mathcal{N} $ =1 Deformations
  and RG flows of $ \mathcal{N} $ =2 SCFTs, part II: non-principal
  deformations}}, \href{http://dx.doi.org/10.1007/JHEP12(2016)103,
  10.1007/JHEP04(2017)113}{\emph{JHEP} {\bf 12} (2016) 103},
  [\href{https://arxiv.org/abs/1610.05311}{{\tt 1610.05311}}].

\bibitem{Gadde:2015wta}
A.~Gadde, S.~S. Razamat and B.~Willett, \emph{{On the reduction of 4d $
  \mathcal{N}=1 $ theories on ${\mathbb{S}}^2 $}},
  \href{http://dx.doi.org/10.1007/JHEP11(2015)163}{\emph{JHEP} {\bf 11} (2015)
  163}, [\href{https://arxiv.org/abs/1506.08795}{{\tt 1506.08795}}].

\bibitem{Amariti:2017cyd}
A.~Amariti, L.~Cassia and S.~Penati, \emph{{Surveying 4d SCFTs twisted on
  Riemann surfaces}},
  \href{http://dx.doi.org/10.1007/JHEP06(2017)056}{\emph{JHEP} {\bf 06} (2017)
  056}, [\href{https://arxiv.org/abs/1703.08201}{{\tt 1703.08201}}].

\bibitem{Aharony:2016jki}
O.~Aharony, S.~S. Razamat, N.~Seiberg and B.~Willett, \emph{{The long flow to
  freedom}}, \href{http://dx.doi.org/10.1007/JHEP02(2017)056}{\emph{JHEP} {\bf
  02} (2017) 056}, [\href{https://arxiv.org/abs/1611.02763}{{\tt 1611.02763}}].

\bibitem{Fredrickson:2017yka}
L.~Fredrickson, D.~Pei, W.~Yan and K.~Ye, \emph{{Argyres-Douglas Theories,
  Chiral Algebras and Wild Hitchin Characters}},
  \href{https://arxiv.org/abs/1701.08782}{{\tt 1701.08782}}.

\bibitem{Gukov:2010sw}
S.~Gukov, \emph{{Quantization via Mirror Symmetry}}, {\emph{Jpn. J. Math.} {\bf
  6} (2011) 65--119}, [\href{https://arxiv.org/abs/1011.2218}{{\tt
  1011.2218}}].

\bibitem{Ganor:2014pha}
O.~J. Ganor, N.~P. Moore, H.-Y. Sun and N.~R. Torres-Chicon, \emph{{Janus
  configurations with SL(2,$\mathbb{Z}$)-duality twists, strings on mapping
  tori and a tridiagonal determinant formula}},
  \href{http://dx.doi.org/10.1007/JHEP07(2014)010}{\emph{JHEP} {\bf 07} (2014)
  010}, [\href{https://arxiv.org/abs/1403.2365}{{\tt 1403.2365}}].

\bibitem{Frenkel:2007tx}
E.~Frenkel and E.~Witten, \emph{{Geometric endoscopy and mirror symmetry}},
  \href{http://dx.doi.org/10.4310/CNTP.2008.v2.n1.a3}{\emph{Commun. Num. Theor.
  Phys.} {\bf 2} (2008) 113--283}, [\href{https://arxiv.org/abs/0710.5939}{{\tt
  0710.5939}}].

\bibitem{Gukov:2015sna}
S.~Gukov and D.~Pei, \emph{{Equivariant Verlinde formula from fivebranes and
  vortices}},  \href{https://arxiv.org/abs/1501.01310}{{\tt 1501.01310}}.

\bibitem{Mukhi:1993zb}
S.~Mukhi and C.~Vafa, \emph{{Two-dimensional black hole as a topological coset
  model of c = 1 string theory}},
  \href{http://dx.doi.org/10.1016/0550-3213(93)90094-6}{\emph{Nucl. Phys.} {\bf
  B407} (1993) 667--705}, [\href{https://arxiv.org/abs/hep-th/9301083}{{\tt
  hep-th/9301083}}].

\bibitem{Ghoshal:1995wm}
D.~Ghoshal and C.~Vafa, \emph{{C = 1 string as the topological theory of the
  conifold}}, \href{http://dx.doi.org/10.1016/0550-3213(95)00408-K}{\emph{Nucl.
  Phys.} {\bf B453} (1995) 121--128},
  [\href{https://arxiv.org/abs/hep-th/9506122}{{\tt hep-th/9506122}}].

\bibitem{Ooguri:1995wj}
H.~Ooguri and C.~Vafa, \emph{{Two-dimensional black hole and singularities of
  CY manifolds}},
  \href{http://dx.doi.org/10.1016/0550-3213(96)00008-9}{\emph{Nucl. Phys.} {\bf
  B463} (1996) 55--72}, [\href{https://arxiv.org/abs/hep-th/9511164}{{\tt
  hep-th/9511164}}].

\bibitem{Giveon:1999zm}
A.~Giveon, D.~Kutasov and O.~Pelc, \emph{{Holography for noncritical
  superstrings}},
  \href{http://dx.doi.org/10.1088/1126-6708/1999/10/035}{\emph{JHEP} {\bf 10}
  (1999) 035}, [\href{https://arxiv.org/abs/hep-th/9907178}{{\tt
  hep-th/9907178}}].

\bibitem{Giveon:1999px}
A.~Giveon and D.~Kutasov, \emph{{Little string theory in a double scaling
  limit}}, \href{http://dx.doi.org/10.1088/1126-6708/1999/10/034}{\emph{JHEP}
  {\bf 10} (1999) 034}, [\href{https://arxiv.org/abs/hep-th/9909110}{{\tt
  hep-th/9909110}}].

\bibitem{Eguchi:2000tc}
T.~Eguchi and Y.~Sugawara, \emph{{Modular invariance in superstring on
  Calabi-Yau n fold with ADE singularity}},
  \href{http://dx.doi.org/10.1016/S0550-3213(00)00150-4}{\emph{Nucl. Phys.}
  {\bf B577} (2000) 3--22}, [\href{https://arxiv.org/abs/hep-th/0002100}{{\tt
  hep-th/0002100}}].

\bibitem{Lerche:2000uy}
W.~Lerche, \emph{{On a boundary CFT description of nonperturbative N=2
  Yang-Mills theory}},  \href{https://arxiv.org/abs/hep-th/0006100}{{\tt
  hep-th/0006100}}.

\bibitem{Eguchi:2004yi}
T.~Eguchi and Y.~Sugawara, \emph{{SL(2,R) / U(1) supercoset and elliptic genera
  of noncompact Calabi-Yau manifolds}},
  \href{http://dx.doi.org/10.1088/1126-6708/2004/05/014}{\emph{JHEP} {\bf 05}
  (2004) 014}, [\href{https://arxiv.org/abs/hep-th/0403193}{{\tt
  hep-th/0403193}}].

\bibitem{Eguchi:2004ik}
T.~Eguchi and Y.~Sugawara, \emph{{Conifold type singularities, N=2 Liouville
  and SL(2:R)/U(1) theories}},
  \href{http://dx.doi.org/10.1088/1126-6708/2005/01/027}{\emph{JHEP} {\bf 01}
  (2005) 027}, [\href{https://arxiv.org/abs/hep-th/0411041}{{\tt
  hep-th/0411041}}].

\bibitem{Ashok:2007ui}
S.~K. Ashok, R.~Benichou and J.~Troost, \emph{{Non-compact Gepner Models,
  Landau-Ginzburg Orbifolds and Mirror Symmetry}},
  \href{http://dx.doi.org/10.1088/1126-6708/2008/01/050}{\emph{JHEP} {\bf 01}
  (2008) 050}, [\href{https://arxiv.org/abs/0710.1990}{{\tt 0710.1990}}].

\bibitem{Ashok:2012qy}
S.~K. Ashok and J.~Troost, \emph{{Elliptic Genera of Non-compact Gepner Models
  and Mirror Symmetry}},
  \href{http://dx.doi.org/10.1007/JHEP07(2012)005}{\emph{JHEP} {\bf 07} (2012)
  005}, [\href{https://arxiv.org/abs/1204.3802}{{\tt 1204.3802}}].

\bibitem{Hori:2001ax}
K.~Hori and A.~Kapustin, \emph{{Duality of the fermionic 2-D black hole and N=2
  liouville theory as mirror symmetry}},
  \href{http://dx.doi.org/10.1088/1126-6708/2001/08/045}{\emph{JHEP} {\bf 08}
  (2001) 045}, [\href{https://arxiv.org/abs/hep-th/0104202}{{\tt
  hep-th/0104202}}].

\end{thebibliography}\endgroup

\end{document}